\documentclass[12pt,a4paper]{article}

\usepackage[T1]{fontenc}
\usepackage[main=english]{babel}
\usepackage[margin=1in]{geometry}

\usepackage{graphicx}
\usepackage{amsmath}
\usepackage{amssymb}
\usepackage{amsthm}
\usepackage{booktabs}
\usepackage{url}
\usepackage{longtable}
\usepackage[figuresright]{rotating}
\usepackage{multirow}
\usepackage{iftex}
\usepackage{libertinus}
\usepackage{microtype} 
\usepackage{tabularx}
\usepackage{array}
\usepackage{float}
\usepackage{caption}
\usepackage{subcaption}
\usepackage{algorithm}
\usepackage{algpseudocode}
\usepackage{underscore}
\usepackage{ragged2e}
\usepackage{placeins}
\usepackage{natbib}

\ifpdf
    \usepackage[plainpages=false,
        hypertexnames=false,
        pdfpagelabels,
        bookmarksnumbered,
        bookmarks=true,
        colorlinks=true,
        linkcolor=black,
        citecolor=black,
        filecolor=black,
        urlcolor=black,
        pdftex,
        unicode]{hyperref}
\else
    \usepackage{hyperref}
\fi

\newcommand{\source}[1]{%
  \par\smallskip
  \begingroup
    \setlength{\parindent}{0pt}%
    \begin{minipage}{\linewidth}%
      \setlength{\parindent}{0pt}%
      \footnotesize\justifying\noindent
      \textit{Note:}~#1%
    \end{minipage}%
  \endgroup
  \par
}

\renewcommand{\topfraction}{0.9}
\renewcommand{\bottomfraction}{0.8}
\renewcommand{\floatpagefraction}{0.7}

\begin{document}

\title{Harvesting the Volatility Risk Premium: A Learning-to-Rank Approach}
\author{Maciej Wysocki$^{1}$\\[0.4em]
  \small $^{1}$Department of Quantitative Finance and Machine Learning,\\
  \small Faculty of Economic Sciences, University of Warsaw,\\
  \small Quantitative Finance Research Group}
\date{}
\maketitle

\begin{quote}\small
  \textbf{Abstract.} This paper develops the first end-to-end application of cross-sectional learning-to-rank to the S\&P 500 weekly options (SPXW) zero-day-to-expiration surface, integrated with margin-aware position sizing, an abstention rule driven by model uncertainty, and a strict out-of-time integrity check. A LightGBM LambdaRank ranker scores a daily nine-strategy cross-section composed of eight delta-targeted short-put positions and a \textit{SKIP} candidate, trained against a path-aware Sortino-on-bars label computed at one-minute resolution. The framework is evaluated under index-option margin requirements, a tiered fee schedule, and bid-to-mid execution assumptions across a four-window walk-forward over 2021--2024 and a strictly held-out 2025 out-of-time slice. Seven sizing methods produce out-of-time annualized Sharpe ratios between $4.31$ and $5.76$, with the headline method reaching a Probabilistic Sharpe Ratio of $0.964$ and a sample-period maximum drawdown of $-2.28\%$, on a single hold-out year against a walk-forward range of $1.90$ to $3.11$. Out of time, every method exceeds three passive benchmarks (CBOE PUT, CBOE WPUT, SPX buy-and-hold) by at least $3.84$ in Sharpe ratio and five internal selection baselines by at least $3.69$. A two-by-two ablation of the confidence gate against the tail-risk features places $5.05$ of the $5.59$ out-of-time Sharpe gap over the CBOE PUT with the ranker and the selection layer, the two risk controls adding $0.54$ between them. On walk-forward, where the gate binds, neither control comes close to the headline alone and their interaction supplies most of the result. A fifteen-group feature ablation shows that removing the multiplicative regime interactions collapses walk-forward statistical confidence.
\end{quote}
\noindent\textbf{Funding:} This research was supported by IDUB University of Warsaw [grant number 501-D124-20-0004410].

\section{Introduction}\label{sec:introduction}
The volatility risk premium (VRP) in S\&P 500 index options is one of the most
persistent empirical regularities in the literature on equity-index
derivatives. \citet{bondarenko_why_2014} reports that historical SPX put prices
were structurally overpriced relative to the realized payoff distribution, with
average excess returns to put buyers strongly negative across the moneyness
range over an August 1987 to December 2000 sample.
\citet{bondarenko_historical_2019} measures the same wedge in volatility terms.
From 1990 to 2018 the VIX averaged $19.3\%$ against $15.1\%$ for the
subsequently realized one-month volatility of the index, a difference of $4.2$
annualized percentage points.
\citet{carr_variance_2009} document the same wedge through synthetic
variance-swap returns. The long-variance position earns a
statistically significant negative average return at one-month horizons.
\citet{bakshi_delta_hedged_2003} confirm the sign by a third route. A long
option position that is dynamically hedged against moves in the underlying
earns a negative average gain, and the shortfall widens with the position's
vega. The
standard interpretation is that option buyers pay an insurance premium against
the catastrophic-loss tail of equity-index movements, with the seller
absorbing that tail risk in exchange for the premium. Passive index harvesters
such as the CBOE PutWrite Index (PUT) and the CBOE Weekly PutWrite Index (WPUT)
capture this premium through a static short-put rule and have produced positive
long-horizon returns. These passive harvesters have a structural limitation, since the position is held
continuously regardless of the prevailing market regime, so capital is
allocated identically on days when the per-trade premium is wide and on days
when it is narrow, and the same fixed exposure is in force during the worst
stress windows. The research question of this paper is whether a more selective
harvester, based on cross-sectional ranking of candidate strategies rather than
a static rule, can extract a larger fraction of the available premium without
incurring the stress-window losses that limit the passive baselines.

The empirical literature on harvesting the volatility risk premium through
option-writing strategies has paid limited attention to two issues that
determine whether a published Sharpe ratio is realizable in deployment. The
first issue is the practical implementation cost.
\citet{santa-clara_option_2009} report that margin requirements impose a
binding limit on short-volatility positions and force liquidations precisely
when realized losses are concentrated, and \citet{do_profitability_2016} and
\citet{hong_profitability_2018} document that the headline Sharpe ratios of
canonical short-volatility strategies fall substantially once explicit bid-ask
spreads and margin rules are included. A related concern is that mean-variance
optimization is poorly suited to the negatively-skewed, fat-tailed payoff
distribution of short-option positions. \citet{faias_optimal_2017} argue that
expected-utility frameworks such as fractional Kelly sizing are required to
allocate capital under these payoffs.

The second issue is methodological. The machine-learning literature on
option-strategy selection is dominated by point-prediction frameworks that
treat the problem as one of forecasting absolute returns, which is a
noise-dominated task at the daily horizon. A more natural formulation of this
task is learning-to-rank, in which a scoring function is trained to order
candidates rather than to forecast their absolute returns.
\citet{duan_learning_2021} and \citet{poh_building_2021} report that
learning-to-rank algorithms applied to the cross-section of equity returns
improve out-of-sample ranking accuracy over absolute-return regressors, and
\citet{linger_unifying_2025} and \citet{barak_deep_2025} extend the framework
to long-short portfolio construction with tailored loss functions. The
learning-to-rank literature has, however, developed almost entirely on equity
cross-sections rather than on the option surface, and the few applications to
options have not been integrated with margin-aware sizing, regime-conditional
feature engineering, or out-of-time hold-out evaluation.

The framework developed in this paper applies a LightGBM LambdaRank ranker
\citep{burges_ranknet_2010, ke_lightgbm_2017} to a cross-section of nine
candidate strategies in the S\&P 500 weekly options (SPXW)
zero-day-to-expiration (0DTE) market. Eight of the candidates are
delta-targeted short-put positions spaced across the put wing of the
volatility surface, and the ninth is a \textit{SKIP} candidate that allows the
ranker to express a zero-position day. A confidence gate sits between the
ranker's daily pick and the executed position and abstains on days when the
ranker has low conviction in its top candidate. Seven independent sizing
methods are evaluated in parallel under explicit margin, fee, and
execution-cost assumptions. The evaluation comprises a four-window
expanding-cycle walk-forward over 2021--2024, followed by a strictly held-out
2025 out-of-time slice that is never used during training, hyperparameter
search, or model selection. To the best of our knowledge, this is the first
end-to-end integration of cross-sectional learning-to-rank on the SPXW 0DTE
option surface with margin-aware sizing, abstention under model uncertainty,
and a strict out-of-time integrity check.

The empirical claims of this paper are organized around three hypotheses.

\begin{itemize}
    \item \textbf{Hypothesis 1 (H1):} A LambdaRank ranker over a cross-section of delta-targeted SPXW short-put strategies, trained on a path-aware Sortino-on-bars label and combined with an abstention gate, produces risk-adjusted returns that exceed those of passive short-volatility benchmarks (CBOE PUT, CBOE WPUT, SPX buy-and-hold) and a panel of internal selection baselines, and the advantage survives explicit margin requirements, fee schedules, and slippage stress on a 2025 out-of-time hold-out.
    \item \textbf{Hypothesis 2 (H2):} The strategy's return is regime-conditional, rising with the level of realized volatility and peaking in the mid range of implied volatility, and the realized-volatility component of this conditioning transfers from the walk-forward training period (2021--2024) to the out-of-time hold-out (2025).
    \item \textbf{Hypothesis 3 (H3):} Engineered multiplicative interactions of per-strategy exposures with volatility-state regime variables carry a substantial part of the walk-forward ranking signal. Removing these interaction features collapses walk-forward statistical confidence.
\end{itemize}

On the 2025 out-of-time slice, the seven sizing methods produce annualized
Sharpe ratios between $4.31$ and $5.76$, exceeding every external benchmark
and internal selection baseline by at least $3.69$ in Sharpe ratio, consistent
with H1. The headline strategy's sample-period maximum drawdown is $-2.28\%$.
Regime-conditional annualized return rises monotonically across
realized-volatility terciles on the out-of-time slice for six of the seven
methods, consistent with H2, while the Sharpe ratio does not follow the same
ordering. A
fifteen-group feature ablation shows that removing the multiplicative
regime-conditional interactions collapses walk-forward statistical confidence,
consistent with H3, while out-of-time Sharpe degrades only modestly. The
two-by-two ablation of the confidence gate against the tail-risk features
places almost all of the out-of-time Sharpe gap over the CBOE PUT benchmark
with the ranker and the selection layer, the two risk controls adding $0.54$ between them. On the walk-forward slice, where the gate binds,
the features alone lower Sharpe, the gate alone adds little, and the two
together account for most of what is recorded there.

The remainder of this paper is organized as follows.
Section~\ref{sec:literature} surveys the related literature on the volatility
risk premium, on machine-learning applications to option markets, and on
learning-to-rank in cross-sectional asset selection.
Section~\ref{sec:methodology} sets out the LambdaRank model, the path-aware
ranking target, the per-window feature selection pipeline, the confidence gate, and the
position-sizing layer.
Section~\ref{sec:data} describes the SPXW options dataset and the feature
engineering process. Section~\ref{sec:empirical} reports the empirical results
across the walk-forward and out-of-time slices. Section~\ref{sec:robustness}
presents the sensitivity, ablation, and execution-drag analysis.
Section~\ref{sec:mechanism} decomposes the headline result into its component
contributions and discusses the economic interpretation.
Section~\ref{sec:conclusions} summarizes the findings against the three
hypotheses and discusses the paper's contributions and limitations.

\section{Literature Review}\label{sec:literature}
The profitability of systematic short S\&P 500 put strategies is rooted in the
``overpriced puts puzzle'' and the variance risk premium.
\citet{bondarenko_why_2014} documents that historical SPX put prices were
significantly overpriced and produced large negative returns for buyers, a
pattern that standard equilibrium models fail to reproduce. The overpricing is
substantially explained by a persistently negative variance risk premium, that
is, investors pay a premium to insure against increases in market volatility
\citep{carr_variance_2009}. \citet{patel_cash-secured_2024} empirically
associate the excess returns of cash-secured put-write (PUTW) strategies with
this VRP and interpret the premium as compensation for bearing disaster risk.
\citet{broadie_understanding_2009} argue, in contrast, that the apparent
mispricing is largely attributable to finite-sample noise and jump risk, and
that explicit modeling of extreme tail risk is essential for any
short-volatility strategy.

Although passive benchmark indices such as the CBOE PUT have historically
captured the VRP and generated long-term positive returns, as documented by
\citet{black_35-year_2022} and \citet{ungar_cash-secured_2009}, these static
rules are capital-inefficient and produce uneven returns.
\citet{burrello_applications_2024} report that dynamically adjusting strike
prices is associated with more consistent income and a more stable risk profile
than fixed-strike short-put strategies.

Capturing the VRP requires modeling a dynamic implied volatility surface.
\citet{cont_dynamics_2002} show that IV surfaces evolve through multi-factor
dynamics in level, skew, and convexity, which motivates models capable of
representing nonlinear interactions rather than purely static shifts. The
surface also changes substantially during crises:
\citet{constantinides_puzzle_2013} argue that severe tail risks, such as market
jumps and volatility spikes, account for the bulk of short-put pricing errors.
Consistent with this view, \citet{andersen_risk_2015} identify a left-tail jump
factor in the SPX option surface that is not explained by market volatility
alone but is essential for forecasting future risk premia. Furthermore,
\citet{da_fonseca_variance_2019} report that the VIX itself contains separate
variance and skew risk premiums.

Predictive signals are central to timing short-put exposures.
\citet{malkiel_option_2018} document that conditioning short-put entry on the
level of the VIX is associated with higher absolute returns and lower downside
risk. The shape of the volatility term structure is also informative:
\citet{vasquez_equity_2017} reports that its slope predicts the cross-section
of future option returns. \citet{sheu_effective_2011} show that adding
investor-sentiment proxies, such as the VIX, improves both volatility forecasts
and trading profitability. \citet{tannous_expected_2008} analyze theta decay
under jump-diffusion dynamics and find that deterministic theta curves are
misleading in realistic price paths.

An important limitation of the existing option-strategy literature is that it
typically abstracts from practical implementation constraints.
\citet{santa-clara_option_2009} show that margin requirements impose a binding
limit to arbitrage and force investors to liquidate positions at substantial
losses precisely during sharp market declines. \citet{do_profitability_2016}
and \citet{hong_profitability_2018} further report that the theoretical Sharpe
ratios of short-volatility strategies largely vanish once bid-ask spreads and
margin rules are taken into account.

Standard mean-variance optimization is inadequate for option portfolios.
\citet{faias_optimal_2017} argue that expected-utility frameworks, such as
Kelly-type sizing, are necessary to size positions dynamically and to
accommodate extreme tail risk. \citet{israelov_which_2017} show that sizing
positions to loss capacity under stress scenarios is a more informative risk
measure than return volatility alone. \citet{schwalbach_analysis_2018} argue
for 30-day OTM puts, but their fully collateralized constraint nearly
eliminates absolute returns, which is the capital inefficiency that
explicit margin modeling can address.

To navigate these changing market regimes, recent research has increasingly
turned to machine learning. \citet{gunnarsson_prediction_2024} report in their
systematic review that tree-based ensemble methods such as LightGBM often
outperform traditional econometric models for volatility prediction.
\citet{huang_novel_2020} show that shallow econometric models cannot capture
the non-stationary dynamics of option pricing, which motivates the use of
richer nonlinear architectures. Within supervised machine learning,
learning-to-rank is the formulation closest to the cross-sectional
strategy-selection problem studied in this paper.

Learning-to-rank (LTR) algorithms are well matched to cross-sectional asset
selection and outperform standard classification or regression on this task. \citet{duan_learning_2021} and
\citet{poh_building_2021} report that LTR models improve ranking accuracy
because they directly optimize the relative ordering of assets rather than
absolute return predictions. The advantage grows with the heterogeneity of the
input data, as documented by \citet{wu_momentum_2024} and
\citet{song_stock_2017}.

Extending LTR to long-short portfolio construction requires loss functions
designed for that purpose. \citet{linger_unifying_2025} and
\citet{zhang_constructing_2022} argue for tailored loss functions such as
ListFold and ListMLE-weighted, which give symmetric weight to the top and the
bottom of the ranking rather than only the top. Recent frameworks in
\citet{linger_enhancing_2025} and \citet{barak_deep_2025} combine LTR
rankings with dynamic portfolio weighting and sizing and report
higher risk-adjusted returns than each component in isolation. These integrated
LTR frameworks have, however, been developed and tested almost entirely on
equity cross-sections, not on the option surface.

The volatility risk premium has been studied extensively, and LTR algorithms
have performed well in equity applications, yet existing work has rarely
integrated these strands. Theoretical option-pricing studies typically abstract
from margin limits and execution constraints, while ML and LTR research has
paid little attention to the non-stationary volatility surface of S\&P 500
Index options. This paper addresses that gap by applying a LightGBM LambdaRank
model directly to the SPX option surface. The model ranks strategies on regime
features and is evaluated under explicit margin constraints, fee schedules, and
dynamic sizing rules, so that the reported returns are net of the modeled
frictions.

\section{Methodology}\label{sec:methodology}
The framework's purpose is to identify, on each rebalancing day $t$, the single
SPXW short-put strategy whose risk-adjusted forward return over the trading day
is expected to be highest. The framework does not attempt to predict each
candidate's absolute point return, which is a noise-dominated task in
short-horizon financial time series. The selection problem is instead
formulated as a learning-to-rank problem in the sense of
\citet{burges_ranknet_2010}. A scoring function is trained to order candidates
by relative quality on each day, not to estimate each candidate's expected
return on an absolute scale.

Let $\mathcal{S}$ denote the parameterized universe of nine candidate
strategies, defined in Section~\ref{sec:universe} as eight delta-targeted short
puts and a \textit{SKIP} option. At each trading day $t$, the candidate set
$C_t = \mathcal{S}$ is refreshed by re-resolving each delta target to its
closest-matching strike on the SPXW chain. A scoring function $f(\mathbf{x}_{s,
        t})$ assigns a real-valued relevance score to each $s \in C_t$ from the
candidate's feature vector $\mathbf{x}_{s, t}$. Capital is then allocated to
the highest-scoring candidate, $s^\star_t = \arg\max_{s \in C_t}
    f(\mathbf{x}_{s, t})$, with the abstention overlay detailed in
Section~\ref{sec:gate} acting between the score and the final position size.

\subsection{Universe of Options Subject to Ranking}\label{sec:universe}
On every trading day at 10:00 ET, the ranker faces a candidate set of nine
alternatives: eight short SPXW put positions, one per delta bucket, and a SKIP
option, which is the choice not to trade. The eight delta targets are
\begin{equation}
    \Delta^{\text{tgt}} \in \{0.05,\ 0.10,\ 0.15,\ 0.20,\ 0.25,\ 0.30,\ 0.40,\ 0.45\},
\end{equation}
spaced at $0.05$ through the out-of-the-money wing, with a single wider step between $0.30$ and $0.40$ as the strike approaches the at-the-money forward. The grid is constructed this way because the volatility risk premium expressed in SPX put prices is concentrated in the OTM wing. The model, therefore, needs finer resolution in the region where the strike-quality decision is most sensitive.

The candidate set is put-only. \citet{bondarenko_why_2014} and
\citet{israelov_klein_2016} document an asymmetric variance risk premium in SPX
index options: put writers earn a larger expected return per unit of risk than
call writers, because put buyers pay for crash insurance whose statistical-loss
profile differs from the upside-participation profile call buyers seek. The
carry profiles also differ. A short put benefits from the unconditional
positive drift of the index. A short call absorbs that drift, so calls expire
in the money more frequently than the probability implied by the option's own
volatility quote would suggest. We tested an all-options variant of the same architecture during the
trial phase, and the calls leg was a consistent drag on performance across
sizing methods. The presented configurations, therefore, restrict the candidate
set to puts.

The \textit{SKIP} option is a synthetic ninth candidate carrying a fixed score of zero, and zero gross profit and
loss by construction. \textit{SKIP} enables the model to abstain on days when
no candidate strike is more attractive than not trading at all. Including
\textit{SKIP} in the ranker turns the per-day decision from \emph{which strike
    to sell} into \emph{whether to sell at all and which one}, which is the
economically appropriate framing for a strategy whose edge is
regime-conditional rather than uniform. The \textit{SKIP} row enters the
LambdaRank label assignment with a fixed score of zero rather than a graded
outcome of its own. The model then learns to rank
\textit{SKIP} above unattractive strikes and below attractive ones.

The expiration choice is deterministic. Each day the universe resolver scans
the SPXW chain and selects the shortest available DTE. What that rule returns
depends on the listing calendar. SPXW carried Monday, Wednesday, and Friday
expirations from the start of the sample and added Tuesdays on 18 April 2022
and Thursdays on 11 May 2022 \citep{cboe2022spxwdaily}, so the rule returns a
same-session expiry on $62.0\%$ of trading days over 2018 to 2021, on $87.2\%$
in 2022, and on every trading day from 2023 onward. On the remaining days it
returns a one-day expiry, and no day in the sample resolves to a tenor longer
than one session. The walk-forward test years 2023 and 2024 and the 2025
out-of-time slice therefore consist entirely of same-session contracts, while
the earlier training windows and the 2021 test year mix same-session and
one-day expiries in roughly three to two.

Every listed SPXW strike for the selected expiration is eligible.
The strike-selection criterion uses the Black--Scholes model with the effective
federal funds rate (EFFR) as the risk-free rate. For each delta target
$\Delta^{\text{tgt}}$, the resolver picks the strike that minimizes
\begin{equation}
    \big| |\Delta^{\text{BS}}(K)| - \Delta^{\text{tgt}} \big|,
\end{equation}
where $\Delta^{\text{BS}}(K)$ is the Black--Scholes put delta evaluated at strike $K$ using the implied volatility recovered from the 10:00 ET option mid quote. Time to expiration $\tau$ is measured in trading time on the NYSE calendar. The same Black--Scholes model supplies the option Greeks consumed by the downstream feature pipeline and the position-sizing layer.

The selected position is held to expiration and closed against the official PM
cash settlement. The day's gross profit and loss for a short put position is
\begin{equation}
    \text{P\&L}_{\text{gross}} = Q \cdot \text{multiplier} \cdot \big[\, p_{\text{entry}} - \max(K - S_{\text{settle}}, 0)\,\big],
\end{equation}
where $p_{\text{entry}}$ is the per-share entry premium received, $Q$ is the contract count selected by the sizing layer, and the SPXW multiplier is 100.

\subsection{Ranking Target Construction}
The ranker is trained against a per-day per-strategy score that reflects both
realized premium collection and intraday drawdown stress. The score is
constructed in four steps: per-bar profit-and-loss, aggregation into a
Sortino-on-bars score, \textit{SKIP} injection, and ordinal binning into
LambdaRank grades.

For each strategy $s$ on each trade date $t$, the position is tracked at
one-minute resolution from 10:00 ET entry through PM settlement. The mark
series is
\begin{equation}
    \text{mark} = \big[\, p_{\text{entry}},\ a_1,\ a_2,\ \ldots,\ a_{N-1},\ \max(K - S_{\text{settle}},\, 0) \,\big],
\end{equation}
where $p_{\text{entry}}$ is the entry premium received, $a_i$ is the option ask price at bar $i$, and the terminal mark is replaced by the settlement intrinsic. Marking intermediate bars to the ask follows the cost-to-close convention for a short position. At any intraday point, the position is valued at the price one would pay to buy it back, which is the most conservative intraday valuation for a writer. The per-bar profit-and-loss is
\begin{equation}
    \Delta p_i = \text{mark}_{i-1} - \text{mark}_i,
\end{equation}
positive when the mark falls (favorable for the short) and negative when the mark rises. The bar series telescopes to the trade-level gross P\&L, $\sum_i \Delta p_i = p_{\text{entry}} - \max(K - S_{\text{settle}}, 0)$, so the bar-level decomposition does not change the trade-level outcome but does expose the intraday path.

The day's score is the ratio of trade-level gross P\&L to a path-aware downside
measure:
\begin{equation}
    s_{s,t} = \frac{\text{GrossPnL}_{s,t}}{\text{TDD}_{s,t} + \varepsilon}, \qquad \text{TDD}_{s,t} = \sqrt{\sum_i \min(\Delta p_i,\, 0)^2},
    \label{eq:score_sortino_bars}
\end{equation}
where $\text{GrossPnL}_{s,t} = \sum_i \Delta p_i$ is the per-trade gross profit-and-loss, $\text{TDD}_{s,t}$ is the trade downside deviation aggregated only over negative bar moves, and $\varepsilon = 10^{-8}$ guards the ratio numerically. The numerator rewards strategies that finish the day profitably. The denominator penalizes strategies that arrive at the same terminal payoff through a stressed intraday path. Two candidates with identical terminal P\&L but different drawdown profiles therefore receive different scores, so the model learns to rank strategies on realized path quality and not on terminal outcome alone.

For each trade date $t$, one \textit{SKIP} row is appended to the score panel
with $s_{0,t} = 0$ by construction. The fixed-zero score places \textit{SKIP}
in the second-lowest of the five grade bands in every training window. The
ranker therefore grades \textit{SKIP} above the worst-scoring decile of
candidates and below any candidate scoring above the fortieth percentile. The \textit{SKIP}
row is excluded from the percentile computation when the binning thresholds are
derived in the next step, so its synthetic score does not distort the empirical
CDF that defines the grade boundaries.

The continuous score is mapped to an integer grade $g_{s,t} \in \{0, 1, 2, 3,
    4\}$ through four absolute thresholds derived once per walk-forward training
window from the empirical score distribution of non-\textit{SKIP} candidates:
\begin{equation}
    \theta_p = \text{Quantile}_p\!\Big(\{\, s_{j,t}\,:\, j \neq 0,\ t \in \mathcal{T}_{\text{train}}\,\}\Big), \qquad p \in \{10, 40, 60, 90\}.
    \label{eq:thresholds}
\end{equation}
The grade assignment is
\begin{equation}
    g_{s,t} = \begin{cases}
        4 & \text{if } s_{s,t} > \theta_{90}                 \\
        3 & \text{if } \theta_{60} < s_{s,t} \le \theta_{90} \\
        2 & \text{if } \theta_{40} < s_{s,t} \le \theta_{60} \\
        1 & \text{if } \theta_{10} < s_{s,t} \le \theta_{40} \\
        0 & \text{if } s_{s,t} \le \theta_{10}.
    \end{cases}
    \label{eq:grade_assignment}
\end{equation}
The percentile spacing $\{10, 40, 60, 90\}$ produces target grade frequencies of \{10\%, 30\%, 20\%, 30\%, 10\%\} in expectation on the training slice. The distribution places most of its mass in the two intermediate grades and least at the tails, which matches the LambdaRank objective's emphasis on the top of the rank list. The thresholds are absolute and frozen at training time, so the same boundaries apply at inference regardless of the score distribution observed on any individual prediction day.

\subsection{Feature Selection Pipeline}\label{sec:feature-pipeline}
Market regimes exhibit material non-stationarity, and a feature that is
informative during a low-volatility regime can be uninformative during a stress
event. To address this, the framework retrains the LambdaRank model at every
walk-forward step and runs an independent feature selection pipeline inside
each training window, before hyperparameter tuning. Running the pipeline per
window prevents look-ahead leakage from later periods into earlier-window
selections, and the cross-window survival of features is itself reported in
Section~\ref{subsec:features} as a regime-stability diagnostic.

The pipeline (Algorithm \ref{alg:feature_selection}) has five stages. Stages
\textit{S1} and \textit{S2} are unsupervised data-quality filters. Stage
\textit{S3} is a univariate target-relevance filter scoped differently for
cross-sectional and per-strategy features. Stage \textit{S4} clusters
correlated features at the Spearman threshold $|\rho| \geq 0.85$, picks a
deterministic representative per cluster, and then enforces per-group survivor
caps that prevent any single economic block from dominating the feature set.
Stage \textit{S5} is a temporal-stability filter that re-runs stages
\textit{S1-S4} on three expanding sub-windows of the training data and keeps
features surviving in at least two of three sub-windows.

\begin{algorithm}[!ht]
    \caption{Per-window feature selection pipeline (S1--S5).}
    \label{alg:feature_selection}
    \begin{algorithmic}[1]
        \Require Training-window feature matrix $\mathbf{X}$, labels $\mathbf{y}$, scope map $\text{scope}: \mathcal{F} \to \{\text{CS}, \text{PS}\}$, group map $\text{group}: \mathcal{F} \to \mathcal{G}$
        \Ensure Selected feature subset $\mathcal{F}_{\text{final}}$
        \State $\mathcal{F} \gets$ candidate features (approximately 190)

        \Statex \textbf{S1. Null filter}
        \For{each $f \in \mathcal{F}$}
        \State drop $f$ if $\text{NullRate}_{\text{train, real}}(f) > 0.30$
        \EndFor

        \Statex \textbf{S2. Variance and quasi-constant filter}
        \For{each $f \in \mathcal{F}$}
        \State drop $f$ if $\text{Var}(f) < 10^{-6}$ \textbf{or} $\text{ModeFreq}(f) > 0.99$
        \EndFor

        \Statex \textbf{S3. Target-relevance filter}
        \For{each $f \in \mathcal{F}$}
        \If{$\text{scope}(f) = \text{CS}$}
        \State $\rho(f) \gets \text{Spearman}\big(f_t,\, \bar{y}_t\big)$ across training days
        \Else
        \State $\rho(f) \gets \operatorname{median}_t \text{Spearman}\big(f_{\cdot,\, t},\, y_{\cdot,\, t}\big)$ within day
        \EndIf
        \State drop $f$ if $|\rho(f)| < 0.05$
        \EndFor

        \Statex \textbf{S4. Correlation clustering and per-group caps}
        \State Build a within-scope Spearman correlation matrix; cluster features at $|\rho_{ij}| \geq 0.85$
        \For{each cluster $C$}
        \State keep the feature with the largest $|\rho(f)|$ from \textit{S3}, with ties broken by catalog-group priority, then by data quality, then by feature name
        \EndFor
        \For{each group $g \in \mathcal{G}$}
        \State cap at $c_g$: \texttt{position} 8, \texttt{vol\_surface} 4, \texttt{vix} 6, \texttt{calendar} 8, default 5; drop the lowest-$|\rho|$ representatives in over-cap groups
        \EndFor

        \Statex \textbf{S5. Stability filter}
        \State Partition the training window into $K = 3$ expanding-time folds
        \State Run stages \textit{S1-S4} on each fold; let $s(f) \in \{0, \ldots, K\}$ count fold survival
        \State drop $f$ if $s(f) < 2$

        \State \Return $\mathcal{F}_{\text{final}}$
    \end{algorithmic}
\end{algorithm}

Two design choices in the pipeline merit explicit comment because they shape
the headline result. First, the \textit{S3} relevance score is scoped
differently for cross-sectional and per-strategy features. Cross-sectional
features take one value per trading day, so their relevance is computed
cross-day against the day-mean label. Per-strategy features vary across the
eight delta-bucket candidates plus \textit{SKIP} within a single day, so their
relevance is computed as the median across training days of the within-day
Spearman rank correlation between $f$ and $y$. The per-strategy formulation
matches the within-day discriminative signal that the LambdaRank objective
exploits, and prevents cross-day variation from masking economically meaningful
within-day signal.

Second, the \textit{S4} per-group survivor caps prevent the largest catalog
group, the 53-feature implied-volatility surface, from displacing smaller but
economically distinct groups on raw relevance grounds. Without the caps, the
surface would be expected to supply most of the surviving features, leaving little room for
morning-session, intra-strategy, or entry-liquidity blocks that carry
information the surface alone does not. The cap values used are
\texttt{position} = 8, \texttt{vol\_surface} = 4, \texttt{vix} = 6,
\texttt{calendar} = 8, and a default of 5 in each remaining group. They are set
by the relative size of each catalog group and are not separately tuned.

The \textit{S5} stability filter operates inside one training window: it
partitions the training dates into three expanding-time folds, re-runs stages
\textit{S1-S4} on each, and retains a feature only if it survives in at least
two of the three folds. This is structurally different from the
cross-walk-forward feature-survival statistic. That statistic is computed
across the four walk-forward windows and the out-of-time window, while \textit{S5} is computed
inside each one before any other window is observed.

After the pipeline returns the final feature subset for a given walk-forward
window, hyperparameter tuning is run on that subset within the same window: 50
Optuna trials with the Tree-structured Parzen Estimator sampler
\citep{akiba_optuna_2019}, optimizing NDCG@1 on a 6-month inner-validation
slice held out from the end of the training window. The sequence of feature selection followed by
hyperparameter tuning is repeated independently for every walk-forward boundary, so each
window's final model is both feature-selected and hyperparameter-tuned only on
data that ends strictly before its prediction year.

\subsection{Model Architecture and Optimization}

The ranker is a LightGBM gradient-boosted decision-tree model trained under the
LambdaRank objective \citep{burges_ranknet_2010, ke_lightgbm_2017}. The
LambdaRank objective treats learning to rank as a pairwise problem. For each
pair of differently-graded candidates in a query group, the gradient applied to
that pair is scaled by the change in NDCG (Normalized Discounted Cumulative
Gain) that would result from swapping the two candidates' positions in the
predicted ranking. The resulting implicit cost function approximates direct
NDCG optimization without requiring a closed-form differentiable surrogate.
Tree-based gradient boosting frameworks have been reported to outperform
standard econometric baselines on short-horizon volatility and option-related
prediction tasks \citep{gunnarsson_prediction_2024}. This evidence motivates
the choice of LightGBM over linear or shallow-network alternatives for the
candidate-ranking problem at hand.

A subset of LightGBM parameters is held fixed throughout. The objective is
lambdarank. The evaluation metric is NDCG, reported at
$\text{NDCG@1}$ and $\text{NDCG@3}$. The truncation level restricts the gradient computation
to the top $5$ positions of each ranked group, so the optimization focuses on
the head of the rank list. The boosting type is gradient-boosted decision trees (GBDT).

The LambdaRank gain weights associated with the five grades are
\begin{equation}
    \texttt{label\_gain} = [\, 0,\ 1,\ 3,\ 7,\ 15 \,].
    \label{eq:label_gain}
\end{equation}

The geometric progression places super-linear emphasis on the top grade, so
that the ranker's loss is dominated by errors at the rank-1 position. The
per-day decision is a single-pick problem: only the top-ranked candidate is
acted on, so improvements that move a grade-3 candidate ahead of a grade-2
candidate at rank 5 contribute negligibly to the loss, while a single rank-1
misallocation is heavily penalized.

LightGBM's LambdaRank objective requires a query-group structure that
delineates the candidate sets used for pairwise ranking comparisons. One query
group is defined per trading day, so each group contains the day's nine
candidates: eight delta-targeted short puts and \textit{SKIP}. Grouping by trade date keeps the pairwise comparisons
within the cross-section of strategies competing for that day's allocation,
which is the economically appropriate unit of ranking. The model ranks
candidates within a day rather than across days, because the cross-day dimension
is dominated by regime-level noise rather than by within-day strike-quality
information.

The remaining parameters are tuned per walk-forward window using Optuna's
Tree-structured Parzen Estimator sampler \citep{akiba_optuna_2019}. The search
runs $50$ trials per window. The search space is reported in
Table~\ref{tab:hyperparameters}.

\begin{table}[!ht]
    \centering
    \caption{Hyperparameter search space for LightGBM LambdaRank.}
    \label{tab:hyperparameters}
    \begin{tabularx}{\textwidth}{@{} X c c @{}}
        \toprule
        \textbf{Hyperparameter} & \textbf{Search space} & \textbf{Sampling} \\
        \midrule
        Number of leaves & $[16,\ 256]$ & log \\
        Learning rate  & $[0.01,\ 0.10]$ & log \\
        Minimum samples per leaf & $[20,\ 200]$ & log \\
        Subsampling fraction & $[0.5,\ 1.0]$ & uniform \\
        Row subsampling fraction & $[0.5,\ 1.0]$ & uniform \\
        Bagging frequency & $\{0, 1, \ldots, 10\}$ & integer uniform \\
        L1 regularization ($\lambda_{1}$) & $[10^{-3},\ 10]$ & log \\
        L2 regularization ($\lambda_{2}$) & $[10^{-3},\ 10]$ & log \\
        Boosting rounds & up to $2000$, early stopping & -- \\
        \bottomrule
    \end{tabularx}
    \source{Hyperparameters are re-tuned at each walk-forward retraining boundary by the Optuna Tree-structured Parzen Estimator (TPE) sampler \citep{akiba_optuna_2019}. The search runs $50$ trials per window. Boosting rounds are bounded above at $2000$ and resolved by early stopping with $50$ rounds of patience on the most recent three months of the inner-validation slice. The search objective is $\text{NDCG@1}$ on the held-out inner-validation slice.}
\end{table}

Within each walk-forward training window, hyperparameter search and final model
refit draw their validation slices from the end of the ranker's training data.
The ranker does not see the whole window. Its final $6$ months are reserved for
calibrating the confidence gate (Section~\ref{sec:gate}) and are withheld from
feature selection and from ranker training alike. The search slice is the last
$6$ months of what remains. Each Optuna trial samples a hyperparameter
configuration, trains a candidate booster on the data preceding that slice, and
returns the candidate's $\text{NDCG@1}$ on the slice as the optimization
objective. After $50$ trials, the best-found hyperparameters are committed. The
final model is then refit with those hyperparameters, with the last $3$ months
held back for early stopping.

The early-stopping slice is therefore the more recent half of the search slice
rather than a separate period. The split addresses the inner-validation
overfitting that arises when a single slice is used both for hyperparameter
selection and for early stopping in time-series settings. Hyperparameters chosen
to maximize an objective on a slice tend to overfit that slice, so re-using the
whole of it for early stopping compounds the overfit. Restricting early stopping
to the more recent $3$ months limits the compounding without removing it,
because those observations also entered the search objective.

\subsection{Inference-Time Decision}\label{sec:gate}
A model-free overlay sits between the LightGBM ranker's per-day predictions and
the final position size. It is calibrated on training-window data and applied
unchanged at inference.

For each day $t$, the ranker emits a real-valued score per candidate. The
confidence signal is the gap between the top-1 and top-2 scores,
\begin{equation}
    \Delta \hat{y}_t = \hat{y}_{(1), t} - \hat{y}_{(2), t},
    \label{eq:confidence_signal}
\end{equation}
which captures how strongly the ranker prefers its first pick over the next-best alternative. Within each training window, a per-window threshold $\tau^\star_w$ is chosen on the last $6$ months of the training data (the held-out gate-calibration slice) by maximizing the Sortino ratio of the implied top-1 NetPnL series across a grid of candidate $\tau$ values. At inference, the gate abstains on day $t$ whenever $\Delta \hat{y}_t < \tau^\star_w$ for the active walk-forward window $w$. The sizing layer then treats the day as a \textit{SKIP} and assigns $Q_t = 0$.

A forced-prediction model discards information when it emits an uncertainty
signal. The LambdaRank score gap top1 - top2 is exactly such a signal, it
measures how decisively the model separated its top pick from the next-best
alternative. \citet{geifman_selective_2017} formalize this result. A classifier with
a controlled abstention region can dominate its forced-predict counterpart on
calibration-aware metrics. The gate is a domain-specific realization of that
result for a ranker.

A near-tie at the top of the ranking corresponds to days on
which no candidate strike carries a particularly attractive VRP. That indifference is consistent with a surface that is roughly
fair across deltas. Trading those
days pays the realization cost (half-spread, fees, occasional tail loss)
without compensating expected return. The gate converts the model's own
measured uncertainty into a decision not to trade, retaining information that a
rule selecting the top-ranked candidate unconditionally would discard.

Sortino penalizes downside variability only, so the objective biases $\tau$ toward aggressively abstaining on days dominated by negative-skew realizations. That is consistent with the asymmetric loss profile of short-put writing, whose relevant risk is the left tail. A globally fixed $\tau$ would be expected to bind unevenly across volatility regimes. Per-window $\tau$ instead re-fits on each window's own hold-out, and the calibrated trade rate ranges from 30.6\% (2023) to 100\% (2021 and 2025). The variation in caution emerges from the hold-out calibration rather than being imposed.

\subsection{Walk-Forward Backtesting Simulation}
A common shortcoming in volatility-trading research is the presentation of
theoretical returns that overlook market frictions. Reported short-option
profits are routinely overstated when margin requirements, bid-ask spreads, and
exchange fees are not modeled \citep{hong_profitability_2018,
    santa-clara_option_2009}. To address this, every reported daily profit-and-loss
in this paper applies (i) the Reg-T put-margin model\footnote{See:
    \href{https://www.interactivebrokers.com/en/trading/margin-requirements.php}{Interactive
        Brokers Margin Requirements} for more information.}, (ii) the tiered fees
schedule defined according to the Interactive Brokers fees\footnote{See:
    \href{https://www.interactivebrokers.com/en/pricing/commissions-home.php}{Interactive
        Brokers Commissions}.}, and (iii) entry fills at the bid-ask midpoint
(50\% spread coverage), with the execution sensitivity of
Section~\ref{subsec:execution_drag} probing the impact of less favorable fills. Position sizing is delegated to the calibrated sizing
layer of Section~\ref{sec:sizing}, whose free parameter is re-fit on each
walk-forward training window.

The model is fit and re-fit on a four-window expanding walk-forward (WF) plus a
one-window out-of-time (OOT) hold-out. The schedule is summarized in
Table~\ref{tab:walkforward}.

\begin{table}[!ht]
    \centering
    \caption{Walk-forward and out-of-time schedule.}
    \label{tab:walkforward}
    \begin{tabular}{ccc}
        \toprule
        \textbf{Window} & \textbf{Training years} & \textbf{Prediction year} \\
        \midrule
        WF 1 & 2018--2020 & 2021 \\
        WF 2 & 2018--2021 & 2022 \\
        WF 3 & 2018--2022 & 2023 \\
        WF 4 & 2018--2023 & 2024 \\
        OOT  & 2018--2024 & 2025 \\
        \bottomrule
    \end{tabular}
    \source{The initial training window covers $3$ calendar years ($2018$--$2020$), so that every training set contains at least one significant volatility-spike episode (the February $2018$ Volmageddon episode, the Q4 $2018$ selloff, and the March $2020$ COVID stress). Retraining cadence is $12$ months. Year $2025$ is held entirely out of every walk-forward fit and serves as a single-shot integrity check against the rest of the design.}
\end{table}

The expanding-window construction grows the training set by one year at each
retraining boundary, so the WF $4$ training set ($2018$--$2023$) is a strict
superset of the WF $1$ training set ($2018$--$2020$). At each retraining
boundary, the full procedure is repeated on the new training data. Feature
selection is re-run, the Optuna search re-tunes the LightGBM hyperparameters, the
model is refit, the confidence-gate threshold $\tau^\star$ is recalibrated on a
held-out slice (Section~\ref{sec:gate}), and the per-method sizing parameter
$\theta^\star$ is re-fit by single-parameter grid search (Section~\ref{sec:sizing}). No information from a future year leaks into a past
year's model. The annual retraining cadence is the main configuration on
bias-variance grounds, and the sensitivity to a $2$-year initial training
window is reported in Section~\ref{sec:robustness}.

The single-shot OOT prediction in $2025$ is the integrity check on the entire
pipeline. The model trained on $2018$--$2024$ is used to predict every trading
day of $2025$ without further retraining or hyperparameter search.

The day-by-day evaluation loop is summarized in
Algorithm~\ref{alg:walk_forward}. Phase~A repeats the entire training stack at
each retraining boundary. Phase~B walks forward through the out-of-sample days
in the active window, generating one P\&L observation per trading day under the
headline configuration (model picks, margin cap, and tiered exchange fees).

\begin{algorithm}[!ht]
    \caption{Walk-forward backtesting simulation.}
    \label{alg:walk_forward}
    \begin{algorithmic}[1]
        \Require Universe panel $\mathcal{U}$ over $T$ trading days, walk-forward schedule $\mathcal{W}$, sizing method $\mathcal{M}$, sizing configuration $\mathcal{C}_{\text{size}}$
        \State Initialize $\text{NAV}_0 \gets \mathcal{C}_{\text{size}}$ initial capital ($\$5{,}000{,}000$)

            \vspace{1.5mm}
            \Statex \textbf{\textit{Phase A: Per-window training and calibration.}}
            \For{each window $w \in \mathcal{W}$}
            \State $\mathcal{F}_w \gets$ feature selection on the training years of $w$ (Algorithm~\ref{alg:feature_selection})
            \State $\boldsymbol{\Phi}_w \gets$ Optuna search over $50$ trials on the inner-validation slice of $w$
            \State $f_w \gets$ refit LightGBM ranker with $\boldsymbol{\Phi}_w$ on the training window, excluding the $6$-month gate hold-out and a final $3$-month early-stopping slice
            \State $\tau^\star_w \gets$ calibrate confidence-gate threshold on the held-out slice (Section~\ref{sec:gate})
            \State $\theta^\star_w \gets$ single-parameter grid search to minimize $\big|\sigma_{\text{train}}(\theta) - 0.16\big|$ for $\mathcal{M}$
            \EndFor

            \vspace{1.5mm}
            \Statex \textbf{\textit{Phase B: Daily out-of-sample loop.}}
            \For{day $t = 1$ to $T$}
            \State $w \gets$ active walk-forward window for day $t$
            \State $\hat{y}_{\cdot, t} \gets f_w(\mathbf{x}_{\cdot, t})$ over the day's nine candidates
            \State $s^\star_t \gets \arg\max_s\, \hat{y}_{s, t}$, with confidence $\Delta \hat{y}_t = \hat{y}_{(1), t} - \hat{y}_{(2), t}$
            \If{$s^\star_t = \text{SKIP}$ \textbf{or} $\Delta \hat{y}_t < \tau^\star_w$}
            \State $Q_t \gets 0$, $\text{NetPnL}^{\$}_t \gets 0$
            \Else
            \State $M_t \gets$ per-contract Reg-T put margin
            \State $Q^{\text{des}}_t \gets \mathcal{M}(s^\star_t,\, \theta^\star_w,\, \text{NAV}_{t-1},\, \mathcal{C}_{\text{size}})$
            \State $Q_t \gets \min\!\big(Q^{\text{des}}_t,\ \lfloor \text{NAV}_{t-1} / M_t \rfloor\big)$ \Comment{margin cap}
            \State $\text{Gross}_t \gets Q_t \cdot \text{NetPnL}^{(\text{share})}_t \cdot 100$
            \State $\text{Fees}_t \gets F(Q_t,\, p_{\text{entry}, t})$
            \State $\text{NetPnL}^{\$}_t \gets \text{Gross}_t - \text{Fees}_t$
            \EndIf
            \State $\text{NAV}_t \gets \text{NAV}_{t-1} + \text{NetPnL}^{\$}_t$
        \EndFor
    \end{algorithmic}
\end{algorithm}

The architecture in Algorithm~\ref{alg:walk_forward} approximates a live
production environment under conservative friction assumptions. The model's
information set grows monotonically across walk-forward boundaries, so the
LightGBM ranker, the gate calibration, and the sizing calibration all adapt to
evolving market microstructure and volatility regimes without temporal leakage.
NAV is chained across windows by passing $\text{NAV}_{T_w}$ from the close of
window $w$ as the starting NAV of window $w+1$, so the reported equity curve is
continuous over $2021$--$2025$ rather than reset at each retraining boundary.

\subsection{Position Sizing}\label{sec:sizing}

The sizing layer maps a daily prediction stream into a daily contract count
$Q_t$ through one of seven candidate methods. Each method takes the gate-cleared
top-1 strategy, the equal-volatility calibration objective, and the trade-day's
universe data, and returns a non-negative integer contract
count. The seven methods share a common scaffold: each is parameterized by
exactly one free scalar $\theta$, fitted per walk-forward window. They differ
only in how $\theta$ and the day's data combine to produce the desired pre-cap
contract count $Q^{\text{des}}_t$. The margin cap and tiered fee schedule are
applied identically across methods (Algorithm~\ref{alg:walk_forward}).

\subsubsection{Margin, Fees, and Calibration Objective}
Every short put incurs an upfront margin requirement under the Reg-T
schedule\footnote{See
    \href{https://www.interactivebrokers.com/en/trading/margin-requirements.php}{Interactive
        Brokers Margin Requirements} for the full schedule.}. For an SPX short put with
entry premium $P_t$, underlying spot $S_t$, and strike $K$, the per-contract
margin is
\begin{equation}
    M_t = 100 \cdot \Big[\, P_t + \max\!\big(0.15\, S_t - \max(0,\, S_t - K),\ 0.10\, K\big) \,\Big],
    \label{eq:put_margin}
\end{equation}
where $\max(0, S_t - K)$ is the put's out-of-the-money amount. 

The contract count is capped at the maximum the current NAV can collateralize,
$Q_t \le \lfloor \text{NAV}_{t-1} / M_t \rfloor$. The cap is rarely binding under the headline configuration because the
equal-volatility calibration produces target utilization well below the cap on most
days.

Entry-only fees follow the IBKR index-options tier:
\begin{equation}
    \texttt{fee\_per\_contract} =
    \begin{cases}
        \$0.25 & \text{if } P_t < \$0.05            \\
        \$0.50 & \text{if } \$0.05 \le P_t < \$0.10 \\
        \$0.65 & \text{if } P_t \ge \$0.10,
    \end{cases}
    \label{eq:fee_tier}
\end{equation}
with a $\$1.00$ minimum total fee per non-empty trade. SPXW 0DTE positions are held to PM cash settlement, so no exit fee applies in such a case.

    For every method, the free parameter $\theta$ is selected by single-parameter
    grid search on the training window of each walk-forward fold. The headline
    objective is \emph{equal-volatility}: pick the $\theta^\star$ that minimizes the
    deviation of the training-window NAV's realized annualized volatility from the
$16\%$ risk anchor,
    \begin{equation}
        \theta^\star = \arg\min_{\theta \in \mathcal{G}_{\mathcal{M}}}\, \big| \sigma_{\text{train}}(\theta) - 0.16 \big|,
        \label{eq:equal_vol_calibration}
    \end{equation}
    where $\mathcal{G}_{\mathcal{M}}$ is the per-method grid (Table~\ref{tab:sizing_grids}) and $\sigma_{\text{train}}(\theta)$ is the annualized standard deviation of training-window daily NAV returns produced by running the full backtest at $\theta$. Equal-volatility is intended to commit each method to the same target risk budget, so that cross-method Sharpe comparisons separate selection and shape differences from raw position-size differences. Two alternative objectives, \emph{sharpe-max} and \emph{sortino-max}, are reported as Section~\ref{sec:robustness} sensitivities.

    Two qualifications apply to that intent. The first concerns what $\sigma_{\text{train}}(\theta)$ is computed on. The backtest that produces it runs on training-window predictions, which the ranker has already fitted, so the calibration inherits whatever optimism that fit carries. The direction of the effect is known: a better in-sample fit lowers training volatility, which raises $\theta^\star$ and carries more leverage out of sample. The confidence gate avoids this by calibrating on a slice withheld from training (Section~\ref{sec:gate}), and calibrating $\theta$ on an inner validation slice would be the consistent treatment. The second qualification is that the grid often binds before the anchor is reached. Across seven methods and five windows, $\theta^\star$ settles on the largest value its grid allows in $24$ of $35$ cells, and the mean realized training volatility at $\theta^\star$ is $0.139$ against the $0.16$ anchor, with a range of $0.081$ to $0.165$. Where $\theta^\star$ pins in this way the first channel cannot operate, since $\theta$ has no room to rise. For Edge Allocation the shortfall is structural rather than a matter of grid width, since matching the anchor would require $u_{\max}$ between $1.07$ and $1.58$ in four of the five windows, and $u_{\max}$ is a margin utilization fraction bounded above by one. The methods are therefore not placed on a common risk budget, in the training window or outside it, and cross-method Sharpe comparisons should be read as controlling for position size only in part. Realized volatility over the evaluation period, reported per method in Table~\ref{tab:headline_metrics}, falls further below the anchor than the calibration does.

    \subsubsection{Sizing Methods}

    Fixed Margin Utilization (FMU) commits a constant fraction of NAV to margin every trading day,
    \begin{equation}
        Q^{\text{des}}_t = \left\lfloor \frac{f_{\text{FMU}} \cdot \text{NAV}_{t-1}}{M_t} \right\rfloor,
        \label{eq:fmu}
    \end{equation}
    with free parameter $\theta = f_{\text{FMU}} \in (0, 1)$. FMU has no regime adaptation and is the baseline against which the other six methods are measured at the equal-volatility calibration point.

    Volatility Targeting (VT) replaces the constant-fraction rule with a dollar-volatility target. The per-contract dollar P\&L volatility is approximated by the delta-based linearization
    \begin{equation}
        \sigma^{\$, c}_t = |\delta_t| \cdot S_t \cdot \sigma^{\text{intra}, 5d}_t \cdot 100,
    \end{equation}
    where $\sigma^{\text{intra}, 5d}_t$ is the 5-day realized intraday volatility of SPX. The day's target dollar P\&L volatility is
    \begin{equation}
        \sigma^{\$, \text{tgt}}_t = f_{\text{VT}} \cdot \frac{\sigma_{\text{ann}}}{\sqrt{252}} \cdot \text{NAV}_{t-1},
    \end{equation}
    with $\sigma_{\text{ann}} = 0.16$. The contract count is
    \begin{equation}
        Q^{\text{des}}_t = \left\lfloor \frac{\sigma^{\$, \text{tgt}}_t}{\sigma^{\$, c}_t} \right\rfloor,
        \label{eq:vt}
    \end{equation}
    and $\theta = f_{\text{VT}}$ is the per-day volatility-target multiplier.

    Short-Richness Scaling (SRS) ties per-day size to a 0DTE ATM IV richness measure that compares short-dated implied volatility against VIX9D,
    \begin{equation}
        \mathcal{R}_t = \sigma^{0\text{DTE}}_{\text{ATM}, t} \,\Big/\, \frac{\text{VIX9D}_t}{100},
    \end{equation}
    where $\sigma^{0\text{DTE}}_{\text{ATM}, t}$ is the at-the-money-forward implied volatility on the 0DTE chain at 10:00 ET. The richness is normalized by its training-set median and capped at the training-set $95$th percentile,
    \begin{equation}
        \mathcal{S}_t = \min\!\big(\mathcal{C}_{\text{train}},\ \max\!\big(1,\ \mathcal{R}_t \,/\, \text{median}(\mathcal{R}_{\text{train}})\big)\big),
    \end{equation}
    which floors the scale at $1$ and prevents extreme richness readings from dominating sizing. The contract count is
    \begin{equation}
        Q^{\text{des}}_t = \left\lfloor \frac{f_{\text{SRS}} \cdot \mathcal{S}_t \cdot \text{NAV}_{t-1}}{M_t} \right\rfloor,
        \label{eq:srs}
    \end{equation}
    with $\theta = f_{\text{SRS}}$. The median and the cap are estimated once per training window and are not tunable.

    Edge Allocation (EA) ranks today's IV-vs-RV edge against the empirical distribution of past edges rather than scaling on its raw level. The day's edge of 0DTE ATM IV over annualized 5-day intraday realized volatility,
    \begin{equation}
        \mathcal{E}_t = \frac{\sigma^{0\text{DTE}}_{\text{ATM}, t} - \sigma^{\text{intra}, 5d}_t \cdot \sqrt{252}}{\sigma^{\text{intra}, 5d}_t \cdot \sqrt{252}},
    \end{equation}
    is mapped through the empirical CDF of the training-window edges,
    \begin{equation}
        u_t = u_{\max} \cdot \widehat{F}_{\mathcal{E}, \text{train}}(\mathcal{E}_t),
    \end{equation}
    producing a per-day margin-utilization fraction. The contract count is
    \begin{equation}
        Q^{\text{des}}_t = \left\lfloor \frac{u_t \cdot \text{NAV}_{t-1}}{M_t} \right\rfloor,
        \label{eq:ea}
    \end{equation}
    with $\theta = u_{\max} \in (0, 1)$. EA is the headline sizing method (Section~\ref{subsec:per_method}) and scales position size monotonically with the percentile rank of today's edge in the training-window edge distribution.

    Gamma-Budgeted Sizing (GB) shifts the conditioning variable from the volatility surface to the position's own Greeks. A scenario-loss bound caps per-trade exposure under a fixed adverse intraday move $k$, frozen at the training-set $95$th-percentile of daily maximum adverse moves. The per-contract Taylor-expansion scenario loss combines the linear and quadratic Greeks,
    \begin{equation}
        L^c_t = \big(|\delta_t| \cdot k \cdot S_t + \tfrac{1}{2} |\Gamma_t| \cdot (k \cdot S_t)^2\big) \cdot 100.
    \end{equation}
    The contract count is
    \begin{equation}
        Q^{\text{des}}_t = \left\lfloor \frac{b_{\text{GB}} \cdot \text{NAV}_{t-1}}{L^c_t} \right\rfloor,
        \label{eq:gb}
    \end{equation}
    with $\theta = b_{\text{GB}}$, the fraction of NAV the budget is willing to lose if the scenario move materializes.

    The half-Kelly (HK) and quarter-Kelly (QK) variants size on the strategy's own realized return distribution rather than on the option surface or position Greeks. Both apply the Kelly criterion \citep{kelly_jr_new_1956, thorp_kelly_2008} to the per-trade return on margin, $r^c_t = \text{NetPnL}_t \cdot 100 \,/\, M_t$. A rolling $252$-day window of past per-day returns on margin is maintained at all times. The mean-variance Kelly fraction is
    \begin{equation}
        f^\star_t = \frac{\hat{\mu}^r_t}{\big(\hat{\sigma}^r_t\big)^2},
    \end{equation}
    where $\hat{\mu}^r_t$ and $\big(\hat{\sigma}^r_t\big)^2$ are the rolling-window mean and variance. The fraction is fractional-Kelly-scaled and capped at the free parameter $f_{\text{CK}}$,
    \begin{equation}
        f_t = \min\!\big(f_{\text{CK}},\ \max(0,\ \alpha \cdot f^\star_t)\big),
    \end{equation}
    with $\alpha = 0.5$ for HK and $\alpha = 0.25$ for QK. The contract count is
    \begin{equation}
        Q^{\text{des}}_t = \left\lfloor \frac{f_t \cdot \text{NAV}_{t-1}}{M_t} \right\rfloor.
        \label{eq:ck}
    \end{equation}
    The free parameter $\theta = f_{\text{CK}}$ is the per-day cap on the deployed Kelly fraction. The two variants share a single calibration grid. The half- and quarter-Kelly multipliers $\alpha$ are not tunable and reflect the standard practitioner adjustments for parameter-estimation noise \citep{thorp_understanding_2011}.

    \subsubsection{Per-Method Calibration Grids}
    The single-parameter grid for each method is bounded by economic plausibility
    (e.g., $f_{\text{FMU}} \in (0, 1)$) and discretized at a step that produces a
    smooth $\sigma_{\text{train}}(\theta)$ curve under the equal-volatility objective.
    Table~\ref{tab:sizing_grids} reports the grids.

    \begin{table}[!ht]
    \centering
    \caption{Per-method calibration grids for the equal-volatility objective.}
    \label{tab:sizing_grids}
    \begin{tabular*}{\textwidth}{@{\extracolsep{\fill}} l c c @{}}
        \toprule
        \textbf{Method} & \textbf{Free parameter} & \textbf{Grid} \\
        \midrule
        FMU & $f_{\text{FMU}}$  & $\{0.02,\ 0.04,\ \ldots,\ 0.60\}$ \\
        VT  & $f_{\text{VT}}$   & $\{0.1,\ 0.2,\ \ldots,\ 2.5\}$ \\
        SRS & $f_{\text{SRS}}$  & $\{0.02,\ 0.04,\ \ldots,\ 0.50\}$ \\
        EA  & $u_{\max}$        & $\{0.05,\ 0.10,\ \ldots,\ 0.80\}$ \\
        GB  & $b_{\text{GB}}$   & $\{0.005,\ 0.010,\ \ldots,\ 0.100\}$ \\
        HK, QK & $f_{\text{CK}}$   & $\{0.05,\ 0.10,\ \ldots,\ 0.80\}$ \\
        \bottomrule
    \end{tabular*}
    \source{Grids are discretized in equal increments. The training-window backtest is run at every grid point. The point that minimizes $|\sigma_{\text{train}}(\theta) - 0.16|$ is committed as $\theta^\star$ for that walk-forward window. The HK and QK variants share the same grid. They differ only in the half- versus quarter-Kelly multiplier $\alpha \in \{0.5,\ 0.25\}$ applied inside the sizing formula.}
\end{table}

    The sizing layer is therefore a composite function $\mathcal{M}: (s^\star_t,\allowbreak\ \boldsymbol{x}_t,\allowbreak\ \theta^\star_w,\allowbreak\ \text{NAV}_{t-1}) \mapsto Q_t$ that
    combines the chosen method, the per-window calibrated parameter, the day's
    universe data, and the running NAV into a single integer contract count. The
    seven methods explore distinct economic intuitions for what should govern
    position size (fixed margin draw, dollar-volatility target, IV-vs-RV richness,
    ATM-vs-VIX9D edge, Greek-budgeted scenario loss, or fractional Kelly on rolling
    returns), but operate under a common calibration discipline so that
    cross-method comparisons are interpretable.

    Without a common calibration, cross-method comparison conflates sizing-logic
    differences with risk-level differences, which is the role of the equal-volatility
    anchor of Equation~\eqref{eq:equal_vol_calibration}. The 16\% target reflects the long-run realized
    volatility of the SPX. The 2018--2020 training window contains the March 2020
    COVID spike, so the in-sample realized volatility is approximately 21\%.
    Anchoring to in-sample volatility would be expected to tilt sizing toward
    levels that anticipate continued elevated volatility. The long-run 16\%
    anchor is insulated from volatility regimes specific to the training window. Sizing
    every method to that anchor would also place the strategies at the volatility
    of the buy-and-hold S\&P 500, though the calibration does not reach it.
    Sensitivity analysis at additional targets of 12\% and 20\% is reported as a
    robustness check.

    \subsection{Strategy Performance and Risk Metrics}
    Evaluating a systematic short-volatility strategy requires both a baseline
    assessment of risk-adjusted profitability and an explicit treatment of the
    asymmetric, fat-tailed return distribution that premium-collection strategies
    typically generate. The out-of-sample performance of each ranked portfolio is
    evaluated against a suite of metrics that combines four canonical descriptors
    of return and risk, an asymmetric risk-adjusted ratio that targets the
    downside, and two probabilistic measures that account
    for distributional non-normality and multiple testing.

    The four descriptors are the Annualized Compounded Return (aRC), Annualized
    Standard Deviation (aSD), Maximum Drawdown (MD), and the Annualized Sharpe
    Ratio (SR) \citep{sharpe_sharpe_1994}. Because short-volatility return
    distributions are negatively skewed and leptokurtic, the standard deviation
    underweights left-tail risk, so the Sharpe ratio overstates the realized
    compensation per unit of downside risk. To complement it, we report the Sortino
    ratio \citep{sortino_downside_1994}, which divides the annualized return by an
    annualized downside deviation rather than total volatility.

    To assess whether the realized Sharpe reflects a true Sharpe at or above the
    external benchmark rather than a sample-period or multiple-testing artifact, we
    use the Probabilistic Sharpe Ratio (PSR) of \citet{bailey_sharpe_2012} and the
    Deflated Sharpe Ratio (DSR) of \citet{bailey_deflated_2014}. The PSR converts
    the realized Sharpe into the probability that the true Sharpe exceeds an
    external benchmark given the observed sample skewness and kurtosis. Negative
    skewness and fat tails lower the PSR for a given point estimate. The DSR
    additionally deflates the benchmark by the expected maximum Sharpe under the
    null of $N_t$ independently tested candidates, controlling for the
multiple-configuration search reported in Section \ref{sec:methodology}. The
detailed mathematical expressions for all metrics used in this study, together
with the Diebold--Mariano test used for benchmark comparisons, are provided in
Appendix \ref{app:metrics}.

\section{Data}\label{sec:data}
\subsection{The S\&P 500 Index and SPXW Options}
The empirical analysis in this study uses minute-frequency price data on SPXW
options written on the S\&P 500 Index. The dataset includes 1-minute OHLC bars
and intraday quote data, obtained directly from the Chicago Board Options
Exchange (CBOE).

SPXW options are European-style, cash-settled contracts on the S\&P 500 that
expire on every trading day. In contrast to standard SPX options, which use the
open of the third Friday for AM settlement and cease trading the day before,
SPXW contracts settle on the close of the expiration day (PM settlement) and
remain tradable through that close. Settlement is in cash, computed as the
difference between the strike and the closing level of the S\&P 500, multiplied
by a \$100 contract multiplier\footnote{See
    \href{https://cdn.cboe.com/resources/spx/spx-fact-sheet.pdf}{CBOE SPX Fact
        Sheet} and
    \href{https://www.cboe.com/tradable-products/sp-500/spx-options/spx-specifications}{CBOE
        SPX Product Specification}.}.

Two considerations motivate the use of SPXW contracts as the trading
instrument. First, PM settlement combined with intraday tradability through
expiration makes SPXW options well-suited to short-horizon volatility
strategies that monetize close-to-expiry premium decay. Second, trading volume
in the segment has expanded sharply, with a pronounced shift toward daily and
zero-day-to-expiration (0DTE) cycles. This shift has been accompanied by
deeper liquidity for the strategies analyzed in this paper
\citep{anderse_short_term_2017, vasquez_0dte_2025, bozovic_intraday_2025}.

The CBOE S\&P 500 PutWrite Index (PUT) and the CBOE S\&P 500 WeeklyPutWrite
Index (WPUT) are the passive short-volatility benchmarks used in this paper.
The PUT index reflects a continuous strategy of selling at-the-money one-month
SPX puts on the third Friday of each month, holding the position to expiration,
and rolling into the next monthly contract. The collateral is invested in
zero-coupon U.S. Treasury bills with maturity matched to the option, so the
index earns the corresponding T-bill yield on top of the option premium. The
WPUT index applies the same construction to a weekly cycle, selling
at-the-money SPX puts every Friday and rolling weekly. WPUT is therefore the
closer passive analogue to the short-tenor SPXW writing strategies studied in
this paper, although PUT is the longer-established of the two indices.

\begin{table}[!ht]
\centering
\caption{Summary statistics of daily log returns of the S\&P 500 Index, VIX, CBOE PUT Index, and CBOE WPUT Index.}
\label{tab:summary-stats}
\resizebox{\textwidth}{!}{%
\begin{tabular}{@{}lccccccccccccccc@{}}
\toprule
 & \textbf{Mean} & \textbf{Median} & \textbf{Std Dev (daily)} & \textbf{Std Dev (ann)} & \textbf{Skewness} & \textbf{Excess Kurtosis} & \textbf{Min} & \textbf{Max} & \textbf{MD} & \textbf{VaR 95\%} & \textbf{CVaR 95\%} & \textbf{Autocorrelation (1)} & \textbf{JB} & \textbf{ADF} & \textbf{LB(10)} \\ \midrule
\textbf{S\&P 500} & 0.0005 & 0.0009 & 0.0124 & 0.20 & -0.65 & 14.77 & -0.1277 & 0.0909 & -0.34 & -1.83\% & -3.05\% & -0.15 & 18325*** & -14.15*** & 231.37*** \\
\textbf{VIX} & 0.0002 & -0.0073 & 0.0810 & 1.29 & 1.40 & 8.88 & -0.4424 & 0.7682 & -0.86 & -10.81\% & -15.23\% & -0.08 & 7321*** & -18.62*** & 26.45*** \\
\textbf{PUT} & 0.0003 & 0.0005 & 0.0089 & 0.14 & -1.96 & 45.04 & -0.1218 & 0.0903 & -0.29 & -1.16\% & -2.31\% & -0.26 & 170315*** & -14.04*** & 400.51*** \\
\textbf{WPUT} & 0.0001 & 0.0007 & 0.0084 & 0.13 & -1.90 & 31.52 & -0.1014 & 0.0836 & -0.26 & -1.28\% & -2.34\% & -0.20 & 83857*** & -14.32*** & 206.85*** \\ \bottomrule
\end{tabular}%
}
\source{The dataset spans from January 1, 2018 to December 31, 2025. Logarithmic returns were computed based on daily closing prices. The column \textit{MD} contains the Maximum Drawdown statistic. The \textit{JB} column contains the results of the Jarque-Bera test for normality of the series. The \textit{ADF} column reports the results of the Augmented Dickey-Fuller test for stationarity of the series. The \textit{LB} column contains the results of the Ljung-Box test for autocorrelation up to 10 lags. The asterisks *, **, and *** indicate statistical significance at the 0.1, 0.05, and 0.01 levels, respectively.}
\end{table}

Table \ref{tab:summary-stats} reports descriptive statistics and standard test
results for the daily log returns of the S\&P 500 Index, the VIX, the PUT, and
the WPUT over 2018-01-01 to 2025-12-31. This eight-year sample spans the
periods used for model training (2018--2020), walk-forward evaluation
(2021--2024), and the 2025 out-of-time hold-out, as detailed in
Section~\ref{subsec:setup}. It is preceded by a 2017 warm-up year used only to
initialize rolling features and per-strategy statistics. All four series have positive mean daily log
returns, although the levels differ markedly. The S\&P 500 earns 0.05\% per day
on average (median 0.09\%), the PUT index 0.03\%, and the WPUT index 0.01\%.
The lower realized mean of WPUT relative to PUT is consistent with the more
frequent rolling cycle of the weekly index, which captures smaller absolute
premia at higher turnover and so concentrates more of its daily distribution
near zero. Annualized volatility is 20\% for the S\&P 500, 14\% for PUT, and
13\% for WPUT, in line with the documented variance-reducing effect of
systematic put writing relative to direct long-equity exposure. The VIX itself
has a daily mean of 0.02\% but a sharply negative median of -0.73\%. This
pattern is consistent with the mean-reverting behavior of implied volatility.
The index typically drifts downward in calm regimes and spikes upward in
shocks. The result is a positive average despite a negative central tendency.

None of the four series is well approximated by a Gaussian distribution. The
S\&P 500 has moderate negative skew (-0.65) and excess kurtosis of 14.77, in
line with the pattern that equity drawdowns are sharper than recoveries. The
two short-volatility benchmarks display materially stronger left-tail
behavior, with skewness of -1.96 (PUT) and -1.90 (WPUT) and excess kurtosis of
45.04 (PUT) and 31.52 (WPUT). This shape is mechanically inherited from the
convexity of short-put payoffs. Gains are bounded above by the option premium
collected, while losses can rise sharply in stress regimes. The ordering of
kurtosis between PUT and WPUT, with the monthly index more leptokurtic than the
weekly one, is consistent with the diversification of premium-collection events
across multiple weekly expirations, which dampens the influence of any single
tail observation on the realized return distribution. The VIX, by contrast, is
positively skewed (1.40), in line with its propensity for sudden upward spikes
followed by gradual reversion. Jarque--Bera statistics reject normality at the
1\% level for every series, with values ranging from 7,321 (VIX) to 170,315
(PUT). The rejection of normality across every series motivates non-parametric and
machine-learning methods that do not impose Gaussianity.

\begin{figure}[!ht]
    \centering
    \caption{S\&P 500 Index, VIX, PUT Index, and WPUT Index series from 2018-01-01 to 2025-12-31.}
    \label{fig:c3:indexvalue}
    \includegraphics[width=\columnwidth]{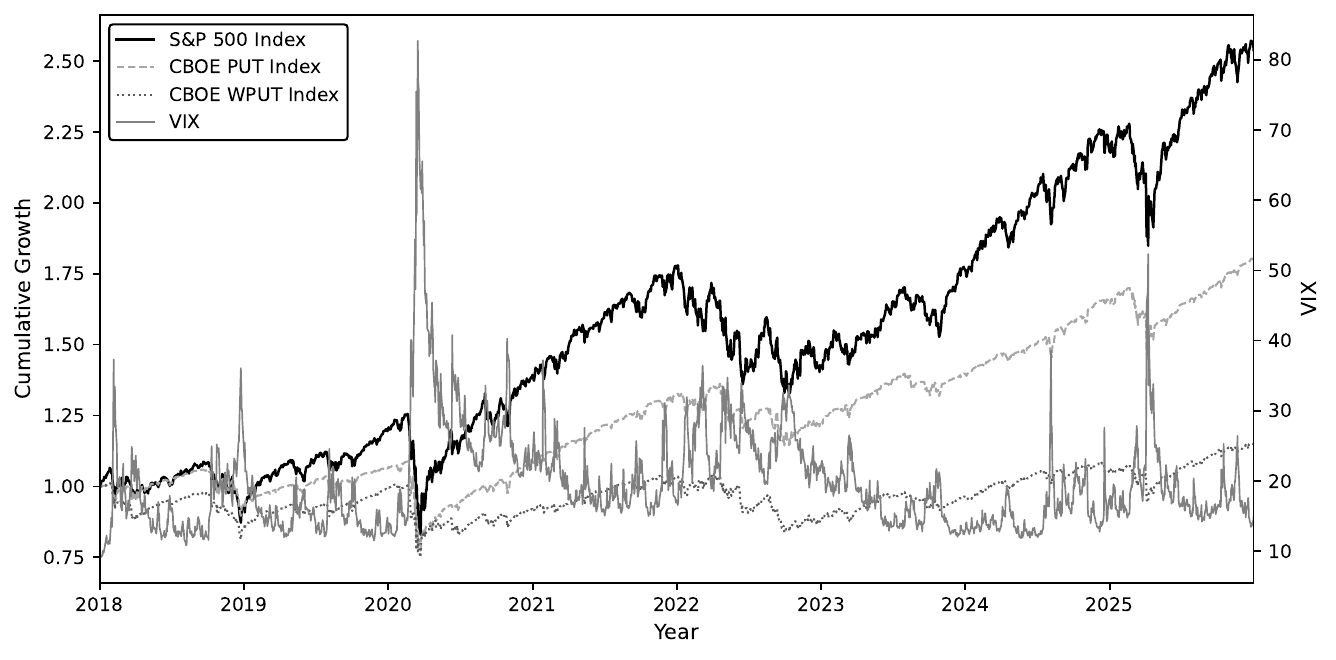}
    \source{The chart shows the daily closing levels of the S\&P 500 Index, the CBOE PUT Index, and the CBOE WPUT Index, all rescaled to a starting value of 1.0 and read against the left-hand axis (Cumulative Growth), together with the daily closing level of the VIX in raw index points on the right-hand axis. The sample period is 2018-01-01 to 2025-12-31.}
\end{figure}

Tail risk metrics are consistent with this characterization. The S\&P 500
incurred a maximum drawdown of 34\% over the sample, driven by the COVID-19
sell-off of March 2020 and clearly visible in Figure \ref{fig:c3:indexvalue}. The
PUT and WPUT indices realized smaller, but still substantial, drawdowns of 29\%
and 26\%, respectively. The one-day 95\% Value-at-Risk is -1.83\% for the S\&P
500, with an associated Conditional VaR of -3.05\%. The put-write benchmarks
have tighter daily-loss profiles, with VaR / CVaR of -1.16\% / -2.31\% for PUT
and -1.28\% / -2.34\% for WPUT. Systematic put writing is therefore associated
with lower typical daily volatility and a smaller magnitude of the most adverse
single-day outcomes relative to direct equity exposure, although the tail
thickness documented above indicates that these reductions do not extend to the
most extreme stress events.

All four series display statistically significant negative first-order
autocorrelation, ranging from -0.08 for the VIX to -0.26 for PUT, with WPUT at
-0.20 and the S\&P 500 at -0.15. The negative sign points to short-horizon mean
reversion in daily returns and is strongest in the put-write benchmarks, whose
returns are dominated by alternating phases of premium collection and recovery
after expiry. The Ljung--Box test at ten lags rejects the joint hypothesis of
zero autocorrelation at the 1\% level for every series, with test statistics
ranging from 26.45 (VIX) to 400.51 (PUT). Lagged return information therefore
has measurable predictive content, which is consistent with the autoregressive
features used in subsequent modeling stages carrying information. The Augmented Dickey--Fuller test, with the lag order selected by the
Akaike information criterion and a constant-only deterministic term, rejects the
unit-root null at the 1\% level for all four series. The daily log returns are
therefore stationary over the sample period.

\subsection{Feature Engineering}
The candidate feature catalog is organized into groups spanning approximately
190 features. Exact definitions, data
sources, and rolling-window choices are documented in Appendix
\ref{app:features}, while this subsection motivates the taxonomy and grounds it
in the literature.

The catalog separates \emph{cross-sectional} (CS) features, which take one
value per trading day and are broadcast across all candidates, from
\emph{per-strategy} (PS) features, which vary across candidates on a given day.
The LambdaRank objective exploits only within-day variation across candidates,
so PS features supply the discriminative signal for ranking, whereas CS
features are regime conditioners. The following groups are categorized as
cross-sectional: calendar and event flags, morning-session dynamics, SPX index
features, the VIX family, macroeconomic series, volatility-surface levels and skew,
RV-versus-IV differentials, realized higher moments, VIX term-structure
curvature, and medium-frequency trend. The per-strategy groups are: option
Greeks at entry, rolling P\&L statistics, entry
liquidity, intra-strategy term context, and regime-conditional sensitivities.

The motivation for the principal blocks follows the volatility-risk-premium
literature.

\begin{itemize}
    \item \textbf{Implied-volatility surface (CS).} ATMF IV at multiple DTEs, per-delta IV at the 10$\Delta$ and 25$\Delta$ buckets, 252-day rolling percentiles, the 25$\Delta$ risk reversal as the primary skew measure, and forward-volatility differentials. The model ranks the eight delta buckets each day by their mispricing on this surface relative to expected realized volatility \citep{cont_dynamics_2002, constantinides_puzzle_2013}.

    \item \textbf{Realized-minus-implied volatility differentials (CS).} 5-, 21-, and 63-day RV-versus-IV spreads and ratios. These differentials provide a direct in-sample measure of the volatility risk premium. Consistent with the documented persistence of the equity-index VRP \citep{bondarenko_why_2014, carr_variance_2009}, they are one of the largest contributors to model performance in the ablation panel of Section~\ref{subsec:ablation}.

    \item \textbf{VIX family and term-structure curvature (CS).} Levels and percentile statistics of VIX, VIX1D, VIX9D, VIX3M, VIX6M, and VVIX, the front-month VIX future together with its slope and roll yield, and three term-curvature combinations. The VIX level conditions the premium that short-volatility strategies collect, and the slope of the term structure has been reported to predict the cross-section of subsequent option returns \citep{vasquez_equity_2017, da_fonseca_variance_2019, malkiel_option_2018}.

    \item \textbf{Morning-session features (CS).} Computed from 1-minute SPX bid-ask data and 09:35 ET volatility-surface snapshots, these capture the only fresh post-open information available at the 10:00 ET entry: morning log-return, intraday range, annualized realized volatility, the overnight gap and a gap-filled flag, and 09:35-to-10:00 changes in ATMF IV, 25$\Delta$ risk reversal, VIX, and VVIX. The block matters most for 0DTE, since the entry decision must reflect the regime that the morning auction has just revealed.

    \item \textbf{SPX index features and put-call ratios (CS).} Multi-horizon SPX log returns, intraday realized volatility, and the SPX and VIX put-call ratios. The PCR series carry sentiment-driven information beyond what is contained in the implied-volatility surface itself \citep{sheu_effective_2011}.

    \item \textbf{Macroeconomic indicators (CS).} The effective Federal Funds rate (level, 1-year percentile, rate of change) and seasonally non-adjusted initial jobless claims (level, year-on-year change, 1-year percentile). These are low-frequency regime conditioners on the realized volatility risk premium \citep{bollerslev_dynamic_2011, corradi_macroeconomic_2013}.

    \item \textbf{Calendar and event flags (CS).} Day-of-week, day-of-month, and month-of-year integers, monthly and quarterly OPEX flags, post- and pre-holiday session flags, and \texttt{days\_until\_next} / \texttt{is\_X\_day} pairs for FOMC, CPI, NFP, and PCE releases. FOMC days are excluded upstream from the candidate universe. The surrounding macro-event days carry distinct volatility behavior and are included in the candidate feature set.

    \item \textbf{Per-strategy blocks (PS)}. Entry-time Black--Scholes Greeks (delta, gamma, theta, vega), with the dollar-delta, gamma exposure, log-moneyness, and per-strategy leverage measure derived from them, supply the per-candidate risk profile. Per-strategy rolling statistics supply each candidate's recent realized performance. Entry liquidity (bid-ask spread in dollars and as a percentage of mid, log entry premium, all measured at the 10:00 ET snapshot) flags whether a pick is executable at near-mid quality. Intra-strategy term context bridges each candidate's specific point on the volatility surface back to the cross-sectional regime signals.
\end{itemize}

The combined catalog of approximately 190 candidate features is reduced per
walk-forward window by the selection pipeline described in Section
\ref{sec:feature-pipeline} to roughly 50 to 60 surviving features. Twenty-seven
of these survive across all four walk-forward windows and the out-of-time
window, and are common to every fitted model. The remainder are window-specific or window-frequent, and
Section~\ref{subsec:ablation} quantifies the marginal contribution of each base
group.

\section{Empirical Results}
\label{sec:empirical}

\subsection{Setup}
\label{subsec:setup}

The strategy is evaluated over the period 2017--2025. The 2017 calendar year
provides warm-up data for rolling features and per-strategy statistics, and the
initial training window covers 2018--2020. The four-window walk-forward (WF)
with annual retraining on an expanding window spans 2021--2024, and the 2025
calendar year is reserved as an out-of-time (OOT) hold-out, never used during
training, hyperparameter search, or model selection. The LightGBM LambdaRank
ranker described in Section~\ref{sec:methodology} selects one position per day
from the nine-candidate universe (eight delta-targeted short puts plus SKIP),
entered at 10:00~ET, with sizing calibrated to a 16\,\% annual volatility
target across the seven sizing methods of Section~\ref{sec:sizing}. The seven
methods share the same model, selection layer, and confidence gate, and differ only in how the per-day position size $Q$ is determined. The
DSR multi-test correction reported in
Section~\ref{subsec:multi_test_correction} uses $N_t=75$ trials.

\subsection{Per-method performance}
\label{subsec:per_method}

Table~\ref{tab:headline_metrics} reports annualized Sharpe ratio, Sortino
ratio, annualized return, annualized volatility, and maximum drawdown under
each of the seven sizing methods, separately for the four-window walk-forward
(2021--2024) and the 2025 out-of-time hold-out. Annualized WF Sharpe ranges from $1.897$ for the
quarter-Kelly variant to $3.105$ for Edge Allocation, and annualized OOT
Sharpe ranges from $4.308$ to $5.761$ across the same two methods. Every
method has positive Sharpe on both slices, and every method's OOT Sharpe
exceeds its WF Sharpe by more than $2.4$ units. The three highest WF Sharpe
ratios belong, in order, to Edge Allocation, Short-Richness Scaling, and
Fixed Margin Utilization, and the OOT ranking has the same three at the top.

PSR values, computed against the worst of the three external benchmarks per
slice, range from $0.906$ to $1.000$ on the WF slice and from $0.949$ to
$0.994$ on the OOT slice. Six of the seven methods clear $0.95$ on the OOT
slice, Volatility Targeting missing it narrowly at $0.949$.
Three clear it on both: Edge
Allocation ($1.000 / 0.964$), Short-Richness Scaling ($0.992 / 0.994$), and
Fixed Margin Utilization ($0.981 / 0.984$). The remaining four
(Volatility Targeting, Gamma-Budgeted Sizing, and the two fractional-Kelly
variants) fall short on WF, with WF PSR in the range $0.906$ to
$0.926$. The corresponding multi-test-corrected confidence
under the Deflated Sharpe Ratio is reported in
Section~\ref{subsec:multi_test_correction}.

\begin{figure}[!ht]
        \centering
        \caption{Cumulative net asset value by sizing method, 2021-01-04 to 2025-12-31.}
        \label{fig:equity}
        \includegraphics[width=\textwidth]{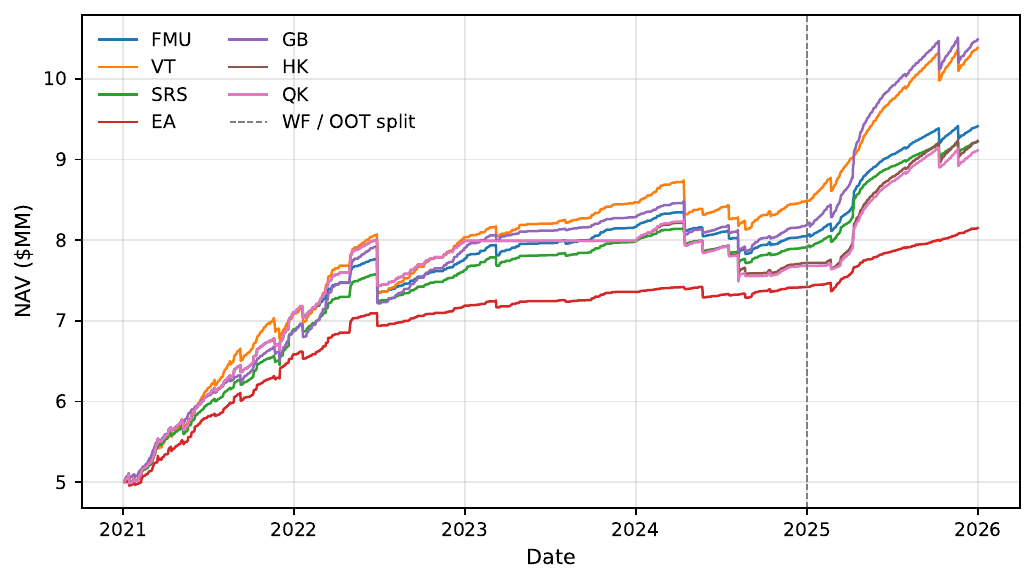}
        \source{Cumulative net asset value for the seven sizing methods over the full 2021-01-04 to 2025-12-31 sample, starting from a common initial capital of \$5{,}000{,}000. The vertical dashed line marks the WF/OOT split at 2025-01-01. The walk-forward (WF) period covers 2021--2024 across the four expanding-window training cycles. The 2025 calendar year is the out-of-time (OOT) hold-out, never used during training, hyperparameter search, or model selection. All seven curves use the same model, selection layer, and confidence gate, and differ only in the per-day position-size rule.}
    \end{figure}

\begin{figure}[!ht]
        \centering
        \caption{Underwater drawdown trajectory by sizing method, 2021-01-04 to 2025-12-31.}
        \label{fig:drawdown}
        \includegraphics[width=\textwidth]{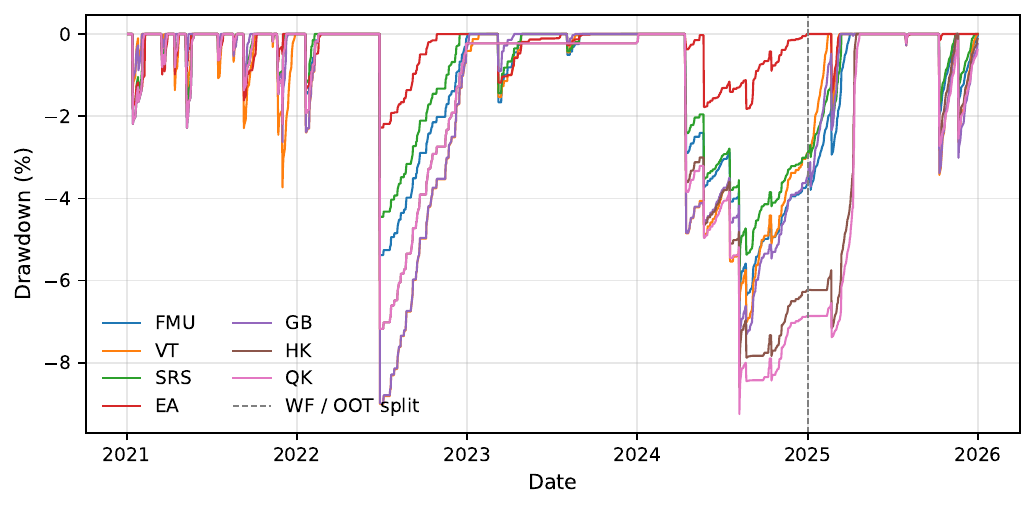}
        \source{Underwater drawdown trajectory per sizing method over the full 2021-01-04 to 2025-12-31 sample. Drawdown is computed as the running NAV deficit from the historical high-water mark, expressed in percent. The vertical dashed line marks the WF/OOT split at 2025-01-01.}
    \end{figure}

Figure~\ref{fig:equity} plots cumulative net asset value for all seven methods
from 2021-01-04 through 2025-12-31, starting from a common \$5{,}000{,}000
initial capital. Over the four WF years the seven curves' annualized
volatilities range from $3.5\%$ (Edge Allocation) to $6.7\%$ (Volatility
Targeting). Method-specific volatility differences appear as parallel offsets rather
than as divergent trends. Each curve steepens after the WF/OOT split at
2025-01-01, consistent with the higher OOT Sharpes in
Table~\ref{tab:headline_metrics}. Edge Allocation finishes the five-year sample
at \$8.15M (a $63.0\,\%$ cumulative gain, a geometric annualized return of
$10.8\,\%$). Volatility Targeting and Gamma-Budgeted Sizing reach the highest
terminal NAVs, $\$10.38$M and $\$10.49$M ($107.7\,\%$ and $109.5\,\%$
    cumulative). Their higher terminal NAVs reflect higher annualized volatility,
    not higher risk-adjusted return, relative to Edge Allocation.
    Figure~\ref{fig:drawdown} plots the drawdown trajectory per method. Edge
    Allocation's deepest drawdown over the full sample is $-2.28\,\%$ on
    2022-06-28. The deepest drawdown across the seven methods reaches $-9.25\,\%$
    for the quarter-Kelly variant on 2024-08-07. None of the seven methods
    experiences a drawdown in the $-30\,\%$ to $-50\,\%$ range that an unhedged
    passive short-volatility baseline produces over the same sample
    (Table~\ref{tab:headline_metrics}).

    \begin{table}[!ht]
  \centering
  \caption{Headline performance, internal baselines, and external benchmarks.}
  \label{tab:headline_metrics}
  \resizebox{\textwidth}{!}{%
    \begin{tabular}{lrrrrrrrrrr}
      \toprule
                            & \multicolumn{5}{c}{Walk-forward (2021--2024)} & \multicolumn{5}{c}{Out-of-time (2025)}                                                                                                   \\
      \cmidrule(lr){2-6} \cmidrule(lr){7-11}
                            & Sharpe                                        & Sortino                                & Ann.\ return & Ann.\ vol.\  & Max DD & Sharpe  & Sortino & Ann.\ return & Ann.\ vol.\  & Max DD \\
      \midrule
      \multicolumn{11}{l}{\textit{Sizing methods}}                                                                                                                                                                     \\
      EA                    & 3.1048                                        & 4.5234                                 & 0.1091       & 0.0351       & 0.0228 & 5.7612  & 7.0291  & 0.1048       & 0.0182       & 0.0143 \\
      FMU                   & 2.5119                                        & 2.9925                                 & 0.1319       & 0.0525       & 0.0691 & 5.0632  & 6.0447  & 0.1795       & 0.0354       & 0.0196 \\
      SRS                   & 2.7597                                        & 3.4259                                 & 0.1270       & 0.0460       & 0.0614 & 5.2622  & 6.9824  & 0.1755       & 0.0334       & 0.0172 \\
      VT                    & 2.2102                                        & 2.4040                                 & 0.1489       & 0.0674       & 0.0900 & 4.6669  & 4.9068  & 0.2388       & 0.0512       & 0.0343 \\
      GB                    & 2.0554                                        & 2.2676                                 & 0.1366       & 0.0665       & 0.0899 & 5.0147  & 5.8707  & 0.2995       & 0.0597       & 0.0338 \\
      HK                    & 1.9464                                        & 2.2562                                 & 0.1200       & 0.0616       & 0.0860 & 4.3711  & 5.1981  & 0.2091       & 0.0478       & 0.0271 \\
      QK                    & 1.8965                                        & 2.1857                                 & 0.1186       & 0.0625       & 0.0925 & 4.3084  & 5.1671  & 0.1993       & 0.0463       & 0.0273 \\
      \midrule
      \multicolumn{11}{l}{\textit{Internal baselines}}                                                                                                                                                                 \\
      Random                & 0.0844                                        & 0.0912                                 & 0.0131       & 0.1550       & 0.3935 & 0.3415  & 0.3626  & 0.0570       & 0.1669       & 0.1091 \\
      Always-P25d           & -0.0191                                       & -0.0203                                & -0.0030      & 0.1571       & 0.3438 & 0.0005  & 0.0005  & 0.0001       & 0.1663       & 0.1273 \\
      Always-P45d           & -0.0405                                       & -0.0456                                & -0.0064      & 0.1581       & 0.3396 & 0.1577  & 0.1752  & 0.0248       & 0.1572       & 0.1146 \\
      Momentum              & -0.0051                                       & -0.0057                                & -0.0008      & 0.1616       & 0.2894 & -0.3982 & -0.4276 & -0.0669      & 0.1681       & 0.1490 \\
      Rolling-Sharpe        & 0.5569                                        & 0.5991                                 & 0.0698       & 0.1254       & 0.2383 & 0.6128  & 0.6258  & 0.0545       & 0.0890       & 0.0830 \\
      \midrule
      \multicolumn{11}{l}{\textit{External index benchmarks}}                                                                                                                                                          \\
      SPX buy-and-hold (BH) & 0.6221                                        & 0.8871                                 & 0.1011       & 0.1625       & 0.3086 & 0.4622  & 0.6787  & 0.0884       & 0.1912       & 0.2008 \\
      CBOE PUT index        & 0.6795                                        & 0.9091                                 & 0.0675       & 0.0993       & 0.1994 & 0.1754  & 0.2605  & 0.0252       & 0.1436       & 0.1642 \\
      CBOE WPUT index       & 0.1449                                        & 0.1879                                 & 0.0145       & 0.1003       & 0.2285 & 0.1312  & 0.1593  & 0.0151       & 0.1152       & 0.1339 \\
      \bottomrule
    \end{tabular}%
  }
  \source{Every metric is reported separately for the four-window walk-forward (WF, 2021--2024) 
  and the 2025 out-of-time hold-out (OOT). Strategy returns are daily profit and loss per unit of collateral. 
  The collateral is held in cash and earns the risk-free rate, so these returns are already in excess of it. 
  The CBOE PUT and WPUT indices are total-return series. The SPX series is a price index, so the continuous 
  dividend yield is accrued daily before conversion, otherwise buy-and-hold would be understated by its dividend stream. 
  All three are then converted to excess of EFFR to place both sides of the comparison on one basis. Sizing methods share the same model,
   selection layer, and confidence gate, and differ only in the per-day position-size rule.}
\end{table}

    \subsection{Multi-test-corrected confidence}
    \label{subsec:multi_test_correction}

    Table~\ref{tab:deployment_confidence} reports the Probabilistic Sharpe Ratio
    (PSR) and Deflated Sharpe Ratio (DSR) of \citet{bailey_deflated_2014} for the
    three sizing methods that occupy the top of both the walk-forward and
    out-of-time Sharpe rankings (Edge Allocation, Fixed Margin Utilization, and
    Short-Richness Scaling). PSR is reported separately against three external
    benchmarks: SPX buy-and-hold (BH), the CBOE PUT index, and the CBOE WPUT index.
    DSR is benchmark-independent and incorporates a multi-test correction over
    $N_t=75$ trials. The count includes every search that ranked its candidates by
    realized performance: the 28 configurations reported in this article (the
    headline strategy, 10 structural sensitivity runs, 15 feature-group
    ablations, and 2 post-processed execution-drag variants), the 40
    confidence-gate calibrations of Section~\ref{sec:gate} (an 8-point
    trade-rate grid scored on Sortino in each of the five windows), the six
    sizing methods other than the one carried as headline, and the all-options
    variant abandoned in the trial phase of Section~\ref{sec:universe}.

    Two larger searches do not enter the count. The 250 hyperparameter trials,
    50 in each of the five windows, are scored on NDCG@1 over the inner
    validation slice. The 740 position-size grid evaluations are scored on the
    gap between realized training volatility and the 16\,\% anchor of
    Equation~\eqref{eq:equal_vol_calibration}. Neither search ranks its
    candidates by realized performance, so neither inflates the expected maximum
    Sharpe ratio that the correction subtracts. For the sizing methods whose
    $\theta$ enters as a proportional scale, the Sharpe ratio is invariant to
    position scale, so those grid points cannot differ in the statistic being
    deflated.

    All three methods produce PSR values above $0.95$ against every external
    benchmark on both the walk-forward and out-of-time slices. The smallest of the
    18 PSR cells is Edge Allocation's $0.964$ against SPX buy-and-hold on the
    out-of-time slice. Under the multi-test correction the DSR falls below every PSR cell for the same method and slice. Edge
    Allocation and Short-Richness Scaling produce walk-forward DSR values above
$0.95$ ($0.996$ and $0.956$ respectively), while FMU's walk-forward DSR is
$0.917$. On the out-of-time slice, the three DSR values are $0.856$ for Edge
    Allocation, $0.864$ for FMU, and $0.913$ for Short-Richness Scaling. The
    out-of-time DSR shortfall follows from the combination of a single calendar
    year of hold-out data and the $N_t=75$ trial penalty. The out-of-time Sharpe
    values before the multi-test deflation remain substantially higher than their walk-forward
    counterparts (Table~\ref{tab:headline_metrics}), and the PSR results indicate
    separation from passive short-volatility benchmarks even after accounting for
    non-Gaussianity in daily returns.

    \begin{table}[!ht]
\centering
\caption{Multi-test-corrected statistical confidence for top-3 sizing methods.}
\label{tab:deployment_confidence}
\begin{tabular}{llrrrr}
\toprule
Method & Slice & PSR vs BH & PSR vs PUT & PSR vs WPUT & DSR \\
\midrule
EA  & WF  & 0.9999 & 0.9998 & 1.0000 & 0.9962 \\
EA  & OOT & 0.9636 & 0.9710 & 0.9721 & 0.8563 \\
\midrule
FMU & WF  & 0.9837 & 0.9806 & 0.9967 & 0.9168 \\
FMU & OOT & 0.9836 & 0.9887 & 0.9893 & 0.8638 \\
\midrule
SRS & WF  & 0.9934 & 0.9919 & 0.9989 & 0.9561 \\
SRS & OOT & 0.9944 & 0.9966 & 0.9968 & 0.9128 \\
\bottomrule
\end{tabular}
\source{Top-3 methods selected by walk-forward Sharpe (Edge Allocation, Fixed Margin Utilization, and Short-Richness Scaling), which is also the top-3 set by out-of-time Sharpe. PSR and DSR follow \citet{bailey_deflated_2014}. The DSR multi-test correction uses $N_t=75$ trials. PSR is reported against each of three external benchmarks: BH = SPX buy-and-hold, PUT = CBOE PUT index, WPUT = CBOE WPUT index. For the four methods outside the top three, PSR against the worst of the three benchmarks is $0.9240$ (WF) and $0.9493$ (OOT) for Volatility Targeting, $0.9063$ and $0.9773$ for Gamma-Budgeted Sizing, $0.9264$ and $0.9722$ for half-Kelly, and $0.9203$ and $0.9684$ for quarter-Kelly.}
\end{table}

    \subsection{Year-by-year heterogeneity}
    \label{subsec:year_heterogeneity}

    Table~\ref{tab:headline_year}, in Appendix~\ref{sec:appendix_year}, reports
    per-year Sharpe ratios across the seven sizing methods, with each of the four walk-forward years (2021--2024) and the
    2025 out-of-time hold-out shown separately. The four-year range is wide. For
    Edge Allocation, per-year Sharpe ranges from $0.41$ in 2024 to $6.52$ in
    2021. The same pattern holds for the other six methods: every method's
    strongest year is 2021 (per-year Sharpe in the range $6.52$ to $9.21$), and
    every method's weakest year is 2024, with only Edge Allocation (at $0.41$) and
    Volatility Targeting (at $0.05$) posting a positive 2024 Sharpe and the
    remaining five methods negative. The 2022 and 2023 years are intermediate.
    All seven methods fall between $1.37$ and $2.51$ in 2022, and the five methods
    that trade in 2023 fall between $1.79$ and $3.93$.

    The 2025 out-of-time year is comparable to but somewhat below 2021 across all
    seven methods, with per-year Sharpe in the range $4.31$ (QK) to
$5.76$ (Edge Allocation). The hold-out year was not used during training, hyperparameter
    search, or model selection, and that separation is what the
    out-of-time design establishes. The credibility of that separation rests on the design of the hold-out, not
    on the size of the Sharpe ratio the hold-out produced. For Edge Allocation the 2025 figure of $5.76$ sits
    closer to the strongest walk-forward year, $6.52$ in 2021, than to the $2.51$,
$1.79$ and $0.41$ recorded in 2022, 2023 and 2024. Every method's out-of-time
    Sharpe also exceeds its own walk-forward figure by more than $2.4$ units, which
    is the reverse of the direction a hold-out year would be expected to move. One hold-out year is
    therefore not sufficient to separate generalization from a favorable regime,
    and the four-year range reported above is the more informative summary of
    performance across volatility states. The out-of-time figures reported in the
    remainder of this article should be read against that range.

    The two fractional-Kelly variants (HK and QK) post a Sharpe of
    exactly zero in 2023. Both rules abstain when the estimated Kelly fraction
    falls below a positive threshold, and the 2023 zero reflects the abstention
    rule binding throughout that year rather than a degenerate trade or data error.
    The two variants trade again in 2024 but at a loss (Sharpe of $-0.58$ for HK
    and $-0.62$ for QK) before recovering in 2025 (Sharpe of $4.37$ and $4.31$).

    \subsection{Performance across regimes}
    \label{subsec:regimes}

    Table~\ref{tab:regime_vix}, in Appendix~\ref{sec:appendix_regimes}, reports
    per-VIX-regime Sharpe ratios across the seven sizing methods on the walk-forward and out-of-time slices. The VIX regime
    is computed from the entry-day close, with thresholds following the conventions
$\mathrm{VIX} < 15$ (low), $15 \leq \mathrm{VIX} \leq 25$ (mid), and
$\mathrm{VIX} > 25$ (high). Regime-conditional figures cover every day whose
    VIX close falls in the regime, with days on which the model abstained or the
    gate blocked entry entering at a return of zero, so they use the same
    convention as Table~\ref{tab:headline_metrics} applied to a subset of days.
    The Days and Trades columns of Table~\ref{tab:regime_vix} report the two
    counts separately. Walk-forward performance is strongly
    conditional on VIX regime across all seven methods, and the conditioning does
    not increase monotonically with the regime. The annualized return peaks in the
    mid-VIX band for six of the seven methods. Edge Allocation is the exception
    and rises across all three regimes, at $0.50\%$ on low-VIX days, $13.90\%$ on
    mid-VIX days, and $15.40\%$ on high-VIX days. For Volatility Targeting and
    Gamma-Budgeted Sizing the high-VIX return sits below the low-VIX return, at
$3.80\%$ against $7.26\%$ and $4.81\%$ against $5.35\%$. The risk-adjusted ratio
    concentrates in the mid band more sharply still. Mid-VIX days carry the
    strongest regime-conditional Sharpe for five of the seven methods, and the
    mid-band ratios run from $3.59$ (QK) to $4.48$ (Gamma-Budgeted Sizing),
    against $0.29$ to $2.77$ on high-VIX days. Volatility Targeting and
    Gamma-Budgeted Sizing peak in the low band instead, at $5.46$ and $6.62$,
    where their own return volatility is smallest at $1.33\%$ and $0.81\%$. For
    Edge Allocation the annualized volatility of the strategy's own returns rises
    from $1.91\%$ to $3.28\%$ to $5.57\%$ across the three regimes, faster than
    the return it collects, which accounts arithmetically for the mid-band peak.
    One reading is that on the most elevated days realized volatility catches up
    with implied, and the additional premium arrives with more than proportional
    risk. The premium is
    therefore harvested most efficiently one regime below the top of the
    implied-volatility distribution, and for Volatility Targeting and
    Gamma-Budgeted Sizing two regimes below.

    The 2025 out-of-time slice reproduces this ordering only in the regime with an
    adequate sample. The mid-VIX regime holds $196$ of Edge Allocation's $237$
    out-of-time days and produces a regime-conditional Sharpe of $4.94$, somewhat above the
$3.59$ to $4.48$ range on the walk-forward slice. The two outer regimes rest on
$20$ and $21$ days. Annualized statistics computed on that many observations
    carry no useful precision. That is why the low-VIX cell reaches $13.71$ for
    Edge Allocation and $83.29$ for Gamma-Budgeted Sizing on an annualized return
    of $23.78\%$. Those cells are reported for completeness and are not
    interpreted here. The VIX-regime evidence rests on the walk-forward panel,
    measured on $148$ high-VIX, $593$ mid-VIX, and $223$ low-VIX days.

    Table~\ref{tab:regime_rv} reports per-tercile figures across the seven
    sizing methods. Realized volatility is the 5-day intraday realized volatility
    of the S\&P 500 measured at the entry-day close, partitioned into
    equal-quantile terciles within each slice, so a tercile label spans different
    volatility levels on the two panels. The conditioning appears most clearly in
    annualized returns. Annualized return in the top tercile exceeds the bottom
    tercile for all seven methods on the out-of-time slice and for six of the
    seven on walk-forward, with the exception of Volatility Targeting at $13.22\%$ against
$13.53\%$. It rises monotonically across the three terciles for six of the
    seven on the out-of-time slice. For Edge
    Allocation it runs $4.11\%$, $10.25\%$ and $17.67\%$ out of time. The Sharpe
    ordering is less regular. On the walk-forward slice the highest
    tercile-conditional Sharpe sits in tercile 2 for six of the seven methods, Edge
    Allocation excepted, whose ratio rises across all three ($2.77$, $2.98$,
$3.55$). On the out-of-time slice no method is monotone across the three
    terciles, although tercile 3 exceeds tercile 1 for six of the seven. The
    pattern parallels the VIX-regime result in the return dimension. Higher
    realized volatility coincides with a wider differential between implied and
    realized volatility, so the strategy earns more per day. The volatility of its
    own returns responds less uniformly across methods, rising with the tercile for
    six of the seven but falling for Edge Allocation out of time. The Sharpe
    ordering inherits that unevenness.

\begin{figure}[!ht]
        \centering
        \caption{Per-regime Sharpe ratios across sizing methods.}
        \label{fig:regime}
        \includegraphics[width=\textwidth]{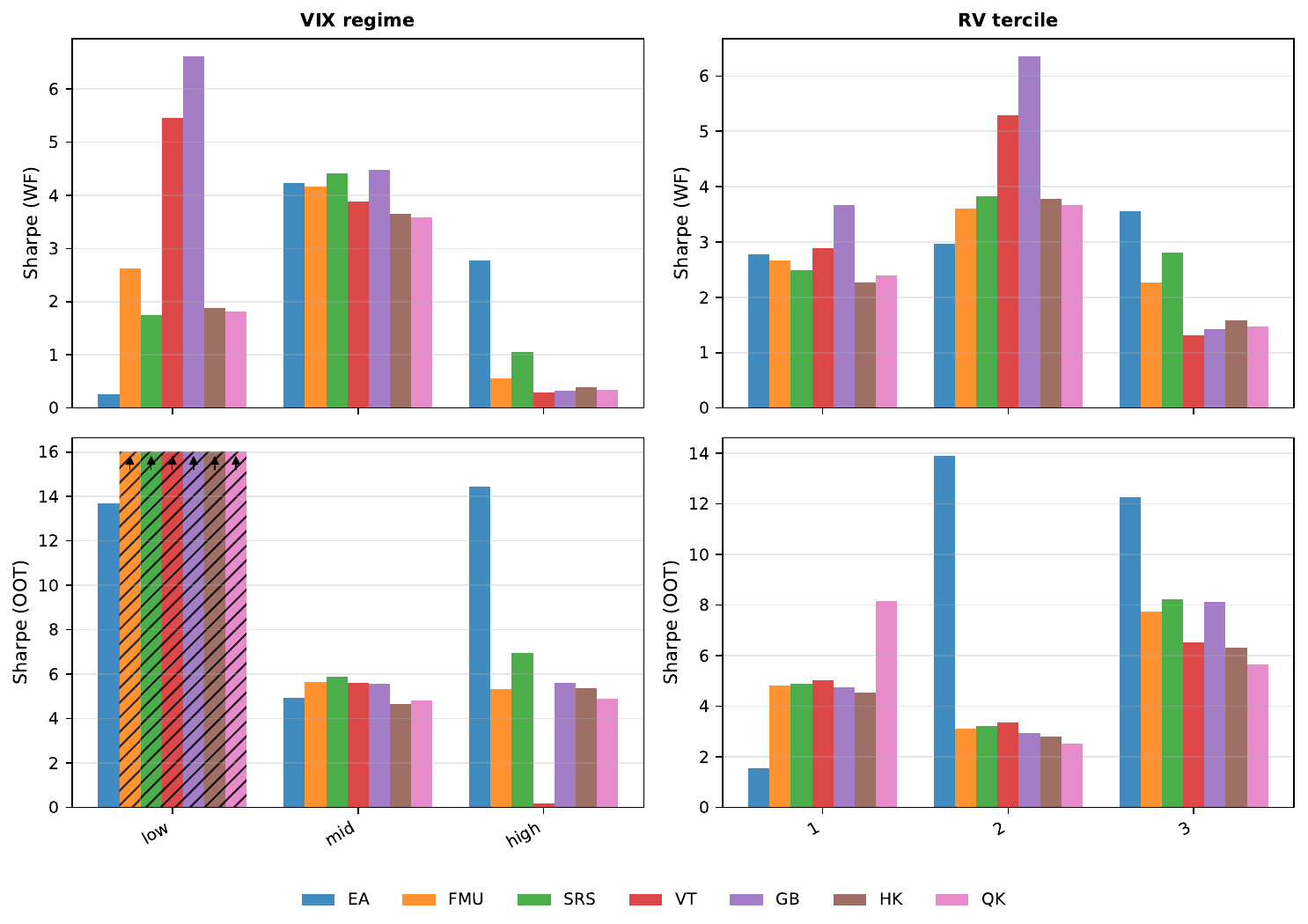}
        \source{Two regime axes (left to right): VIX regime and realized-volatility tercile. Top row reports the four-window walk-forward (2021--2024) regime-conditional Sharpe. Bottom row reports the 2025 out-of-time hold-out regime-conditional Sharpe. Each cell covers every day in the regime, with abstention days entering at a return of zero. Each panel reports seven grouped bars per regime label, one per sizing method, ordered as in Table~\ref{tab:headline_metrics}. The out-of-time VIX panel is capped at $16$, with bars above the cap drawn hatched. Its low-VIX cell rests on $21$ days and reaches a Sharpe of $88.8$, which on an unclipped axis flattens every interpretable bar. Per-year Sharpe ratios are reported in Table~\ref{tab:headline_year}.}
    \end{figure}

    Figure~\ref{fig:regime} reports per-regime Sharpe ratios across all seven
    sizing methods for the two regime axes (VIX and realized-volatility
    tercile), separately for walk-forward and out-of-time. The qualitative
    pattern within each regime axis is similar across the seven methods, with
    magnitude variation that reflects the different leverage levels applied by each
    sizing rule. The 2025 out-of-time bars are larger than the walk-forward bars
    in 34 of the 42 regime cells, consistent with the WF/OOT
    pattern visible in the headline (Table~\ref{tab:headline_metrics}). Six of the
    eight exceptions fall in the middle realized-volatility tercile, where every
    method but Edge Allocation inverts. The
    cross-regime variability dominates the cross-method variability, so the regime
    axis is the more informative dimension for understanding when the strategy's
    regime-conditional Sharpe is largest.

    \subsection{Benchmark comparison}
    \label{subsec:benchmarks}

    Table~\ref{tab:headline_metrics} reports walk-forward and out-of-time Sharpe
    ratios for the seven sizing methods alongside five internal baselines and three
    external index benchmarks. The internal baselines disable the LightGBM ranker,
    the confidence gate, or both, but otherwise share the same option universe and
    execution as the strategy. Three of the five internal baselines produce negative
    Sharpe ratios on at least one slice: Always-short-25-delta-put ($-0.019$ WF,
$0.000$ OOT), Always-short-45-delta-put ($-0.041$ WF, $0.158$ OOT), and
    Momentum ($-0.005$ WF, $-0.398$ OOT). Random earns $0.084$ on walk-forward and
$0.342$ on out-of-time. The fifth
    baseline, a Rolling-Sharpe selector that picks the candidate with the highest
    30-day rolling Sharpe, earns a positive but modest $0.557$ on walk-forward and
$0.613$ on out-of-time. Every model variant exceeds the best internal baseline
    on both slices, with the smallest cross-method gap to Rolling-Sharpe at $1.34$
    on walk-forward (QK) and $3.70$ on out-of-time (QK).

    External benchmarks contextualize the strategy's performance against tradable
    passive comparators. The S\&P 500 buy-and-hold (BH) earns a Sharpe of $0.62$
    on walk-forward and $0.46$ on out-of-time. The CBOE PUT index earns $0.68$ on
    walk-forward and $0.18$ on out-of-time. The CBOE WPUT index earns $0.14$ and
$0.13$ respectively. All seven of the model's sizing methods exceed every
    external benchmark on both slices. Edge Allocation produces the largest gap in
    every (slice $\times$ benchmark) cell, ranging from $2.43$ (PUT, WF) to
$5.63$ (WPUT, OOT). The smallest single (method $\times$ benchmark) gap on the
    walk-forward slice is the quarter-Kelly variant's $1.22$ against the PUT
    index, and the smallest gap on the out-of-time slice is the quarter-Kelly
    variant's $3.85$ against BH. The strategy's gap above the passive short-volatility
    benchmarks (PUT, WPUT) is comparable in magnitude to its gap above outright
    equity exposure (BH). The model's outperformance is therefore not simply
    explained by inheriting the passive volatility risk premium.

    The \citet{diebold_comparing_1995} test for equal predictive accuracy on daily
    profit-and-loss provides a more conservative statistical assessment than the
    Sharpe-based PSR reported in Section~\ref{subsec:multi_test_correction}. On the
    2025 out-of-time slice, the test does not reject the null of equal mean daily
    P\&L for any of the seven methods at conventional levels after a Bonferroni
    correction over the three external benchmarks. The corrected minimum p-value is
$1.000$ for Edge Allocation, Fixed Margin Utilization, and
    Short-Richness Scaling, and the smallest among the seven methods is $0.231$ for
    Gamma-Budgeted Sizing. The contrast with the PSR results reflects the
    difference between the two tests: PSR compares a Sharpe ratio against a
    benchmark's Sharpe and benefits from the strategy's much lower realized
    volatility, while the DM test compares mean daily P\&L levels and is dominated
    by the noise in daily return differences.


    \section{Robustness}
    \label{sec:robustness}

    \subsection{Walk-forward fragility, out-of-time robustness}
    \label{subsec:wf_oot_pattern}
    
    The cross-cutting pattern of walk-forward Sharpe against out-of-time Sharpe for
    each sensitivity perturbation is that walk-forward Sharpe varies far more than out-of-time Sharpe. For Edge
    Allocation, walk-forward Sharpe ranges from $-0.157$ (training window shortened
    to two years) to $3.308$ (hyperparameter search reduced to 25 trials), a
    3.47-unit spread. Across the cells that trade, out-of-time Sharpe ranges from
    $4.817$ (correlation-clustering threshold loosened to $0.95$) to $5.800$
    (hyperparameter search widened to 100 trials), a 0.98-unit spread that is
    roughly three and a half times tighter than the walk-forward spread. One cell
    falls outside that comparison. At a correlation-clustering threshold of $0.90$
    the confidence gate suppresses every 2025 trade, so the account is in cash for
    the full year and its excess return is zero on every day, which leaves the
    ratio undefined. The $0.000$ recorded for that cell in
    Table~\ref{tab:sensitivity_master} is the convention this article applies to a
    year without positions, not a measured performance.

    In economic terms, the underlying signal the model fits in the SPXW
    0DTE market is stable enough to be expressed in 2025 even when training choices
    are perturbed. The walk-forward Sharpe estimator depends materially on which
    years are used for training, how the years are partitioned across the four
    expanding-window cycles, and how aggressively the hyperparameters are searched.
    The 2025 hold-out, which is never used during training or hyperparameter
    search, is the integrity check, and its near-uniform robustness across the
    perturbation matrix is evidence that the out-of-sample Sharpe is insensitive
    to the training choices perturbed here, with the one exception noted below.


    \subsection{Per-dimension structural sensitivities}
    \label{subsec:per_dim_sensitivity}

    Table~\ref{tab:sensitivity_master} reports walk-forward and out-of-time Sharpe
    ratios, headline-relative $\Delta$ Sharpe, and PSR under Edge Allocation for
    each perturbation.

    With the training window shortened from 3 years to 2, EA's walk-forward Sharpe
    falls to $-0.157$ (a $-3.26$-unit move from the headline) and PSR (WF)
    collapses to $0.050$. Out-of-time Sharpe holds at $5.178$, only $-0.58$ below
    the headline. The 2-year window gives the LightGBM ranker too few examples per
    training cycle, so the walk-forward Sharpe estimator becomes noisy, while the
    2025 hold-out result is largely unchanged.
    Across the seven sizing methods, walk-forward Sharpe drops by between $-2.0$
    and $-3.1$ units while out-of-time Sharpe loses less than one unit per method.

    Switching from an expanding-window walk-forward to a 3-year rolling window
    reduces EA's walk-forward Sharpe to $0.564$ (a $-2.54$-unit move) but holds
    out-of-time Sharpe at $5.227$ (only $-0.53$). The rolling 3-year window
    discards earlier training data each cycle, so the training set is consistently
    smaller, but the 2025 hold-out still receives a model whose generalization
    error is comparable to the headline. The seven-method dispersion mirrors the EA
    pattern: walk-forward Sharpe drops by $-1.5$ to $-2.4$ units across methods.

    Adjusting the volatility-target anchor from the headline 16\% to either 12\% or
    20\% leaves Sharpe essentially unchanged. EA's walk-forward Sharpe is $3.036$
    at 12\% (a $-0.07$ move from the headline) and $3.105$ at 20\% (no change),
    and out-of-time Sharpe is $5.739$ at 12\% ($-0.02$) and $5.761$ at 20\% (no
    change). The other
    six sizing methods show similarly small changes, with absolute Sharpe shifts
    under $0.1$ unit on both slices.

    The 20\% arm should be read with care. Raising the anchor leaves $\theta^\star$
    unchanged in $24$ of $35$ method-window cells, and in all five EA windows,
    because $\theta^\star$ already sits at the top of its grid at the headline
    anchor. For EA the 20\% arm therefore reproduces the headline run without
    perturbing it, and its agreement with the headline is not evidence of
    insensitivity to the anchor. The 12\% arm moves $\theta^\star$ in one EA window,
    where it falls from $0.80$ to $0.70$. Where $\theta^\star$ is free to move, the
    calibration rescales exposure roughly proportionately, which leaves the Sharpe
    ratio close to unchanged.

    Reducing the hyperparameter search budget from 50 trials to 25 raises EA's
    walk-forward Sharpe to $3.308$ (a $0.20$ move) and leaves out-of-time Sharpe
    essentially unchanged at $5.736$. Increasing the budget to 100 trials lowers
    walk-forward Sharpe to $0.706$ ($-2.40$) but leaves out-of-time Sharpe at
$5.800$ ($0.04$). The non-monotonic walk-forward response is consistent with
    multiple-testing over a finite hyperparameter space: more trials over-fit the
    walk-forward objective without producing a generalizably better model. The 2025
    out-of-sample result is unaffected in either direction, which confirms that
    walk-forward Sharpe is the more sample-period-sensitive estimator, as discussed
    in Section~\ref{subsec:wf_oot_pattern}.

    The correlation-clustering threshold controls how aggressively the feature
    pipeline collapses redundant features into single representatives. Loosening it
    from $0.85$ to $0.95$ lets more features survive, which lowers EA's
    walk-forward Sharpe to $0.976$ ($-2.13$) and out-of-time Sharpe to $4.817$
    ($-0.94$). The intermediate setting of $0.90$ produces the matrix's most severe
    out-of-time outcome. Walk-forward Sharpe falls to $0.281$, and in 2025 the gate
    places no position on any day, for all seven sizing methods alike. The $0.000$
    recorded out of time therefore reports an account that never traded rather than
    one that traded and earned nothing. Loosened correlation clustering admits
    more redundant features, which destabilizes the top-1-versus-top-2 confidence
    signal that the gate calibrates against. The gate's hold-out $\tau$ calibration
    responds by tightening, and at the 2025 inference time no day's confidence
    clears the tightened threshold, so the gate suppresses every trade. The
    out-of-time failure is a clean diagnosis that the gate's calibration is
    contingent on appropriately-clustered feature representations.

    The perturbation matrix also locates where this design is fragile. The wide
    walk-forward dispersion and the single out-of-time failure have different
    sources. Nine of the ten structural perturbations leave out-of-time Sharpe
    between $4.817$ and $5.800$ while moving walk-forward Sharpe across a range of
    more than three units. The tenth differs from the others in kind, not in
    degree. It does not produce a worse set of positions. It produces none, and the
    component responsible is the confidence gate rather than the ranker or the
    sizing rule. The gate is also the only element of the pipeline governed by a
    single calibrated scalar, which makes its calibration the first place to look
    when the pipeline is perturbed.

    That failure is identifiable without reference to realized profit. The gate's
    threshold is calibrated on the union of the four walk-forward held-out slices,
    965 days in total, and the calibration records the fraction of those days that
    clear the threshold. Under the headline configuration the fraction is $1.0000$
    at $\tau^{\star} = 0.0000$, and the 2025 slice admits every day, so the
    calibrated and realized admission rates agree. Under the 0.90 perturbation the
    same procedure sets $\tau^{\star} = 0.0689$ and a calibrated admission rate of
    $0.3005$, and the 2025 slice admits no day. The divergence between the two
    rates is present on the first morning the gate is consulted, whereas any
    return-based diagnostic requires positions to have been opened and closed
    first. The computation treats successive days as independent, an approximation,
    since the confidence signal is serially correlated. Under a calibrated rate
    of $0.3005$, ten consecutive abstentions have probability $0.028$ and twenty
    have probability below $0.001$. The state the 0.90 cell
    produces is therefore separable from ordinary abstention within two trading
    weeks, on the signal alone.

    Switching the calibration objective from equal-volatility scaling to in-sample
    Sharpe maximization gives EA walk-forward Sharpe of $2.947$ ($-0.16$) and
    out-of-time Sharpe of $5.615$ ($-0.15$). Switching to in-sample Sortino
    maximization gives EA walk-forward Sharpe of $2.902$ ($-0.20$) and out-of-time
    Sharpe of $5.292$ ($-0.47$). For Edge Allocation specifically, the choice of
    calibration objective is a small operational degree of freedom, not a
    structural change to the strategy. Across the seven sizing methods, both
    calibration variants leave the cross-method ordering of out-of-time Sharpe
    ratios unchanged.

    \begin{table}[!ht]
\centering
\caption{Per-dimension sensitivity matrix under Edge Allocation, headline-relative.}
\label{tab:sensitivity_master}
\resizebox{\textwidth}{!}{%
\begin{tabular}{lrrrrrr}
\toprule
Dimension & Sharpe (WF) & Sharpe (OOT) & $\Delta$ Sharpe (WF) & $\Delta$ Sharpe (OOT) & PSR (WF) & PSR (OOT) \\
\midrule
\textit{Headline (volatility anchor 16\%)} & 3.105 & 5.761 & 0.000 & 0.000 & 1.000 & 0.964 \\
\midrule
Training window 2y          & $-0.157$ & 5.178 & $-3.262$ & $-0.583$ & 0.050 & 0.952 \\
Rolling-3y window           &  0.564 & 5.227 & $-2.541$ & $-0.534$ & 0.454 & 0.953 \\
Volatility anchor 12\%             &  3.036 & 5.739 & $-0.069$ & $-0.023$ & 1.000 & 0.964 \\
Volatility anchor 20\%             &  3.105 & 5.761 &  $0.000$ &  $0.000$ & 1.000 & 0.964 \\
HP trials 25                &  3.308 & 5.736 &  $0.203$ & $-0.025$ & 1.000 & 0.963 \\
HP trials 100               &  0.706 & 5.800 & $-2.399$ &  $0.039$ & 0.530 & 0.959 \\
Corr threshold 0.95         &  0.976 & 4.817 & $-2.129$ & $-0.944$ & 0.632 & 0.993 \\
Corr threshold 0.90         &  0.281 & 0.000 & $-2.824$ & $-5.761$ & 0.281 & 0.000 \\
Calib: Sharpe-max           &  2.947 & 5.615 & $-0.158$ & $-0.146$ & 1.000 & 0.966 \\
Calib: Sortino-max          &  2.902 & 5.292 & $-0.202$ & $-0.469$ & 1.000 & 0.960 \\
\midrule
Execution: 75\% spread cov. &  2.826 & 5.426 & $-0.279$ & $-0.335$ & 0.999 & 0.957 \\
Execution: sell-at-bid      &  2.725 & 5.311 & $-0.380$ & $-0.450$ & 0.999 & 0.954 \\
\bottomrule
\end{tabular}%
}
\source{Edge Allocation walk-forward (WF, 2021--2024) and 2025 out-of-time (OOT) Sharpe ratios under each structural and execution sensitivity perturbation, with $\Delta$ Sharpe relative to the headline (volatility anchor 16\%, equal-volatility calibration, expanding-window walk-forward, 3-year training window, 50 hyperparameter trials, correlation-clustering threshold $0.85$). Each row reports one perturbation. Structural sensitivities sit between the two horizontal rules. The two execution rows below the second rule are post-processed re-runs of the headline backtest using the same selection layer, and their results are described in Section~\ref{subsec:execution_drag}. PSR is reported against the worst of the three external benchmarks per slice. The out-of-time Sharpe of $0.000$ at correlation threshold $0.90$ records a year in which the confidence gate placed no position on any day, so no return series exists for that cell.}
\end{table}

    \subsection{Feature-group ablation}
    \label{subsec:ablation}

\begin{figure}[!ht]
        \centering
        \caption{Feature-group ablation impact bars across sizing methods.}
        \label{fig:ablation_bars}
        \includegraphics[width=\textwidth]{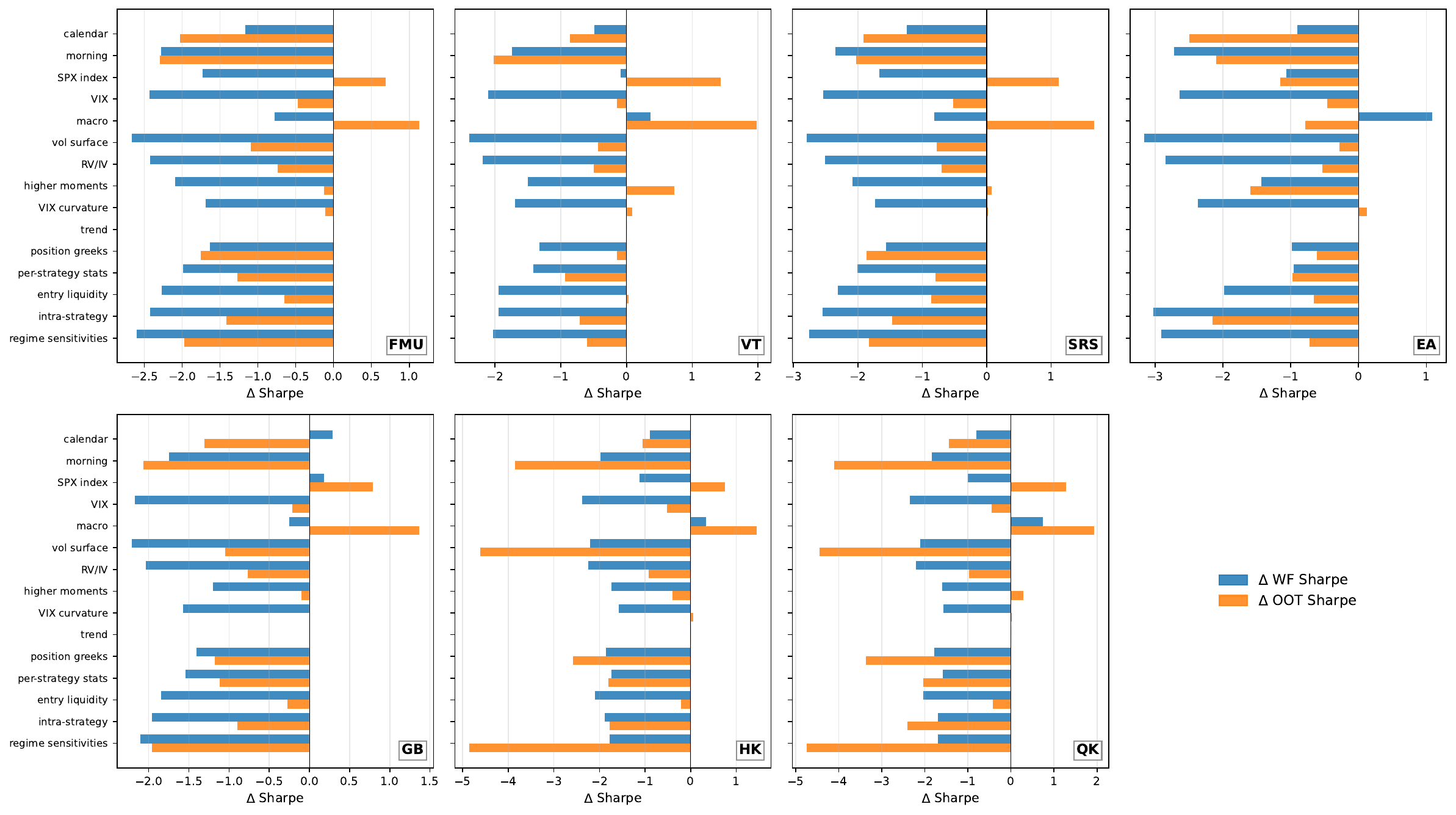}
        \source{Per-method horizontal bar charts of $\Delta$ Sharpe relative to the headline for each of the fifteen feature-group ablations. Each panel reports one sizing method, with two bars per group: $\Delta$ walk-forward Sharpe and $\Delta$ 2025 out-of-time Sharpe. Bars are computed as the per-method Sharpe at each ablation cell minus the headline per-method Sharpe. The trend bar is exactly zero in both slices because none of the four trend features survives the selection pipeline in any walk-forward window. The macro bar is the only positive walk-forward impact for Edge Allocation in the panel.}
    \end{figure}

    We test the marginal contribution of each of fifteen feature groups by dropping
    one group at a time from the candidate feature set fed to the selection
    pipeline. Ten groups are cross-sectional (one value per day, broadcast across
    all candidates): calendar and event flags, morning-session features, SPX index
    features, VIX family and futures, macroeconomic indicators, implied-volatility
    surface, realized-minus-implied volatility differentials, realized higher moments, VIX
    term-structure curvature, and trend. The remaining five are per-strategy (one
    value per day-candidate pair): position greeks, per-strategy rolling
    statistics, entry liquidity, intra-strategy term context, and
    regime-conditional sensitivities (per-strategy quantities multiplied by
    cross-sectional regime variables). Two derived families are source-tagged to
    their primary-input group and dropped alongside the native features in each
    ablation. The within-day ranks are ordinal ranks of six per-strategy features
    across each day's candidates. The tail-risk features are three stress signals
    added to the catalog: the relative gap between 5-day realized volatility and
    the 1-week ATMF implied volatility priced over the preceding week, the
    candidate strategy's 5-day drawdown observed at the prior close, and the
    60-day z-score of the VIX. Each is documented in Appendix~\ref{app:features}.
    Table~\ref{tab:ablation_all7} reports walk-forward and out-of-time Sharpe
    ratios per group for Edge Allocation, and Figure~\ref{fig:ablation_bars}
    plots $\Delta$ Sharpe bars per group for each of the seven sizing methods.

    Three groups stand out as Sharpe-critical for the walk-forward slice:
    volatility surface ($\Delta$WF $= -3.16$), intra-strategy ($\Delta$WF $= -3.02$),
    and regime-conditional sensitivities ($\Delta$WF $= -2.91$). Two further groups produce large
    walk-forward impact but smaller out-of-time impact: vix ($-2.64$ WF, $-0.46$
    OOT) and RV/IV ($-2.84$ WF, $-0.54$ OOT). Two are execution-quality-critical:
    entry liquidity ($-1.98$ WF, $-0.66$ OOT) and morning ($-2.72$ WF, $-2.10$
    OOT). Three further groups have modest walk-forward impact: calendar
    ($\Delta$WF $-0.91$), strategy ($-0.96$), and position ($-0.98$), although
    calendar's out-of-time drop of $-2.50$ is the largest in the matrix. VIX curvature carries a larger walk-forward cost
    ($\Delta$WF $-2.37$), comparable to the VIX and entry-liquidity groups.
    Cross-method agreement is strong on the walk-forward slice and weak out of
    time. Walk-forward group orderings correlate with the Edge Allocation
    ordering at Spearman $0.73$ to $0.94$. Out of time the rank correlations fall
    to between $0.08$ and $0.41$, and per-group differences from the Edge
    Allocation value reach $4.33$ units.

    The trend group has zero impact in either slice. Diagnostic inspection of the per-window feature-survival data
    shows that none of the four features survives the selection pipeline in any of
    the four walk-forward windows or the out-of-time window: their cross-day Spearman correlation with the
    daily mean label sits below the relevance threshold ($\rho < 0.05$) in every
    window. Trend signals are a natural
    hypothesis for short-volatility strategy selection, but the sample provides no
    evidence that they help within-day selection at this threshold. The mechanism
    is that medium-frequency directional signals do not vary enough across
    candidates on a given day to be useful for within-day selection, and they do
    not explain enough day-to-day variation in the daily mean label to clear the
    relevance threshold either.

    Macro is the only group whose removal raises Edge Allocation's walk-forward
    Sharpe. It
    rises from $3.105$ to $4.188$ without macro features ($\Delta$WF $= +1.08$)
    while out-of-time Sharpe drops from $5.761$ to $4.975$ ($\Delta$OOT $=
-0.79$). On a raw-Sharpe reading, dropping macro features from the headline
    would be attractive. The PSR results argue against this. Without macro
    features, Edge Allocation's PSR (OOT) drops from $0.964$ to $0.931$ (just below
    the $0.95$ threshold used in Section~\ref{subsec:multi_test_correction}). That
    trade-off is specific to Edge Allocation. Dropping macro features raises
    out-of-time Sharpe for the other six sizing methods, by between $1.13$ and
    $1.98$ units, and leaves their PSR (OOT) at or above $0.95$. A reading consistent with the
    grid is that the six macro features suppress the third and fourth moments of
    the realized return distribution. Without them, individual trade outcomes
    are more dispersed and the tails are heavier. The realized mean Sharpe is
    higher, but the PSR estimator penalizes the heavier tails enough to push the
    corrected confidence below $0.95$.

    Regime-conditional sensitivities produce one of the steepest walk-forward PSR
    collapses in the matrix despite a modest out-of-time Sharpe drop. The seven multiplicative interactions
    in this group encode ``the same per-strategy quantity carries different weight
    in different macro regimes''. For Edge Allocation, walk-forward Sharpe drops
    from $3.105$ to $0.196$ ($\Delta$WF $= -2.91$) and PSR (WF) falls from
$1.000$ to $0.229$, the third-lowest walk-forward PSR in the matrix behind
    the volatility-surface ($0.092$) and intra-strategy ($0.165$) cells. Out-of-time Sharpe drops only modestly, from
$5.761$ to $5.043$ ($\Delta$OOT $= -0.72$), with PSR (OOT) at $0.933$. The
    asymmetry is interpretable, though it rests on a single out-of-time year and
    should not be over-read. Without the engineered products, the model must
    reconstruct any regime conditioning from the standalone per-strategy and regime
    features that remain in the candidate set. Across the variable 2021--2024
    walk-forward window, which spans the elevated-VIX 2022 year and the calmer
    2023--2024, the engineered interactions are decisive for the walk-forward
    ranking signal. Over the single 2025 out-of-time year the headline result is
    largely reproduced without them, which one hold-out year cannot distinguish
    from sampling variation. The position group ablation drops eight native greeks
    plus five sourced sensitivities, yet it costs only $-0.98$ in walk-forward
    Sharpe. The seven interaction terms therefore carry more of the
    regime-conditioning signal than the eight greeks they are sourced from.

    Group by group, the pattern is consistent with the strategy's
    design. The volatility surface and rv-iv groups jointly describe the volatility risk
    premium across the SPXW IV surface, and are included to indicate which
    delta bucket is mispriced today. The intra-strategy group encodes how close
    each candidate's realized features are to the targeted point on that surface
    (delta drift, IV at strike minus the at-the-money forward, days to expiry). The
    morning group conveys the only intraday information available before the 10:00
    ET entry decision and is particularly important for 0DTE strategies. The vix
    and vix curvature groups supply regime classification for the gate's
    confidence threshold. The entry liquidity group is the model's selection-time
    defense against poor realized fills. The remaining groups
    supply progressively more specialized conditioning that improves either the
    walk-forward Sharpe directly or the realized return distribution's tails as
    captured by the PSR estimator.

    \begin{table}[!ht]
\centering
\caption{Feature-group ablation impact under Edge Allocation, headline-relative.}
\label{tab:ablation_all7}
\resizebox{\textwidth}{!}{%
\begin{tabular}{lrrrrrr}
\toprule
Ablated group & Sharpe (WF) & Sharpe (OOT) & $\Delta$ Sharpe (WF) & $\Delta$ Sharpe (OOT) & PSR (WF) & PSR (OOT) \\
\midrule
\textit{Headline (no group dropped)}      & 3.105 & 5.761 & 0.000  &  0.000 & 1.000 & 0.964 \\
\midrule
Vol surface                                & $-0.051$ & 5.481 & $-3.156$ & $-0.281$ & 0.092 & 0.977 \\
Intra-strategy                             &  0.081 & 3.611 & $-3.024$ & $-2.150$ & 0.165 & 0.970 \\
Regime sensitivities                       &  0.196 & 5.043 & $-2.909$ & $-0.718$ & 0.229 & 0.933 \\
RV/IV                                      &  0.264 & 5.224 & $-2.841$ & $-0.538$ & 0.277 & 0.952 \\
Morning                                    &  0.389 & 3.660 & $-2.716$ & $-2.101$ & 0.348 & 0.895 \\
VIX                                        &  0.469 & 5.299 & $-2.635$ & $-0.462$ & 0.402 & 0.976 \\
VIX curvature                              &  0.735 & 5.887 & $-2.370$ & $0.126$ & 0.539 & 0.963 \\
Entry liquidity                            &  1.124 & 5.102 & $-1.981$ & $-0.660$ & 0.731 & 0.987 \\
Higher moments                             &  1.673 & 4.162 & $-1.431$ & $-1.599$ & 0.912 & 0.915 \\
SPX index                                  &  2.042 & 4.609 & $-1.062$ & $-1.152$ & 0.964 & 0.924 \\
Position greeks                            &  2.121 & 5.148 & $-0.984$ & $-0.614$ & 0.971 & 1.000 \\
Per-strategy stats                         &  2.145 & 4.791 & $-0.959$ & $-0.970$ & 0.958 & 0.935 \\
Calendar                                   &  2.199 & 3.265 & $-0.906$ & $-2.496$ & 0.969 & 0.922 \\
Trend                                      &  3.105 & 5.761 & $0.000$ & $0.000$ & 1.000 & 0.964 \\
Macro                                      &  4.188 & 4.975 & $1.083$ & $-0.786$ & 0.998 & 0.931 \\
\bottomrule
\end{tabular}%
}
\source{Edge Allocation walk-forward (WF, 2021--2024) and 2025 out-of-time (OOT) Sharpe ratios under each of the fifteen feature-group ablations, with $\Delta$ Sharpe relative to the headline (no group dropped). Each row reports one ablation. Rows are sorted by walk-forward $\Delta$ Sharpe, from the largest negative impact (volatility surface) to the only positive impact (macro). Ten of the fifteen groups are cross-sectional (one value per day broadcast to all candidates) and five are per-strategy (one value per day-candidate pair). Each ablation drops the named group's native features along with all source-tagged derived features (within-day ranks and tail-risk features), so a single ablation cell tests the joint marginal contribution of native and sourced features. PSR is reported against the worst of the three external benchmarks per slice. The trend row is exactly zero in both slices because none of the four trend features survives the selection pipeline in any walk-forward window, so that ablation leaves the model unchanged. Per-method values for the other six sizing methods are plotted in Figure~\ref{fig:ablation_bars}.}
\end{table}

    \subsection{Execution drag}
    \label{subsec:execution_drag}

    We measure execution drag by re-pricing the headline strategy's realized trades
    at three entry-fill assumptions while holding the model's selections,
    gate decisions, and per-day position size $Q$ fixed. Re-pricing post hoc,
    without retraining the model under each fill assumption, isolates the
    execution effect from any change in the trained ranker. The three fills are mid
    (50\% spread coverage, the headline assumption), 75\% spread coverage (sell at
$0.75 \cdot \text{bid} + 0.25 \cdot \text{ask}$), and sell at bid (100\% spread
    coverage). Locking $Q$ holds the dollar position size at its headline level, so
    the reported drag does not benefit from any sizing re-calibration that a live
    deployment would apply under a worse fill regime.

    Table~\ref{tab:execution} reports walk-forward and out-of-time Sharpe ratios
    under the three fill assumptions across the seven sizing methods. The 2025
    out-of-time Sharpe degradation is monotonic in fill aggressiveness (mid $\to$
    75\% coverage $\to$ bid) for every method. At 75\% spread coverage, the OOT
    percentage drag ranges from $-4.7\%$ (SRS) to $-8.2\%$ (VT). At sell-at-bid,
    the OOT drag widens to $-6.5\%$ (SRS) at the low end and $-10.9\%$ (VT) at the
    high end. EA loses 5.8\% of its OOT Sharpe at 75\% spread coverage (5.761 to
    5.426) and 7.8\% at sell-at-bid (5.761 to 5.311). FMU and SRS, the other two
    top-Sharpe sizing methods, lose 6.5\% and 4.7\% respectively at 75\% spread
    coverage and 8.8\% and 6.5\% at sell-at-bid. Walk-forward percentage drag is
    larger, even though the absolute Sharpe move is smaller there, at $0.380$
    against $0.450$ for Edge Allocation at sell-at-bid. The walk-forward Sharpe is
    lower in absolute magnitude, so the same per-trade dollar drag translates into
    a larger percentage. The rank ordering of methods by Sharpe
    is preserved across all three fills.


    For a short option held to expiration, the cross-trade mean P\&L scales
    linearly with the entry premium captured, while the variance is dominated by
    the long left tail of the settlement payoff. A reduction in entry premium
    therefore reduces the mean roughly proportionally and leaves the variance
    largely unchanged, so Sharpe declines roughly proportionally to the premium
    reduction. The $4.7$--$8.2\%$ OOT Sharpe drag at 75\% spread coverage and the
    $6.5$--$10.9\%$ drag at sell-at-bid are broadly consistent with this scaling. The per-share
    premium reduction at the bid fill is approximately twice that at the
    75\%-coverage fill, while the bid-fill Sharpe drag is roughly $1.4$ times the
    75\%-coverage drag, somewhat less than the constant-variance benchmark would
    imply. Trading costs in U.S. options markets are documented directly in the
    microstructure literature \citep{battalio_pay_2016}. The drags reported here
    are computed from the fill assumptions set out above rather than taken from
    that literature.

    \begin{table}[!ht]
\centering
\caption{Execution-drag sensitivity.}
\label{tab:execution}
\begin{tabular}{lrrrrrrrr}
\toprule
& \multicolumn{2}{c}{Mid} & \multicolumn{2}{c}{75\,\% spread cov.} & \multicolumn{2}{c}{Sell at bid} & \multicolumn{2}{c}{OOT \% drag} \\
\cmidrule(lr){2-3}\cmidrule(lr){4-5}\cmidrule(lr){6-7}\cmidrule(lr){8-9}
Method & WF & OOT & WF & OOT & WF & OOT & 75\,\% & Bid \\
\midrule
EA         & 3.105 & 5.761 & 2.826 & 5.426 & 2.725 & 5.311 & $-5.8$  & $-7.8$  \\
FMU        & 2.512 & 5.063 & 2.207 & 4.736 & 2.104 & 4.616 & $-6.5$  & $-8.8$  \\
SRS        & 2.760 & 5.262 & 2.453 & 5.017 & 2.349 & 4.918 & $-4.7$  & $-6.5$  \\
VT         & 2.210 & 4.667 & 1.882 & 4.286 & 1.772 & 4.158 & $-8.2$  & $-10.9$ \\
GB         & 2.055 & 5.015 & 1.776 & 4.657 & 1.682 & 4.527 & $-7.1$  & $-9.7$  \\
HK    & 1.946 & 4.371 & 1.675 & 4.070 & 1.583 & 3.960 & $-6.9$  & $-9.4$  \\
QK & 1.896 & 4.308 & 1.627 & 4.018 & 1.537 & 3.911 & $-6.7$  & $-9.2$  \\
\bottomrule
\end{tabular}
\source{Walk-forward (WF, 2021--2024) and 2025 out-of-time (OOT) Sharpe ratios for each sizing method, evaluated under three entry-fill assumptions: mid (50\% spread cov., the headline assumption), 75\% spread cov. (sell at $0.75 \cdot \text{bid} + 0.25 \cdot \text{ask}$), and sell at bid (100\% spread cov., worst case). The percentage indicates the share of the bid-ask spread given up to the market. Per-method daily P\&L is recomputed at each stressed entry premium with the model's selections, gate decisions, and per-day position size $Q$ held fixed at their headline values. The OOT \% drag columns report the percentage change in OOT Sharpe relative to the mid-fill OOT Sharpe.}
\end{table}

    \section{Mechanism Analysis}
    \label{sec:mechanism}

    \subsection{Confidence gate}
    \label{subsec:gate}

    The confidence gate is a per-window calibrated abstention threshold $\tau$. The
    model produces a real-valued confidence signal per day, defined as the score
    gap between the highest-ranked and second-highest-ranked candidate (top-1 minus
    top-2). The threshold $\tau$ is calibrated on the last six months of each
    training window, held out from feature engineering and ranker training, by a
    one-dimensional grid search that maximizes the Sortino ratio on the held-out
    slice. At inference time, on any prediction day with confidence signal below
$\tau^{\star}$, the per-day position size is set to zero regardless of the model's
    pick, and the strategy abstains for the day.

    Table~\ref{tab:gate}, in Appendix~\ref{sec:appendix_gate}, reports the calibrated
    $\tau^{\star}$ and the implied trade rate per window. Two walk-forward windows
    (2021 and 2024) calibrate to a low $\tau^{\star}$ near zero with implied trade rates of $100\%$ and $40.0\%$ respectively. The
    2022 and 2023 windows calibrate to $\tau^{\star}$ values of $0.20$ and $0.16$, with
    implied trade rates of $50.4\%$ and $30.6\%$. The 2025 out-of-time slice and
    the union calibration over all walk-forward held-out slices both calibrate to
$\tau^{\star} = 0$ with a $100\%$ trade rate. In both cases the held-out Sortino was
    strictly higher under no abstention than under any positive threshold
    considered. Across the four walk-forward windows the calibrated trade rate
    spans $30.6\%$ to $100\%$, a $3.3\times$ range. The gate binds only in the
    windows whose calibration returned a positive threshold, 2022 through 2024,
    and admits every day in 2021 and in the 2025 out-of-time slice, where
$\tau^{\star} = 0$.

    The gate is one realization of selective prediction. Its advantage over a
    forced-prediction baseline depends on the abstention region tracking the
    model's epistemic uncertainty rather than the unconditional difficulty of the
    prediction task.
    The top-1-versus-top-2 score gap is a proxy for the ranker's epistemic
    uncertainty over the candidate set on a given day, and the per-window Sortino
    calibration locates the threshold at the value that maximizes the held-out
    Sortino, which by construction rewards downside reduction more strongly than it
    penalizes forfeited upside.

    \subsection{Feature survival across selection windows}
    \label{subsec:features}

    The selection pipeline runs independently in each walk-forward window, applying
    the same five stages to that window's training data. The out-of-time window
    runs through the same pipeline on data through 2024, so its feature set is
    fixed before any 2025 observation enters the evaluation.
    Because each window draws on a different sample period, the surviving feature
    set varies by window. Across the four walk-forward windows and the out-of-time
    window, 85 unique features are selected at least once. The full distribution is
    concentrated at the extremes: 27 features survive in all five, 10 in four,
    16 in three, 13 in two, and 19 in only one (the long tail of window-specific
    features that pass the relevance threshold for a particular sample period and
    then drop out elsewhere).

    The 27 features that survive in every window map cleanly onto economic
    mechanisms. Position greeks and exposures dominate with eight features: the
    four base greeks (delta, gamma, theta, and vega) plus dollar-delta, gamma
    exposure, log-moneyness, and the leverage measure. These are included to
    describe the risk profile each candidate carries. Volatility-surface features contribute three (the 252-day percentile
    of 1DTE ATMF IV, the intraday range of 10DTE ATMF IV, and the IV-to-spot-change
    ratio at 1DTE close), which supply same-session context on the level of the
    surface and its recent movement. Three intra-strategy features
    (delta-distance-from-target and the IV-at-strike-minus-ATMF measures in level
    and percent form) tell the model how far each candidate is from the ideal point
    on the surface. Three entry-liquidity features (bid-ask spread in dollars and
    percent, and log entry premium) prevent the ranker from selecting candidates
    that will be expensive to execute. Two VIX features capture short-horizon
    implied-volatility momentum. Two index features carry SPX 1-day returns and
    252-day realized volatility. Two calendar features (the post-holiday-session
    indicator and the days-until-next-PCE counter) capture the macro-event-day
    patterns the model uses for sizing-related discrimination. The remaining
    four comprise two regime-conditional interaction terms (delta interacted
    with the VIX level, and gamma interacted with annualized morning realized
    volatility), one within-day-rank feature (the cross-sectional rank of the
    IV-at-strike-minus-ATMF measure), and one per-strategy statistic (the
    30-day win rate). This always-selected core is a small subset, about 14\%
    of the roughly 190 candidate features.

    The cross-window stability of the selection itself is summarized by the
    pairwise Jaccard similarity of the surviving feature sets. The mean pairwise
    Jaccard across the four walk-forward windows and the OOT 2025 window is 0.58,
    with values ranging from 0.47 (the most-distant pair, 2022 vs OOT 2025) to 0.67
    (the closest pair, 2023 vs 2024). Figure~\ref{fig:heatmap}, in Appendix~\ref{sec:appendix_features}, shows the
    per-window survival pattern across the top 60 features as a heatmap. The
    27 always-surviving features appear as the contiguous top block of
    always-on cells. Interpreted in the framework of
    \citet{nogueira_stability_2018}, who formalize selection stability as an
    expected-Jaccard-based statistic, a mean Jaccard of 0.58 across the five expanding
    training windows indicates a moderately stable selection. This is consistent with
    a persistent underlying signal, not a window-specific artifact.

    \subsection{Mechanism contribution decomposition}
    \label{subsec:mechanism_decomp}

    Table~\ref{tab:mechanism_decomposition} separates the contributions of the two
    risk controls added to the base pipeline: the confidence gate, the abstention
    rule of Section~\ref{sec:gate} applied to the ranker's daily output, and the
    three tail-risk features of Section~\ref{subsec:ablation}, model inputs that
    mark days of elevated stress through the realized-versus-implied volatility
    surprise, the candidate's recent drawdown, and the VIX z-score. In the
    headline configuration only the prior-close drawdown signal survives the
    selection pipeline, in four of the five windows, so the family also acts
    through the composition of the candidate catalog that the selection stages
    see. Four
    configurations are generated from the headline by toggling the confidence gate
    and the tail-risk features and changing nothing else, so the ranker, the
    selection pipeline, the training split, the sizing rule and the execution
    assumptions are common to all four. The gate is disabled by restricting its
    calibration grid to a trade rate of one, which holds the held-out-slice
    training discipline fixed and leaves the threshold admitting every day.

    Out of time the result rests on the configuration with neither risk control. With both
    controls disabled the strategy earns an out-of-time Sharpe of $5.221$ against
    $0.175$ for the CBOE PUT comparator, so $5.046$ of the $5.586$ gap is present
    before either control is added. The tail-risk features add $0.652$, the
    confidence gate adds $0.012$, and the interaction is $-0.124$. The gate's main
    effect is negligible because its 2025 calibration returns $\tau^{\star} = 0$ in
    every configuration and admits every day of the hold-out, as
    Section~\ref{subsec:gate} reports.

    The walk-forward slice, where the gate does bind, gives the opposite pattern.
    With both controls disabled walk-forward Sharpe is $0.623$. Adding the
    tail-risk features alone lowers it to $0.160$, adding the gate alone raises it
    to $0.759$, and the two together produce $3.105$. The interaction of $+2.809$
    exceeds either main effect, so the two controls are complementary rather than
    additive. The pattern is consistent with the gate withholding capital on days
    the tail-risk features mark as adverse, though the grid does not itself
    establish that day-level overlap, and neither device does much without the
    other.

    Read together, the two panels place the out-of-time performance in the
    pipeline with neither control, a baseline that bundles the ranker with the
    sizing rule, the margin model, and the execution assumptions, and the
    walk-forward performance in the joint action of the two controls, whose
    value appears in the windows where they bind rather than in the hold-out
    year. The gate withholds capital on
    low-confidence days in 2022 through 2024 and on no day of 2025. The grid also
    shows that measuring either control alone would misstate it, since on
    walk-forward the tail-risk features lower Sharpe when the gate is absent and
    raise it when the gate is present.

    \begin{table}[!ht]
\centering
\caption{Mechanism contribution decomposition for Edge Allocation.}
\label{tab:mechanism_decomposition}
\begin{tabular}{lrr}
\toprule
 & Sharpe (WF) & Sharpe (OOT) \\
\midrule
\multicolumn{3}{l}{\textit{Panel A. Configuration grid}} \\
Neither control                 &  0.623 & 5.221 \\
Tail-risk features only         &  0.160 & 5.873 \\
Confidence gate only            &  0.759 & 5.233 \\
Both controls (headline)        &  3.105 & 5.761 \\
\midrule
\multicolumn{3}{l}{\textit{Panel B. Additive decomposition}} \\
Baseline, neither control       &  0.623 & 5.221 \\
Tail-risk features, main effect & $-0.463$ & $+0.652$ \\
Confidence gate, main effect    & $+0.136$ & $+0.012$ \\
Interaction                     & $+2.809$ & $-0.124$ \\
Total (headline)                &  3.105 & 5.761 \\
\midrule
CBOE PUT comparator             &  0.680 & 0.175 \\
\bottomrule
\end{tabular}
\source{Two-by-two ablation of the two risk controls added to the base pipeline, the confidence gate and the three tail-risk features (the realized-versus-implied volatility surprise, the per-strategy prior-close drawdown, and the 60-day VIX z-score, Appendix~\ref{app:features}), under Edge Allocation. The four configurations are generated from the headline by toggling those two elements and differ from it in nothing else, so the training window, the label scheme, the selection pipeline, the sizing rule and the execution assumptions are common to all four. The gate is disabled by restricting its calibration grid to a trade rate of one, which holds the held-out-slice training split in force and leaves the calibrated threshold admitting every day. Panel B measures each main effect at the other factor's disabled level and reports the interaction as the residual of the grid, so the four entries sum to the headline exactly. On the out-of-time slice the gate calibrates to a zero threshold in all four configurations and admits every day, which is why its main effect there is negligible. On the walk-forward slice the gate binds, at calibrated trade rates of 0.50, 0.31 and 0.40 in 2022, 2023 and 2024, and the interaction exceeds both main effects, so the walk-forward main effects should not be read on their own. Walk-forward covers 2021 to 2024, out-of-time is the 2025 hold-out.}
\end{table}

    \subsection{Discussion}
    \label{subsec:discussion}

    The strategy harvests an empirical wedge between the implied volatility
    embedded in equity-index option prices and the volatility that is subsequently
    realized over a comparable horizon. \citet{bondarenko_historical_2019} reports
    an average VIX of $19.3\%$ against an average subsequently realized one-month
    volatility of $15.1\%$ from 1990 to 2018, a wedge of $4.2$ annualized
    percentage points.
    \citet{carr_variance_2009} document the same wedge through synthetic
    variance-swap returns. In their sample the long-variance-swap position earns a
    statistically significant negative average return at one-month horizons,
    equivalent in sign to a positive expected return on the short side of the same
    trade. The economic interpretation is that option buyers pay an insurance
    premium against the catastrophic-loss tail of equity-index movements, with the
    marginal seller absorbing that tail in exchange for the premium.

    The strategy reported in this paper does not harvest the volatility risk
    premium uniformly across days. The LightGBM ranker selects, on each trading
    day, the delta bucket where the gap between implied volatility and expected
    realized volatility is widest in the cross-section of nine candidates, conditional on
    a feature set that combines the volatility-surface state, the prior-day realized
    distribution, and a set of regime indicators. The confidence gate of
    Section~\ref{subsec:gate} abstains on days when the model's top-1-versus-top-2
    score gap is small, which is a proxy for elevated uncertainty over
    the candidate ranking. Taken together, the strategy operates as a
    regime-conditional and uncertainty-aware harvester of the volatility risk
    premium rather than an unconditional short-volatility
    exposure.

    The closest passive short-volatility comparators are unhedged short-put
    strategies of the type characterized empirically by
    \citet{israelov_which_2017}. The headline strategy's seven sizing methods
    produce 2025 out-of-time Sharpe ratios in the range $4.31$ to $5.76$, against
$0.175$ for the CBOE PUT index over the same hold-out period
    (Section~\ref{subsec:benchmarks}). The deepest sample-period drawdown across
    the seven methods is $-9.25\%$ (Section~\ref{subsec:per_method}), considerably
    below the magnitudes documented for unhedged short-put strategies during 2008,
    2018, and 2020 stress events. The decomposition reported in
    Table~\ref{tab:mechanism_decomposition} attributes the gap between the headline
    strategy and the passive comparator mainly to the pipeline without the two
    risk controls, which supplies $5.046$ of out-of-time Sharpe before either
    control is added, with the tail-risk features adding $0.652$ and the gate,
    inactive on this slice, $0.012$. The empirical advantage over the passive short-volatility comparator
    therefore reflects regime-conditional harvesting of the
    volatility risk premium rather than higher leverage applied to the same
    exposure, with the uncertainty-aware component contributing on the
    walk-forward windows where the gate binds.

    \section{Conclusions}
    \label{sec:conclusions}

    The volatility risk premium in S\&P 500 index options is one of the most
    persistent empirical regularities in the literature on equity-index
    derivatives, and passive
    harvesters such as the CBOE PUT and WPUT indices capture this premium through
    static short-put rules that allocate capital identically across regimes and
    absorb the same fixed exposure during stress windows. This paper asks whether a
    more selective harvester, based on cross-sectional ranking of candidate
    strategies in place of a static rule, can extract a larger fraction of the
    available premium without incurring the stress-window losses that limit the
    passive baselines. The framework applies a LightGBM LambdaRank ranker to a
    cross-section of nine candidate strategies in the SPXW zero-day-to-expiration
    market, comprising eight delta-targeted short-put positions and a \textit{SKIP} option,
    with a confidence gate that abstains under elevated model uncertainty. The strategy is evaluated under
    index-option margin requirements, a tiered fee schedule, and
    bid-to-mid execution assumptions, with a four-window walk-forward over
    2021--2024 and a strictly held-out 2025 out-of-time slice that is never used
    for training, hyperparameter search, or model selection.

    On the 2025 hold-out the ranker delivers risk-adjusted returns that exceed both
    the passive short-volatility benchmarks and the panel of internal selection
    baselines, and the advantage survives explicit margin requirements, fee
    schedules, and slippage stress. The seven sizing methods produce annualized
    Sharpe ratios between $4.31$ and $5.76$, and every method exceeds the three
    external benchmarks (CBOE PUT, CBOE WPUT, SPX buy-and-hold) by at least $3.84$
    and the five internal selection baselines by at least $3.69$. The headline
    strategy's Sharpe of $5.76$ corresponds to a Probabilistic Sharpe Ratio of
    $0.964$ against the worst external benchmark and an out-of-time maximum
    drawdown of $-1.43\%$, against $-2.28\%$ over the walk-forward. Those out-of-time figures rest on a single calendar
    year. Every method's out-of-time Sharpe exceeds its own walk-forward figure by
    more than $2.4$ units, and
    Section~\ref{subsec:year_heterogeneity} reports a per-year range for the
    headline method running from $0.41$ in 2024 to $6.52$ in 2021. The hold-out
    result is one draw from that spread, not a stable operating level. The advantage holds on both the risk-adjusted and the
    absolute margin, but the two are established with different strength. On the
    out-of-time slice Edge Allocation earns $10.48\%$ in excess of the risk-free
    rate, against $8.84\%$ for the SPX buy-and-hold and $2.52\%$ for the CBOE PUT
    index, at a realized volatility of $1.82\%$ against $19.12\%$ and $14.36\%$
    respectively. The same ordering holds on walk-forward, although the margin over
    buy-and-hold narrows there to $10.91\%$ against $10.11\%$. The risk-adjusted
    comparison is the more secure of the two. As
    Section~\ref{subsec:benchmarks} reports, the Diebold--Mariano test does not
    reject equality of mean daily P\&L against any external benchmark after a
    Bonferroni correction, so the level difference, positive in every cell, is not
    statistically resolved on a single hold-out year.

    The size of that ratio has no counterpart in the option-writing literature, and
    the comparison is worth stating explicitly. \citet{whaley_return_2002} reports a
    monthly Sharpe ratio of $0.234$ for the CBOE BuyWrite index over June 1988 to
    December 2001 against $0.172$ for the S\&P 500 portfolio, which annualize to
    $0.81$ and $0.59$, and \citet{bondarenko_historical_2019} reports $0.65$ for the
    PUT index against $0.49$ for the index over June 1986 to December 2018. The
    passive benchmarks of this paper land at or below that band on the present
    sample, at $0.68$ for PUT and $0.14$ for WPUT on the walk-forward and $0.18$ and
    $0.13$ out-of-time. A headline of $5.76$ is therefore roughly seven times the
    highest of those figures, and the excess comes mainly from the denominator. Edge
    Allocation earns $10.48\%$ in excess of the risk-free rate on the hold-out
    against $8.84\%$ for the buy-and-hold, while its denominator is $1.82\%$
    against $19.12\%$. Two features of the design produce that denominator. Returns
    are measured per unit of posted collateral on positions the calibration sizes
    toward a $16\%$ volatility anchor that no method reaches
    (Section~\ref{sec:sizing}), and abstention combined with the zero-day
    structure leaves a large share of sessions at or near zero profit and loss. The
    level is therefore not the same quantity as an index Sharpe ratio, since
    matching the PUT index's out-of-time volatility would require scaling Edge
    Allocation by a factor of nearly eight, which margin, capacity, and overnight
    gap risk do not permit, and which is the reason the hold-out figure is read
    above as an upper bound on live deployment. \citet{whaley_return_2002} also
    supplies the reason for reading any such ratio cautiously, since he finds the
    BuyWrite distribution negatively skewed at $-1.44$ with excess kurtosis of
    $4.98$ and rejects normality, and standard mean-variance measures reward that
    shape. The assessment in Section~\ref{subsec:multi_test_correction} rests on
    the Probabilistic and Deflated Sharpe ratios, which price that shape and
    discount the headline, with the out-of-time
    deflated Sharpe ratio falling between $0.856$ and $0.913$ and the
    Diebold--Mariano test failing to reject equality of mean daily profit and loss.
    What the passive literature supports is the ordering of these methods on a
    common sample and a common excess basis, not the multiple between the two
    levels.

    The strategy's return is regime-conditional, and the conditioning transfers
    from the walk-forward training period (2021--2024) to the 2025 out-of-time
    hold-out. Annualized return in the top realized-volatility tercile exceeds the
    bottom tercile for all seven methods out of time and for six of the seven on
    walk-forward. On the walk-forward VIX panel the return peaks in the mid-VIX
    band for six of the seven methods, and rises from $0.50\%$ on low-VIX days to
$15.40\%$ on high-VIX days for the headline method alone. The risk-adjusted
    ratio peaks in the mid-VIX regime for five of the seven, because the
    volatility of the strategy's own returns rises faster than the premium it
    collects. The
    out-of-time VIX extremes rest on twenty-odd days each and are not
    interpreted.

    Engineered multiplicative interactions of per-strategy exposures with
    volatility-state regime variables carry a substantial part of the walk-forward
    ranking signal. The fifteen-group feature ablation identifies the seven-feature
    multiplicative-interaction group as one of the largest contributors to walk-forward
    statistical confidence: removing it collapses walk-forward Probabilistic Sharpe
    from $1.000$ to $0.229$, while out-of-time Sharpe loses about $0.72$.

    A two-by-two ablation of the confidence gate against the tail-risk features
    places $5.05$ of the $5.59$ out-of-time Sharpe gap over the CBOE PUT benchmark
    with the LightGBM ranker and the selection layer. The two risk controls add $0.54$
    between them, and the gate's share of that is negligible because it admits
    every day of the hold-out year. On the walk-forward slice, where the gate
    binds, the ordering reverses. Neither control reproduces the headline on its own and
    their interaction supplies $2.81$ of the $3.11$ recorded there. In conclusion, the
    harvesting is conditional. The regime panels place the premium capture in the
    mid-volatility band, and the gate withholds exposure on walk-forward days
    when the top-1-versus-top-2 gap falls below its calibrated threshold. The paper makes
    two contributions. First, it provides an empirical demonstration that a
    selection architecture built around a learning-to-rank model can select
    profitable short-volatility positions on
    SPXW zero-day-to-expiration options, assessed against a stricter standard of
    evidence than the raw Sharpe ratios usual in this literature. The
    demonstration concerns the ranking model together with the selection layer,
    which carry the out-of-time gap between them, while the two risk controls earn
    their place on the walk-forward slice, where they act jointly. Every
    Probabilistic Sharpe Ratio in Table~\ref{tab:deployment_confidence} exceeds
    $0.95$. The Deflated Sharpe
    Ratio, which prices the breadth of the configuration search back out, reaches
    that level on the walk-forward slice for Edge Allocation and Short-Richness
    Scaling and for no method on the out-of-time slice, a shortfall that
    Section~\ref{subsec:multi_test_correction} attributes to the single calendar
    year of hold-out data and the trial penalty acting together. Second, it sets out an
    architectural pattern combining a model, a confidence gate,
    and volatility-targeted sizing, and the two-by-two ablation evaluates each
    component's contribution separately.

    Two of the study's limitations deserve particular mention. The 2018--2025
    sample contains no sustained 2008-style systemic stress. The deepest in-sample
    shock is the COVID-19 episode of March 2020, which was short and coincided with
    a rapid central-bank response. A future regime with a different stress
    microstructure would test the confidence gate more severely than anything in
    the sample, and it could produce drawdowns larger than the headline maximum. The backtest assumes immediate fills at the configured price
    field and does not model adverse selection, limit-order queue position and fill
    probability, or market impact at higher capital. The reported headline
    out-of-time Sharpe should therefore be read as an upper bound on what a live
    deployment would achieve at the same capital base.

\clearpage
\bibliographystyle{myapalike}
\bibliography{rank}

\clearpage
\phantomsection
\noindent{\Huge\bfseries Appendix}\par
\vspace{0.5em}
\hrule height 0.5pt
\vspace{1em}
\appendix
\begingroup
  \renewcommand{\thesection}{\Alph{section}}
  \renewcommand{\thetable}{\thesection.\arabic{table}}
  \renewcommand{\thefigure}{\thesection.\arabic{figure}}
  \renewcommand{\topfraction}{0.95}
  \renewcommand{\bottomfraction}{0.95}
  \renewcommand{\textfraction}{0.05}
  \renewcommand{\floatpagefraction}{0.6}

  \let\origunderscore\_%
  \renewcommand{\_}{\origunderscore\allowbreak}%
  \setlength{\emergencystretch}{3em}%

    \section{Performance and Risk Metrics Definitions}
    \label{app:metrics}
    \setcounter{table}{0}
    \setcounter{figure}{0}
    This appendix lists the formulas and conventions used to evaluate the in- and out-of-sample performance of the strategies analyzed in this study. All annualized metrics assume 252 trading days per calendar year. The Sharpe, Sortino, Probabilistic, and Deflated Sharpe ratios are computed on excess returns, and the two sides of every comparison reach that basis by different routes. A strategy return is the day's profit and loss per unit of collateral. The collateral is held in cash and earns the risk-free rate, so the series is already in excess of it and no further deduction is applied. Deducting the rate as well would charge it twice and would make an abstention day, on which the strategy holds only cash, register as a loss. The three external index benchmarks are total-return series, so each is converted to excess by subtracting the EFFR rate for the day, expressed in decimal form and divided by 252. Drawdown-based metrics and trade counts use unadjusted strategy returns.

    \subsection{Return and risk components}
    \begin{itemize}
        \item \textbf{Daily returns ($r_i$):} the simple return of the equity curve from one trading day to the next. With $P_i$ the portfolio value at the close of day $i$:
              \begin{equation}
                  r_i = \frac{P_i - P_{i-1}}{P_{i-1}}
              \end{equation}

        \item \textbf{Excess daily returns ($\tilde{r}_i$):} $r_i$ expressed net of the risk-free rate, which the strategy and benchmark series reach by different routes. With $r^{f}_i$ the EFFR rate on day $i$ in decimal form:
              \begin{equation}
                  \tilde{r}_i = \begin{cases}
                      r_i                        & \text{strategy series}  \\
                      r_i - \dfrac{r^{f}_i}{252} & \text{external benchmark series.}
                  \end{cases}
              \end{equation}

        \item \textbf{Annualized compounded return (aRC):} the geometric annualized growth rate of the strategy. For a sample of $N$ trading days:
              \begin{equation}
                  aRC = \left( \prod_{i=1}^{N} (1 + r_i) \right)^{\!\frac{252}{N}} - 1
              \end{equation}

        \item \textbf{Annualized standard deviation (aSD):} the annualized sample volatility of daily returns. With $\bar{r}$ the arithmetic sample mean of the $N$-day series:
              \begin{equation}
                  aSD = \sqrt{252} \cdot \sqrt{\frac{1}{N-1} \sum_{i=1}^{N} (r_i - \bar{r})^2}
              \end{equation}

        \item \textbf{Annualized downside deviation (aDD):} the annualized root-mean-square of the negative-return tail, with target zero. Positive returns are floored at zero before squaring, and the average is taken over all $N$ days:
              \begin{equation}
                  aDD = \sqrt{252} \cdot \sqrt{\frac{1}{N} \sum_{i=1}^{N} \big[\min(r_i, 0)\big]^2}
              \end{equation}

        \item \textbf{Maximum drawdown (MD):} the largest peak-to-trough percentage decline in the equity curve:
              \begin{equation}
                  MD = \max_{\tau \in [0, T]} \left( \frac{\max_{t \in [0, \tau]} P_t - P_\tau}{\max_{t \in [0, \tau]} P_t} \right)
              \end{equation}
    \end{itemize}

    \subsection{Risk-adjusted return ratios}
    \begin{itemize}
        \item \textbf{Sharpe ratio (SR):} the standard risk-adjusted return measure of \citet{sharpe_sharpe_1994}, computed on excess returns:
              \begin{equation}
                  SR = \frac{aRC(\tilde{r})}{aSD(\tilde{r})}
              \end{equation}

        \item \textbf{Sortino ratio:} an asymmetric variant that penalizes only downside volatility \citep{sortino_downside_1994}:
              \begin{equation}
                  \text{Sortino} = \frac{aRC(\tilde{r})}{aDD(\tilde{r})}
              \end{equation}
    \end{itemize}

    The numerator of each ratio above is the compounded annual growth rate of the
    excess-return series, not its annualized arithmetic mean. For a series of
    simple returns with a positive mean and daily volatility as low as the series
    reported here, compounding over 252 days grows faster
    than multiplying the daily mean by 252, so the Sharpe ratio reported here sits
    slightly \emph{above} its counterpart under the arithmetic convention. For the
    headline method on the out-of-time slice the two conventions give $5.761$ and
$5.489$ respectively. The gap widens with the mean and is second-order at the
    return levels reported here. The probabilistic assessment below operates on the
    same daily excess-return series, but forms its Sharpe input as the ratio of the
    daily arithmetic mean to the daily standard deviation. The point estimate
    behind a reported PSR or DSR is therefore the arithmetic analogue of the Sharpe
    ratio tabulated beside it, and is the smaller of the two.

    \subsection{Probabilistic performance assessment}
    Short-volatility strategies generate non-normal, negatively skewed return
    distributions for which point-estimate Sharpe ratios mis-state estimator
    uncertainty. To assess whether the realized Sharpe reflects a true Sharpe at or
    above an external benchmark, rather than an artifact of finite-sample noise,
    distributional asymmetry, or selection from a search over candidate
    configurations, we use the Probabilistic Sharpe Ratio of
    \citet{bailey_sharpe_2012} and the Deflated Sharpe Ratio of
    \citet{bailey_deflated_2014}.

    \begin{itemize}
        \item \textbf{Probabilistic Sharpe Ratio (PSR):} the probability that the true Sharpe ratio of the strategy exceeds a benchmark $SR^{*}$, given the observed sample skewness $\hat{\gamma}_3$ and kurtosis $\hat{\gamma}_4$:
              \begin{equation}
                  \widehat{PSR}(SR^{*}) = \Phi \left( \frac{(\widehat{SR}-SR^{*}) \sqrt{N-1}}{\sqrt{1 - \hat{\gamma}_3 \widehat{SR} + \frac{\hat{\gamma}_4 - 1}{4} \widehat{SR}^2}} \right)
              \end{equation}
              where $\Phi(\cdot)$ is the standard-normal cumulative distribution function, $N$ is the number of return observations, $\hat{\gamma}_4$ is the sample kurtosis (not excess kurtosis, so $\hat{\gamma}_4 = 3$ under Gaussianity), and $\widehat{SR}$ and $SR^{*}$ are expressed at the same return horizon. $SR^{*}$ is set to the realized daily Sharpe ratio of the named benchmark over the same evaluation slice, on the excess basis defined above. Negative skewness and fat tails enlarge the denominator and lower the PSR for a given point estimate.

        \item \textbf{Deflated Sharpe Ratio (DSR):} the PSR with the benchmark replaced by the expected maximum Sharpe ratio under the null of $N_t$ independent candidate strategies tested in parallel, an adjustment that accounts for selection bias from a multi-configuration search \citep{bailey_deflated_2014}. The expected maximum is approximated as
              \begin{equation}
                  \begin{split}
                      \mathbb{E}\!\left[\max_{n=1,\dots,N_t} \widehat{SR}_n\right] &\approx \sqrt{\widehat{V}\!\left(\widehat{SR}\right)} \cdot \Big[(1-\gamma)\,\Phi^{-1}\!\left(1-\tfrac{1}{N_t}\right) \\
                      &\hphantom{{}\approx{}} + \gamma\,\Phi^{-1}\!\left(1-\tfrac{1}{N_t \cdot e}\right)\Big]
                  \end{split}
              \end{equation}
              where $\gamma \approx 0.5772$ is the Euler--Mascheroni constant, $\widehat{V}\!\left(\widehat{SR}\right) = 1/(N-1)$ is the variance of the daily Sharpe estimator under the null, and $\Phi^{-1}(\cdot)$ is the standard-normal quantile function. The DSR is then
              \begin{equation}
                  \begin{aligned}
                      \widehat{DSR} &= \widehat{PSR}\!\left(SR^{*}_{\text{eff}}\right), \\
                      SR^{*}_{\text{eff}} &= \max\!\left(SR^{*},\, \mathbb{E}\!\left[\max_{n=1,\dots,N_t} \widehat{SR}_n\right]\right)
                  \end{aligned}
              \end{equation}
              By raising the benchmark in proportion to the size of the search space, the DSR is a multiple-testing-aware analogue of the PSR.
    \end{itemize}

    Both ratios are reported as values rather than reduced to a pass-or-fail
    classification, the PSR for every method and evaluation slice and the DSR
    for the three top-ranked methods on both slices. The PSR is computed separately
    against each external benchmark (CBOE PUT, CBOE WPUT, and the SPX
    buy-and-hold reference). The DSR is benchmark-independent, because its
    reference point is the expected maximum Sharpe ratio across the $N_t$ trials
    rather than a benchmark return. Each value is read as a confidence level
    attached to that cell, and Section~\ref{subsec:multi_test_correction} states
    where those levels fall short of the $0.95$ level conventionally treated as
    strong evidence.

    \subsection{Significance testing for benchmark comparisons}
    Pairwise comparisons between the strategy and each external benchmark use the
    \citet{diebold_comparing_1995} test for equal predictive accuracy applied to
    daily P\&L differentials, with a Bonferroni correction across the family of
    benchmark tests.
    \begin{itemize}
        \item \textbf{Diebold--Mariano statistic:} with $d_i = \pi^{A}_i - \pi^{B}_i$ the difference between the day-$i$ P\&L of strategies $A$ and $B$, $\bar{d}$ its sample mean, and $\hat{\sigma}^2_d$ an estimator of the long-run variance of $d_i$,
              \begin{equation}
                  DM = \frac{\bar{d}}{\sqrt{\hat{\sigma}^2_d / N}}, \qquad \text{p-value} = 2\left[1 - \Phi\!\left(|DM|\right)\right]
              \end{equation}
              For a one-step-ahead horizon ($h=1$) the sample variance of $d_i$ is used. For $h > 1$ the first $h-1$ Newey--West-style autocovariance terms are added.

        \item \textbf{Bonferroni correction:} each raw $p$-value is multiplied by the family size $m$ (the number of benchmark comparisons reported jointly) and capped at one. The Diebold--Mariano comparisons are evaluated at family-wise significance $\alpha = 0.05$.
    \end{itemize}
    \FloatBarrier
    \clearpage

    \section{Definitions of Candidate Features}
    \label{app:features}
    \setcounter{table}{0}
    \setcounter{figure}{0}
    This appendix documents every feature in the candidate catalog of approximately 190 columns. Each feature group is annotated as \emph{cross-sectional} (CS, one value per trading day, broadcast across all candidates) or \emph{per-strategy} (PS, one value per (date, strategy) pair).

    Feature timing is set by the 10:00 ET entry decision rather than by the calendar day. A timestamp falling on day $t$ is not sufficient on its own, because the closing values of day $t$ also carry a day-$t$ timestamp and are not observable when the position is opened at 10:00 ET that morning. Every predictor built from day-$t$ closing prices, futures settlements, full-day volume or the realized outcome of the day's trade is therefore held back by at least one trading day, so that the row dated $t$ carries values observed no later than the close of $t-1$. Predictors already fixed by 10:00 ET on day $t$ enter without a lag: the calendar and event flags, the 09:30--09:59 morning session window, the 09:35 and 10:00 volatility-surface snapshots, and the entry-time Greeks and liquidity measured at the 10:00 ET snapshot. All rolling statistics are causal within these bounds. Table \ref{tab:feature_timing} states the availability and the lag for each block.

    \begin{table}[!ht]
	\centering
	\caption{Predictor availability and the lag applied to each feature block.}
	\label{tab:feature_timing}
	\resizebox{\textwidth}{!}{%
		\begin{tabular}{lll}
			\toprule
			Feature block                                                          & Earliest availability  & Lag applied          \\
			\midrule
			Calendar and event flags (CS)                                          & Known in advance       & None                 \\
			Morning-session features (CS), 09:30--09:59 SPX window                 & 10:00 ET, day $t$      & None                 \\
			Morning-session features (CS), 09:35-to-10:00 surface changes          & 10:00 ET, day $t$      & None                 \\
			Morning-session features (CS), VIX and VVIX closes                     & Close, day $t$         & $\geq 1$ trading day \\
			SPX index features (CS)                                                & Close, day $t$         & $\geq 1$ trading day \\
			VIX family and futures (CS), spot family                               & Close, day $t$         & $\geq 1$ trading day \\
			VIX family and futures (CS), futures front complex                     & Settlement, day $t$    & $\geq 1$ trading day \\
			Macroeconomic indicators (CS)                                          & Release, day $t$       & $\geq 1$ trading day \\
			Implied-volatility surface (CS), 10:00 snapshot                        & 10:00 ET, day $t$      & None                 \\
			Implied-volatility surface (CS), closing snapshot and intraday changes & Close, day $t$         & $\geq 1$ trading day \\
			Realized-minus-implied volatility differentials (CS)                          & Close, day $t$         & $\geq 1$ trading day \\
			Realized higher moments (CS)                                           & Close, day $t$         & $\geq 1$ trading day \\
			VIX term-structure curvature (CS)                                      & Close, day $t$         & $\geq 1$ trading day \\
			Trend (CS)                                                             & Close, day $t$         & $\geq 1$ trading day \\
			Position Greeks and exposures (PS)                                     & 10:00 ET, day $t$      & None                 \\
			Per-strategy rolling statistics (PS)                                   & Trade outcome, day $t$ & $\geq 1$ trading day \\
			Entry liquidity (PS)                                                   & 10:00 ET, day $t$      & None                 \\
			Intra-strategy term context (PS)                                       & 10:00 ET, day $t$      & None                 \\
			Regime-conditional sensitivities (PS)                                  & Close, day $t$         & $\geq 1$ trading day \\
			\bottomrule
		\end{tabular}%
	}
	\source{Availability and lag for every feature block in the candidate catalog. Lags are counted in trading days rather than calendar days,
    so a one-day lag spans a weekend or a market holiday whenever one intervenes. Three blocks appear on more than one row, since the lag differs across their sub-blocks.
    CS denotes cross-sectional and PS per-strategy.}
\end{table}

    \subsection{Conventions}
    \begin{itemize}
        \item $P_t$ denotes the SPX closing price on trading day $t$, and $S_t$ the intraday SPX mid quote.
        \item $r_{N,t} = P_t / P_{t-N} - 1$ denotes the $N$-day simple return, $\ell_{N,t} = \ln(P_t / P_{t-N})$ the corresponding log return.
        \item $\sigma$ denotes implied or realized volatility in annualized decimal form, $\tau$ denotes days to expiration (DTE).
        \item All annualized volatilities use 252 trading days per year.
        \item Implied volatility, the entry-day Greeks, and the ATMF interpolation are
              computed under Black--Scholes with the EFFR rate as the risk-free rate and no
              dividends. Omitting the dividend yield raises the assumed forward by
              a factor of $e^{q\tau}$, an error of under $0.01\%$ of spot at the tenors that govern strike
              selection and the entry-time anchors and a few tenths of one percent at the
              90-day tenor, the longest in the catalog. The selected strike is therefore
              unchanged in practice, and because the bias is one-signed and moves only with the
              dividend yield, it shifts the level of the longer-dated ATMF series rather
              than the daily variation the ranker scores on.
        \item Per-strategy features do not take any value on the \textit{SKIP}
              candidate where no option is sold. Cross-sectional
              features are populated for every trading day.
        \item Per-strategy features ending in the suffix \texttt{\_wd\_rank} are the
              within-day ordinal rank of the named source feature across the eight
              delta-bucket candidates. \textit{SKIP} candidates are assigned rank 8.5, which
              places them below all eight ranked alternatives.
    \end{itemize}

    \subsection{Calendar and event flags (CS)}
    Pure date arithmetic on the NYSE trading calendar plus a release-date table for
    the major macro announcements.
    \begin{itemize}
        \item \texttt{day\_of\_week}, \texttt{day\_of\_month}, \texttt{month\_of\_year}: integer day-of-week (Mon $= 1$, Fri $= 5$), day-of-month (1--31), and month-of-year (1--12).
        \item \texttt{is\_monthly\_opex}, \texttt{is\_quarterly\_opex}: indicator for the third Friday of any month, and for the third Friday of March, June, September, or December.
        \item \texttt{is\_post\_holiday\_session}, \texttt{is\_pre\_holiday\_session}: indicator for the first trading session after, or the last session before, a market holiday.
        \item \texttt{days\_until\_next\_fomc}, \texttt{days\_since\_last\_fomc}: trading-day counts to the next and from the previous FOMC announcement.
        \item \texttt{is\_cpi\_day}, \texttt{days\_until\_next\_cpi}: release-day indicator and forward count for the BLS CPI announcement.
        \item \texttt{is\_nfp\_day}, \texttt{days\_until\_next\_nfp}: same for the BLS Employment Situation (Nonfarm Payrolls) report.
        \item \texttt{is\_pce\_day}, \texttt{days\_until\_next\_pce}: same for the BEA Personal Consumption Expenditures release.
    \end{itemize}

    \subsection{Morning-session features (CS)}
    Features describing the 09:30 to 10:00 ET window. The 10:00 mark is taken as
    the 09:59 closing mid to avoid leakage from the entry minute itself. Let
$S_{09:30}, \ldots, S_{09:59}$ be 1-minute SPX mid prices over the window and
$S^{(t-1)}_{16:00}$ the prior-day close.
    \begin{itemize}
        \item \texttt{morning\_spx\_log\_return}: $\ln(S_{09:59} / S_{09:30})$.
        \item \texttt{morning\_spx\_range\_pct}: $(\max_t S_t - \min_t S_t) / S_{09:30}$ over the 30-minute window.
        \item \texttt{morning\_spx\_rv\_annualized}: annualized realized volatility from minute log-returns,
              \begin{equation*}
                  \sigma_{\text{morn}} = \sqrt{252 \cdot 390 \cdot \sum_{j} \left(\ln\frac{S_{t_j}}{S_{t_{j-1}}}\right)^2}
              \end{equation*}
              where 390 is the number of one-minute bars in a regular trading day.
        \item \texttt{morning\_spx\_directionality}: close-to-range ratio in $[-1, 1]$, $(S_{09:59} - S_{09:30}) / (\max_t S_t - \min_t S_t)$, set to zero when the range is zero.
        \item \texttt{morning\_gap\_size}: overnight gap, $(S_{09:30} - S^{(t-1)}_{16:00}) / S^{(t-1)}_{16:00}$.
        \item \texttt{morning\_gap\_filled}: indicator that the prior-day close lies within the 09:30 to 09:59 high-low range.
        \item \texttt{morning\_atmf\_iv\_change}: change in 1-DTE ATMF IV between the 09:35 and 10:00 volatility-surface snapshots.
        \item \texttt{morning\_atmf\_iv\_pct\_change}: same change relative to the 09:35 level.
        \item \texttt{morning\_skew\_change}: change in the 1-DTE 25-delta risk reversal $\sigma^C_{25} - \sigma^P_{25}$ between the 09:35 and 10:00 snapshots.
        \item \texttt{morning\_vix\_level}: VIX close lagged by one trading day. The same-day VIX level is not observable until close, so the row dated $t$ carries the prior-day close.
        \item \texttt{morning\_vix\_change}, \texttt{morning\_vvix\_change}: one-day changes in the VIX and VVIX closes, lagged by one trading day for the same observability reason.
    \end{itemize}

    \subsection{SPX index features (CS)}
    Multi-horizon SPX returns, intraday realized volatility, and the daily put-call
    ratios.
    \begin{itemize}
        \item \texttt{spx\_\{N\}\_returns} for $N \in \{\text{1d, 5d, 10d, 1m, 3m, 6m, 1y}\}$: $N$-period simple return $r_N = P_t / P_{t-N} - 1$.
        \item \texttt{spx\_returns\_roll\_avg\_30d}, \texttt{spx\_returns\_roll\_std\_30d}: 30-day rolling mean and standard deviation of \texttt{spx\_1d\_returns}.
        \item \texttt{spx\_\{N\}d\_rv} for $N \in \{5, 21, 63, 252\}$: annualized realized volatility from daily log returns,
              \begin{equation*}
                  \mathrm{RV}_N = \sqrt{252} \cdot \sqrt{\frac{1}{N} \sum_{i=1}^{N} \left(\ln \frac{P_{t-i+1}}{P_{t-i}}\right)^2}.
              \end{equation*}
        \item \texttt{spx\_1y\_percentile}: 252-day rolling percentile rank of the current SPX close, the fraction of past 252 closes below $P_t$.
        \item \texttt{spx\_position\_52w}: normalized position within the 252-day high-low range,
        \begin{equation*}
                  (P_t - \min_{252}) / (\max_{252} - \min_{252}).
              \end{equation*}
        \item \texttt{spx\_put\_call\_ratio}, \texttt{spx\_pcr\_21d\_avg}: daily SPX put-call volume ratio and its 21-day rolling mean.
        \item \texttt{spx\_pcr\_5d\_pct\_change}, \texttt{spx\_pcr\_21d\_pct\_change}: 5- and 21-day percentage changes of the SPX PCR.
        \item \texttt{vix\_put\_call\_ratio}, \texttt{vix\_pcr\_21d\_avg}: daily VIX option PCR and its 21-day mean.
        \item \texttt{vix\_pcr\_5d\_pct\_change}, \texttt{vix\_pcr\_21d\_pct\_change}: 5- and 21-day percentage changes of the VIX PCR.
    \end{itemize}

    \subsection{VIX family and futures (CS)}
    Daily closes for the spot VIX family and the front of the VIX futures complex.
    \begin{itemize}
        \item \texttt{VIX}, \texttt{VIX1D}, \texttt{VIX9D}, \texttt{VIX3M}, \texttt{VIX6M}, \texttt{VVIX}: the 30-day VIX, the 1-day, 9-day, 3-month, and 6-month implied volatility indices, and the volatility-of-VIX index VVIX.
        \item \texttt{VIX\_VIX1D\_ratio}, \texttt{VIX\_VIX9D\_ratio}, \texttt{VIX\_VIX3M\_ratio}, \texttt{VIX\_VIX6M\_ratio}: pairwise ratios of the 30-day VIX to the named tenor. A ratio above one for the short-tenor variants signals near-term term-structure inversion.
        \item \texttt{VIX\_VIX1D\_spread}, \texttt{VIX\_VIX9D\_spread}, \texttt{VIX\_VIX3M\_spread}, \texttt{VIX\_VIX6M\_spread}: the corresponding absolute spreads $\text{VIX} - \text{VIX}_T$.
        \item \texttt{VIX\_5d\_pct\_change}, \texttt{VIX\_21d\_pct\_change}: 5- and 21-day percentage changes of the 30-day VIX.
        \item \texttt{VIX1D\_5d\_pct\_change}, \texttt{VIX1D\_21d\_pct\_change}: same for VIX1D.
        \item \texttt{VVIX\_5d\_pct\_change}, \texttt{VVIX\_21d\_pct\_change}: same for VVIX.
        \item \texttt{\{$X$\}\_1y\_percentile} for $X \in \{\text{VIX}, \text{VIX1D}, \text{VVIX}\}$: 252-day rolling percentile ranks within their own histories.
        \item \texttt{vix\_futures\_front\_price}: settlement price of the front-month VIX future.
        \item \texttt{vix\_front\_slope}: front-curve slope $M_2 - M_1$, where $M_i$ denotes the settlement of the $i$-th nearest monthly future.
        \item \texttt{vix\_front\_curvature}: second-difference proxy for curvature, $2 M_2 - M_1 - M_3$. Positive values indicate a concave (humped) curve.
        \item \texttt{front\_roll\_yield}: daily roll yield from holding the front contract,
              \begin{equation*}
                  \text{RollYield} = \frac{M_2 - M_1}{D_1},
              \end{equation*}
              where $D_1$ is the number of trading days to expiry of the front contract.
        \item \texttt{vix\_z\_60d} (CS): z-score of today's VIX against its trailing 60-day mean and standard deviation,
              \begin{equation*}
                  z_{60d} = \frac{\text{VIX}_t - \overline{\text{VIX}}_{60}}{\sigma_{\text{VIX}, 60}}.
              \end{equation*}
              One of the three tail-risk features.
    \end{itemize}

    \subsection{Macroeconomic indicators (CS)}
    Daily effective Federal Funds rate (EFFR) from FRED, and seasonally
    non-adjusted weekly initial unemployment claims forward-filled to the trading
    calendar.
    \begin{itemize}
        \item \texttt{effr\_rate}: daily EFFR in decimal form (e.g., 0.0525 for 5.25\%).
        \item \texttt{effr\_rate\_1y\_pct\_change}: 252-day percentage change of EFFR.
        \item \texttt{effr\_rate\_1y\_percentile}: 252-day rolling percentile rank of EFFR.
        \item \texttt{jobless\_claims}: weekly initial unemployment claims (level).
        \item \texttt{jobless\_claims\_1y\_pct\_change}, \texttt{jobless\_claims\_1y\_percentile}: 252-day percentage change and percentile rank of the weekly claims series.
    \end{itemize}

    \subsection{Implied-volatility surface (CS)}
    Features derived from the SPXW volatility-surface snapshots taken at 09:35 ET, 10:00
    ET, and the official close. ATMF IV is the implied volatility of the put
    closest to the forward strike $F = S \cdot e^{rT}$. For each timestamp and DTE
    target $\tau$, the chain expiration closest to $\tau$ is selected and an
    IV-vs-delta interpolation is fitted, per-delta IV at $\delta \in \{10, 25\}$
    refers to the IV at $|\Delta_{\text{BSM}} \mp \delta/100| = 0$ for puts and
    calls respectively.

    \subsubsection*{ATM-forward IV levels and intraday levels}
    \begin{itemize}
        \item \texttt{atmf\_iv\_\{$\tau$\}dte\_close} for $\tau \in \{1, 5, 20, 30, 60, 90\}$: closing ATMF IV at the named DTE target.
        \item \texttt{atmf\_iv\_1dte\_1000}, \texttt{atmf\_iv\_5dte\_1000}: 10:00 ET ATMF IV at the 1-DTE and 5-DTE targets, used as the intraday entry-time anchors.
    \end{itemize}

    \subsubsection*{Per-delta IV levels}
    \begin{itemize}
        \item \texttt{call\_25d\_iv\_\{$\tau$\}dte\_close}, \texttt{put\_25d\_iv\_\{$\tau$\}dte\_close} for $\tau \in \{5, 30\}$: closing IV of the 25-delta call and put at the named DTE target.
    \end{itemize}

    \subsubsection*{IV percentile}
    \begin{itemize}
        \item \texttt{atmf\_iv\_\{$\tau$\}dte\_percentile\_252d} for $\tau \in \{1, 5, 30\}$: fraction of past 252 closes on which the ATMF IV at DTE $\tau$ was below today's value,
              \begin{equation*}
                  \mathrm{IVPct}_{252} = \frac{\#\left\{i : \sigma_{t-i} < \sigma_t,\; i = 1, \ldots, 252\right\}}{252}.
              \end{equation*}
    \end{itemize}

    \subsubsection*{Skew}
    With $\sigma_C^{\Delta}$, $\sigma_P^{\Delta}$, and $\sigma_{\text{ATM}}$ the IVs of the $\Delta$-delta call, $\Delta$-delta put, and ATMF option:
    \begin{itemize}
        \item \texttt{risk\_reversal\_delta25\_\{$\tau$\}dte\_close} for $\tau \in \{1, 5, 10, 20, 30\}$: 25-delta risk reversal $\sigma_C^{25} - \sigma_P^{25}$. A more negative value indicates a steeper downside skew (puts more expensive than equidistant calls).
        \item \texttt{risk\_reversal\_delta10\_\{$\tau$\}dte\_close} for $\tau \in \{0, 1, 5, 10, 20, 30\}$: 10-delta risk reversal $\sigma_C^{10} - \sigma_P^{10}$.
        \item \texttt{put\_skew\_delta25\_\{$\tau$\}dte\_close} for $\tau \in \{1, 5, 20, 30\}$: ATM-normalized put-side skew at 25-delta,
              \begin{equation*}
                  \mathrm{PutSkew}_{25} = \frac{\sigma_{\text{ATM}} - \sigma_P^{25}}{\sigma_{\text{ATM}}}.
              \end{equation*}
        \item \texttt{call\_skew\_delta25\_\{$\tau$\}dte\_close} for $\tau \in \{1, 5, 20, 30\}$: analogous call-side measure $(\sigma_{\text{ATM}} - \sigma_C^{25}) / \sigma_{\text{ATM}}$.
    \end{itemize}

    \subsubsection*{Forward implied volatility}
    Forward variance is extracted under the additivity assumption that total variance grows linearly with maturity \citep{Natenberg2015}. For two DTE targets $a < b$ with maturities $t_i = \mathrm{DTE}_i / 252$,
    \begin{equation*}
        \sigma_{a \to b}^{2} = \frac{\sigma^{2}(t_b) \cdot t_b - \sigma^{2}(t_a) \cdot t_a}{t_b - t_a}.
    \end{equation*}
    \begin{itemize}
        \item \texttt{atmf\_forward\_vol\_\{$a$\}d\_\{$b$\}d\_close} for $(a, b) \in \{(5, 10),\allowbreak (5, 30),\allowbreak (10, 20),\allowbreak (20, 30),\allowbreak (30, 60)\}$: forward volatility between the two DTE targets, computed from ATMF IV at each leg.
        \item \texttt{put\_25d\_forward\_vol\_\{$a$\}d\_\{$b$\}d\_close} for $(a, b) \in \{(5, 30),\allowbreak (10, 20)\}$: analogous forward volatility computed from 25-delta put IV at each leg, capturing the forward-volatility slope along the downside skew.
        \item \texttt{rv\_surprise\_5d}: realized-vs-implied surprise at the 5-day horizon, 
        \begin{equation*}
            \frac{\mathrm{RV}_{5} - \overline{\sigma}_{\text{ATMF}, 5\,\text{DTE}, 5\text{d}}}{\overline{\sigma}_{\text{ATMF}, 5\,\text{DTE}, 5\text{d}}}
        \end{equation*}
        where the comparator is the 5-day rolling mean of \texttt{atmf\_iv\_5dte\_close}. One of the three tail-risk features.
    \end{itemize}

    \subsubsection*{IV-to-spot change ratio}
    \begin{itemize}
        \item \texttt{iv\_to\_spot\_change\_ratio\_\{$\tau$\}dte\_close} for $\tau \in \{0, 1, 5,\allowbreak 10, 20,\allowbreak 30\}$: daily ratio of the percentage change in ATMF IV at DTE $\tau$ to the percentage change in $P_t$,
              \begin{equation*}
                  \mathrm{IVSpotRatio} = \frac{\Delta\%\, \sigma_{\text{ATM}, \tau}}{\Delta\%\, P_t}.
              \end{equation*}
              Strongly negative values are characteristic of the leverage effect (IV up when spot down).
    \end{itemize}

    \subsubsection*{Intraday ATMF IV dynamics}
    For $\tau \in \{0, 10, 30\}$:
    \begin{itemize}
        \item \texttt{atmf\_iv\_\{$\tau$\}dte\_intraday\_pct\_change}: close-vs-09:35 percentage change of ATMF IV at DTE $\tau$.
        \item \texttt{atmf\_iv\_\{$\tau$\}dte\_intraday\_range}: daily high-minus-low range of ATMF IV at DTE $\tau$.
    \end{itemize}

    \subsection{Realized-minus-implied volatility differentials (CS)}
    Derived purely from columns already on the index and volatility-surface frames.
    \begin{itemize}
        \item \texttt{rv\_iv\_spread\_5d}: \texttt{spx\_5d\_rv} $-$ \texttt{atmf\_iv\_5dte\_close}.
        \item \texttt{rv\_iv\_ratio\_5d}: \texttt{spx\_5d\_rv} $/$ \texttt{atmf\_iv\_5dte\_close}.
        \item \texttt{rv\_iv\_spread\_21d}, \texttt{rv\_iv\_ratio\_21d}: same construction at the 21-day RV vs 20-DTE ATMF IV pair (20 DTE is the closest grid match to 21 trading days).
        \item \texttt{iv\_term\_slope\_5d\_30d}: \texttt{atmf\_iv\_30dte\_close} $-$ \texttt{atmf\_iv\_5dte\_close}.
        \item \texttt{rv\_term\_slope\_5d\_21d}: \texttt{spx\_21d\_rv} $-$ \texttt{spx\_5d\_rv}.
        \item \texttt{iv\_minus\_rv\_term\_slope}: \texttt{iv\_term\_slope\_5d\_30d} $-$ \texttt{rv\_term\_slope\_5d\_21d}.
    \end{itemize}

    \subsection{Realized higher moments (CS)}
    Sample skewness and excess (Fisher) kurtosis of daily SPX log returns over
    rolling windows.
    \begin{itemize}
        \item \texttt{spx\_realized\_skew\_21d}, \texttt{spx\_realized\_kurtosis\_21d}: 21-day rolling skewness and excess kurtosis of $\ell_{1, t}$.
        \item \texttt{spx\_realized\_skew\_63d}, \texttt{spx\_realized\_kurtosis\_63d}: same statistics at the 63-day window.
    \end{itemize}

    \subsection{VIX term-structure curvature (CS)}
    Three convexity-style combinations of VIX-family levels.
    \begin{itemize}
        \item \texttt{vix\_curvature\_9d\_30d\_3m}: $\text{VIX9D} - 2 \cdot \text{VIX} + \text{VIX3M}$.
        \item \texttt{vix\_curvature\_1d\_9d\_30d}: $\text{VIX1D} - 2 \cdot \text{VIX9D} + \text{VIX}$.
        \item \texttt{vix\_curvature\_1d\_30d\_3m}: $\text{VIX1D} - 2 \cdot \text{VIX} + \text{VIX3M}$.
    \end{itemize}

    \subsection{Trend (CS)}
    Simple-moving-average distances and slope.
    \begin{itemize}
        \item \texttt{spx\_distance\_from\_50dma\_pct}: $(P_t - \mathrm{MA}_{50}) / \mathrm{MA}_{50}$.
        \item \texttt{spx\_distance\_from\_200dma\_pct}: $(P_t - \mathrm{MA}_{200}) / \mathrm{MA}_{200}$.
        \item \texttt{spx\_50\_200\_dma\_signal}: $\operatorname{sign}(\mathrm{MA}_{50} - \mathrm{MA}_{200})$, the classical golden / death-cross indicator.
        \item \texttt{spx\_200dma\_slope\_21d\_pct}: 21-day percentage change in the 200-day moving average.
    \end{itemize}

    \subsection{Position Greeks and exposures (PS)}
    Computed at entry from the Black--Scholes model with EFFR
    \citep{black_pricing_1973}.
    \begin{itemize}
        \item \texttt{delta}, \texttt{gamma}, \texttt{theta}, \texttt{vega}: the four Greeks of the candidate option at entry.
        \item \texttt{dollar\_delta}: $|\Delta| \cdot 100$, the dollar-delta exposure of one contract (each candidate corresponds to a one-contract position for label purposes).
        \item \texttt{gamma\_exposure}: $-|\Gamma| \cdot 100 \cdot S_{\text{entry}}$, where $S_{\text{entry}}$ is the SPX mid at the 10:00 ET entry. Captures the dollar P\&L per one-point move in the underlying for one short contract.
        \item \texttt{log\_moneyness}: $\ln(K / S_{\text{entry}})$.
        \item \texttt{leverage}: $|S_{\text{entry}} \cdot \Delta / P_{\text{entry}}|$, where $P_{\text{entry}}$ is the entry-mid premium of the option.
    \end{itemize}

    \subsection{Per-strategy rolling statistics (PS)}
    Statistics of the candidate strategy's own realized return-on-margin series,
    computed per strategy and lagged by one trading day. Returns are
    defined as
    \begin{equation*}
        \mathrm{ROM}_{t} = \frac{\mathrm{Gross\,P\&L}_t}{P_{\text{entry}, t}},
    \end{equation*}
    the gross P\&L of the trade scaled by the entry premium (notional-of-margin convention). On the \textit{SKIP} candidate ROM is defined to be zero.
    \begin{itemize}
        \item \texttt{returns\_on\_margin}: $\mathrm{ROM}_t$ on day $t$.
        \item \texttt{rom\_diff\_to\_spx}: $\mathrm{ROM}_t - r_{1d, t}^{\text{SPX}}$, the strategy's ROM in excess of the day's SPX simple return.
        \item \texttt{rom\_roll\_avg\_30d}, \texttt{rom\_roll\_std\_30d}, \texttt{rom\_roll\_skew\_30d}: 30-day rolling mean, sample standard deviation, and skewness of $\mathrm{ROM}$.
        \item \texttt{sharpe\_ratio\_rom\_60d}: 60-day rolling daily Sharpe ratio of $\mathrm{ROM}$, $\bar{\mathrm{ROM}}_{60} / \sigma_{\mathrm{ROM}, 60}$.
        \item \texttt{sharpe\_ratio\_rom\_std\_60d}: 60-day rolling sample standard deviation of \texttt{sharpe\_ratio\_rom\_60d} itself, a stability proxy for the rolling Sharpe.
        \item \texttt{drawdown\_\{N\}} for $N \in \{63, 126, 252\}$: rolling drawdown of the cumulative gross P\&L over a window of $N$ trading days, defined as $(\max_N C - C_t) / \max_N C$ with $C_t = \sum_{i \le t} \mathrm{Gross\,P\&L}_i$.
        \item \texttt{distance\_from\_max\_\{N\}} for $N \in \{63, 126, 252\}$: absolute distance $C_t - \max_N C$ in dollar terms.
        \item \texttt{win\_rate\_30d}: fraction of the trailing 30 days on which $\mathrm{ROM} > 0$.
        \item \texttt{stability\_coef\_60d}: squared Pearson correlation between $\mathrm{ROM}_t$ and a within-window time index over the past 60 days,
              \begin{equation*}
                  \mathrm{StabCoef}_{60} = \left[\operatorname{corr}\!\left(\mathrm{ROM}_t,\, t\right)\right]^2,
              \end{equation*}
              a smoothness proxy. Values close to one indicate steadily increasing ROM, values near zero indicate erratic returns.
        \item \texttt{tail\_ratio\_rom\_30d}: ratio of the upper to the lower tail of $\mathrm{ROM}$,
              \begin{equation*}
                  \mathrm{TailRatio}_{30} = \frac{\left|Q_{0.95}(\mathrm{ROM}_{30})\right|}{\left|Q_{0.05}(\mathrm{ROM}_{30})\right|}.
              \end{equation*}
        \item \texttt{prior\_day\_strategy\_drawdown\_max\_5d}: 5-day rolling drawdown of the candidate strategy's cumulative gross P\&L, lagged by one trading day so that the row dated $t$ reflects the drawdown observed at the close of $t-1$. One of the three tail-risk features.
        \item \texttt{iv\_at\_strike\_minus\_atmf\_wd\_rank}: within-day rank of \texttt{iv\_at\_strike\_minus\_atmf}.
        \item \texttt{returns\_on\_margin\_wd\_rank}: within-day rank of \texttt{returns\_on\_margin}.
        \item \texttt{sharpe\_ratio\_rom\_60d\_wd\_rank}: within-day rank of \texttt{sharpe\_ratio\_rom\_60d}.
        \item \texttt{drawdown\_63d\_wd\_rank}: within-day rank of \texttt{drawdown\_63d}.
    \end{itemize}

    \subsection{Entry liquidity (PS)}
    Captures the bid-ask cost and the size of the entry premium for the candidate
    option at the 10:00 ET snapshot.
    \begin{itemize}
        \item \texttt{entry\_bid\_ask\_spread}: $(\mathrm{ask} - \mathrm{bid})$ at entry, in dollars.
        \item \texttt{entry\_bid\_ask\_spread\_pct}: $(\mathrm{ask} - \mathrm{bid}) / \mathrm{mid}$ at entry.
        \item \texttt{entry\_log\_premium}: $\ln(\mathrm{mid}_{\text{entry}})$, the log of the entry-mid premium.
        \item \texttt{entry\_bid\_ask\_spread\_pct\_wd\_rank}: within-day rank of \texttt{entry\_bid\_ask\_spread\_pct}.
        \item \texttt{entry\_log\_premium\_wd\_rank}: within-day rank of \texttt{entry\_log\_premium}.
    \end{itemize}

    \subsection{Intra-strategy term context (PS)}
    Per-candidate features that locate the chosen strike on the surface relative to
    the ATM forward.
    \begin{itemize}
        \item \texttt{delta\_distance\_from\_target}: realized entry delta minus the strategy's target delta (signed).
        \item \texttt{iv\_at\_strike\_minus\_atmf}: IV of the selected strike at entry minus ATMF IV at the same DTE.
        \item \texttt{iv\_at\_strike\_minus\_atmf\_pct}: same difference relative to ATMF IV.
        \item \texttt{dte\_of\_selected\_option}: days to expiration of the chosen option (typically zero, with a one- or two-day fallback when no 0-DTE expiration exists).
    \end{itemize}

    \subsection{Regime-conditional sensitivities (PS)}
    Pairwise products of one PS source with one CS source, intended to let the
    model encode regime-conditional sensitivities.
    \begin{itemize}
        \item \texttt{delta\_x\_VIX}: $\Delta \cdot \text{VIX}$.
        \item \texttt{gamma\_x\_morning\_spx\_rv\_annualized}: $\Gamma \cdot \sigma_{\text{morn}}$.
        \item \texttt{vega\_x\_VIX\_5d\_pct\_change}: $\mathcal{V} \cdot \Delta_5 \% \text{VIX}$.
        \item \texttt{iv\_at\_strike\_minus\_atmf\_x\_VIX\_VIX9D\_spread}: strike-vs-ATMF IV gap multiplied by the VIX-minus-VIX9D spread.
        \item \texttt{theta\_x\_atmf\_iv\_5dte\_percentile\_252d}: $\Theta$ multiplied by the 5-DTE ATMF IV percentile.
        \item \texttt{delta\_distance\_from\_target\_x\_VVIX\_1y\_percentile}: delta-target gap multiplied by the VVIX 1-year percentile.
        \item \texttt{gamma\_x\_dte}: $\Gamma$ multiplied by DTE of the chosen option.
    \end{itemize}
    \FloatBarrier
    \clearpage

    \section{Year-by-year Sharpe heterogeneity}
    \label{sec:appendix_year}
    \setcounter{table}{0}
    \setcounter{figure}{0}

    Section~\ref{subsec:year_heterogeneity} discusses year-by-year Sharpe
    heterogeneity for the headline strategy. Table~\ref{tab:headline_year} reports
    the annualized Sharpe ratio for each of the seven sizing methods across the four
    walk-forward years (2021--2024) and the 2025 out-of-time hold-out, with the
    per-year and per-method values that the body summarizes by range and exception.

    \begin{table}[!ht]
	\centering
	\caption{Per-year Sharpe ratios across sizing methods.}
	\label{tab:headline_year}
	\begin{tabular}{lrrrrr}
		\toprule
		Method & 2021   & 2022   & 2023   & 2024      & 2025 (OOT) \\
		\midrule
		EA     & 6.5184 & 2.5066 & 1.7852 & 0.4074    & 5.7612     \\
		FMU    & 7.3126 & 1.5074 & 2.5067 & $-0.2873$ & 5.0632     \\
		SRS    & 7.3571 & 2.0167 & 2.6270 & $-0.2115$ & 5.2622     \\
		VT     & 7.5322 & 1.3736 & 2.8968 & 0.0462    & 4.6669     \\
		GB     & 9.2101 & 1.4893 & 3.9302 & $-0.1927$ & 5.0147     \\
		HK     & 7.3126 & 1.5266 & 0.0000 & $-0.5776$ & 4.3711     \\
		QK     & 7.3126 & 1.5266 & 0.0000 & $-0.6205$ & 4.3084     \\
		\bottomrule
	\end{tabular}
	\source{Each column reports the annualised Sharpe ratio of that calendar year's daily excess returns alone, 
    computed as the year's geometric annualised return divided by its annualised standard deviation. 
    The four 2021--2024 years are drawn from the four-window walk-forward, each year's training window expanding 
    through the prior calendar year, and 2025 is reserved as the out-of-time hold-out, never used during training, 
    model selection, or hyperparameter search. 
    The two fractional-Kelly variants, HK and QK, hold no position on any day of 2023, because the estimated Kelly 
    fraction stays below the abstention threshold throughout the year. The $0.0000$ recorded in that column therefore
     describes a year without positions rather than a year of trading that earned nothing.}
\end{table}

    \FloatBarrier
    \clearpage

    \section{Regime-conditional performance}
    \label{sec:appendix_regimes}
    \setcounter{table}{0}
    \setcounter{figure}{0}

    Section~\ref{subsec:regimes} reports cross-regime patterns for the headline
    strategy across two regime axes: VIX (forward-looking implied volatility,
    partitioned at 15 and 25) and realized volatility (backward-looking 5-day
    intraday, partitioned into equal-quantile terciles).
    Tables~\ref{tab:regime_vix} and~\ref{tab:regime_rv} report the underlying
    per-method, per-regime Sharpe values on the walk-forward and out-of-time slices
    that the body summarizes by per-axis range and monotonicity. Both tables cover
    every day in the regime, with abstention days entering at a return of zero, and
    report the day count and the trade count separately.

    {\scriptsize
\setlength{\tabcolsep}{3pt}
\renewcommand{\arraystretch}{0.85}
\begin{longtable}{lrrrrrrr}
\caption{Headline performance by VIX regime.}\label{tab:regime_vix}\\
\toprule
Regime & Method & Split & Sharpe & Ann.\ return & Ann.\ vol.\ & Days & Trades \\
\midrule
\endfirsthead
\multicolumn{8}{l}{\textit{Table~\ref{tab:regime_vix} (continued)}}\\
\toprule
Regime & Method & Split & Sharpe & Ann.\ return & Ann.\ vol.\ & Days & Trades \\
\midrule
\endhead
\midrule
\multicolumn{8}{r}{\textit{Continued on next page}}\\
\endfoot
\bottomrule
\endlastfoot
low & EA & WF & 0.2611 & 0.0050 & 0.0191 & 223 & 61 \\
mid & EA & WF & 4.2354 & 0.1390 & 0.0328 & 593 & 337 \\
high & EA & WF & 2.7675 & 0.1540 & 0.0557 & 148 & 53 \\
low & EA & OOT & 13.7060 & 0.0549 & 0.0040 & 21 & 15 \\
mid & EA & OOT & 4.9387 & 0.0912 & 0.0185 & 196 & 187 \\
high & EA & OOT & 14.4589 & 0.3100 & 0.0214 & 20 & 18 \\
\midrule
low & FMU & WF & 2.6184 & 0.0417 & 0.0159 & 223 & 68 \\
mid & FMU & WF & 4.1695 & 0.1901 & 0.0456 & 593 & 355 \\
high & FMU & WF & 0.5503 & 0.0527 & 0.0958 & 148 & 56 \\
low & FMU & OOT & 85.6325 & 0.1469 & 0.0017 & 21 & 21 \\
mid & FMU & OOT & 5.6371 & 0.1578 & 0.0280 & 196 & 196 \\
high & FMU & OOT & 5.3274 & 0.4568 & 0.0857 & 20 & 20 \\
\midrule
low & SRS & WF & 1.7529 & 0.0319 & 0.0182 & 223 & 68 \\
mid & SRS & WF & 4.4101 & 0.1763 & 0.0400 & 593 & 355 \\
high & SRS & WF & 1.0494 & 0.0867 & 0.0826 & 148 & 56 \\
low & SRS & OOT & 63.9948 & 0.1258 & 0.0020 & 21 & 21 \\
mid & SRS & OOT & 5.8867 & 0.1454 & 0.0247 & 196 & 196 \\
high & SRS & OOT & 6.9412 & 0.5862 & 0.0845 & 20 & 20 \\
\midrule
low & VT & WF & 5.4640 & 0.0726 & 0.0133 & 223 & 64 \\
mid & VT & WF & 3.8811 & 0.2084 & 0.0537 & 593 & 349 \\
high & VT & WF & 0.2859 & 0.0380 & 0.1329 & 148 & 54 \\
low & VT & OOT & 88.7880 & 0.2575 & 0.0029 & 21 & 21 \\
mid & VT & OOT & 5.5988 & 0.2621 & 0.0468 & 196 & 193 \\
high & VT & OOT & 0.1657 & 0.0164 & 0.0993 & 20 & 20 \\
\midrule
low & GB & WF & 6.6201 & 0.0535 & 0.0081 & 223 & 68 \\
mid & GB & WF & 4.4769 & 0.1944 & 0.0434 & 593 & 355 \\
high & GB & WF & 0.3317 & 0.0481 & 0.1451 & 148 & 56 \\
low & GB & OOT & 83.2863 & 0.2378 & 0.0029 & 21 & 21 \\
mid & GB & OOT & 5.5620 & 0.2632 & 0.0473 & 196 & 196 \\
high & GB & OOT & 5.5906 & 0.8065 & 0.1443 & 20 & 20 \\
\midrule
low & HK & WF & 1.8744 & 0.0366 & 0.0195 & 223 & 55 \\
mid & HK & WF & 3.6559 & 0.1725 & 0.0472 & 593 & 302 \\
high & HK & WF & 0.3839 & 0.0473 & 0.1232 & 148 & 55 \\
low & HK & OOT & 55.5100 & 0.1898 & 0.0034 & 21 & 20 \\
mid & HK & OOT & 4.6566 & 0.1756 & 0.0377 & 196 & 175 \\
high & HK & OOT & 5.3496 & 0.6196 & 0.1158 & 20 & 20 \\
\midrule
low & QK & WF & 1.8152 & 0.0377 & 0.0208 & 223 & 55 \\
mid & QK & WF & 3.5873 & 0.1712 & 0.0477 & 593 & 302 \\
high & QK & WF & 0.3349 & 0.0419 & 0.1250 & 148 & 55 \\
low & QK & OOT & 55.3857 & 0.1846 & 0.0033 & 21 & 20 \\
mid & QK & OOT & 4.8011 & 0.1684 & 0.0351 & 196 & 174 \\
high & QK & OOT & 4.8847 & 0.5697 & 0.1166 & 20 & 20 \\
\end{longtable}
\source{Sharpe ratios are computed on the daily returns of every day whose VIX close falls in the regime, with days on which the model abstained or the confidence gate blocked entry entering at a return of zero. The Days column counts those days and the Trades column counts how many of them carried a position, so the gap between the two is the abstention rate inside the regime. Returns are computed once on the ordered slice and attributed to the day they were earned on, so the three regime cells partition the slice. VIX is measured at the entry-day close, which is after the 10:00 ET entry, so the classification uses information the strategy itself never sees. Regime thresholds follow the convention $\mathrm{VIX} < 15$ (low), $15 \leq \mathrm{VIX} \leq 25$ (mid), and $\mathrm{VIX} > 25$ (high). The walk-forward (WF) slice covers 2021--2024 across the four expanding-window training cycles. The 2025 out-of-time (OOT) slice covers the hold-out year, never used during training, model selection, or hyperparameter search. Cells resting on fewer than about thirty days carry no useful precision and are reported for completeness rather than interpreted, as Section~\ref{subsec:regimes} sets out.}
}

    {\scriptsize
\setlength{\tabcolsep}{3pt}
\renewcommand{\arraystretch}{0.85}
\begin{longtable}{lrrrrrrr}
\caption{Headline performance by realized-volatility tercile.}\label{tab:regime_rv}\\
\toprule
Tercile & Method & Split & Sharpe & Ann.\ return & Ann.\ vol.\ & Days & Trades \\
\midrule
\endfirsthead
\multicolumn{8}{l}{\textit{Table~\ref{tab:regime_rv} (continued)}}\\
\toprule
Tercile & Method & Split & Sharpe & Ann.\ return & Ann.\ vol.\ & Days & Trades \\
\midrule
\endhead
\midrule
\multicolumn{8}{r}{\textit{Continued on next page}}\\
\endfoot
\bottomrule
\endlastfoot
1 & EA & WF & 2.7711 & 0.0965 & 0.0348 & 322 & 167 \\
2 & EA & WF & 2.9750 & 0.0886 & 0.0298 & 320 & 147 \\
3 & EA & WF & 3.5503 & 0.1416 & 0.0399 & 322 & 137 \\
1 & EA & OOT & 1.5499 & 0.0411 & 0.0265 & 80 & 77 \\
2 & EA & OOT & 13.9158 & 0.1025 & 0.0074 & 79 & 72 \\
3 & EA & OOT & 12.2625 & 0.1767 & 0.0144 & 78 & 71 \\
\midrule
1 & FMU & WF & 2.6665 & 0.1021 & 0.0383 & 322 & 176 \\
2 & FMU & WF & 3.6109 & 0.1273 & 0.0352 & 320 & 157 \\
3 & FMU & WF & 2.2680 & 0.1689 & 0.0745 & 322 & 146 \\
1 & FMU & OOT & 4.8249 & 0.1032 & 0.0214 & 80 & 80 \\
2 & FMU & OOT & 3.1131 & 0.1159 & 0.0372 & 79 & 79 \\
3 & FMU & OOT & 7.7448 & 0.3362 & 0.0434 & 78 & 78 \\
\midrule
1 & SRS & WF & 2.4899 & 0.0838 & 0.0336 & 322 & 176 \\
2 & SRS & WF & 3.8341 & 0.1168 & 0.0305 & 320 & 157 \\
3 & SRS & WF & 2.8145 & 0.1838 & 0.0653 & 322 & 146 \\
1 & SRS & OOT & 4.8857 & 0.0864 & 0.0177 & 80 & 80 \\
2 & SRS & OOT & 3.2211 & 0.1052 & 0.0327 & 79 & 79 \\
3 & SRS & OOT & 8.2180 & 0.3567 & 0.0434 & 78 & 78 \\
\midrule
1 & VT & WF & 2.8840 & 0.1353 & 0.0469 & 322 & 171 \\
2 & VT & WF & 5.2927 & 0.1786 & 0.0337 & 320 & 153 \\
3 & VT & WF & 1.3061 & 0.1322 & 0.1012 & 322 & 143 \\
1 & VT & OOT & 5.0290 & 0.1787 & 0.0355 & 80 & 80 \\
2 & VT & OOT & 3.3516 & 0.2135 & 0.0637 & 79 & 78 \\
3 & VT & OOT & 6.5120 & 0.3312 & 0.0509 & 78 & 76 \\
\midrule
1 & GB & WF & 3.6686 & 0.0917 & 0.0250 & 322 & 176 \\
2 & GB & WF & 6.3623 & 0.1657 & 0.0260 & 320 & 157 \\
3 & GB & WF & 1.4276 & 0.1558 & 0.1091 & 322 & 146 \\
1 & GB & OOT & 4.7558 & 0.1685 & 0.0354 & 80 & 80 \\
2 & GB & OOT & 2.9237 & 0.1872 & 0.0640 & 79 & 79 \\
3 & GB & OOT & 8.1170 & 0.5881 & 0.0725 & 78 & 78 \\
\midrule
1 & HK & WF & 2.2623 & 0.0901 & 0.0398 & 322 & 160 \\
2 & HK & WF & 3.7739 & 0.1232 & 0.0326 & 320 & 127 \\
3 & HK & WF & 1.5791 & 0.1474 & 0.0933 & 322 & 125 \\
1 & HK & OOT & 4.5273 & 0.1289 & 0.0285 & 80 & 77 \\
2 & HK & OOT & 2.8110 & 0.1416 & 0.0504 & 79 & 73 \\
3 & HK & OOT & 6.3287 & 0.3752 & 0.0593 & 78 & 65 \\
\midrule
1 & QK & WF & 2.3980 & 0.0948 & 0.0396 & 322 & 160 \\
2 & QK & WF & 3.6736 & 0.1216 & 0.0331 & 320 & 127 \\
3 & QK & WF & 1.4710 & 0.1398 & 0.0951 & 322 & 125 \\
1 & QK & OOT & 8.1524 & 0.1469 & 0.0180 & 80 & 77 \\
2 & QK & OOT & 2.5104 & 0.1271 & 0.0506 & 79 & 72 \\
3 & QK & OOT & 5.6502 & 0.3370 & 0.0596 & 78 & 65 \\
\end{longtable}
\source{Sharpe ratios are computed on the daily returns of every day in the tercile, with days on which the model abstained or the confidence gate blocked entry entering at a return of zero. The Days column counts those days and the Trades column counts how many of them carried a position. Returns are computed once on the ordered slice and attributed to the day they were earned on, so the three tercile cells partition the slice. Realized volatility is the 5-day intraday realized volatility of the S\&P 500, measured at the entry-day close, partitioned into terciles at the $1/3$ and $2/3$ quantiles. The entry-day close falls after the 10:00 ET entry, so the classification uses information the strategy itself never sees. The cut points are computed within each slice separately, so a given tercile label spans different realized-volatility levels on the walk-forward and out-of-time panels. The walk-forward (WF) slice covers 2021--2024. The 2025 out-of-time (OOT) slice covers the hold-out year, never used during training, model selection, or hyperparameter search.}
}

    \FloatBarrier
    \clearpage

    \section{Confidence-gate calibration}
    \label{sec:appendix_gate}
    \setcounter{table}{0}
    \setcounter{figure}{0}

    Section~\ref{subsec:gate} describes the confidence gate, a per-window
    calibrated abstention threshold $\tau$ that suppresses the day's position when
    the model's top-1-versus-top-2 score gap is below $\tau^{\star}$. Table~\ref{tab:gate}
    reports the calibrated $\tau^{\star}$ value and the implied trade rate for each of the
four walk-forward training cycles (2021--2024) plus the 2025 out-of-time slice
and the union calibration over all walk-forward held-out slices. The values
show a threshold that is re-fit rather than fixed, with the calibrated trade
rate spanning $30.6\%$ to $100\%$ across windows.

\begin{table}[!ht]
\centering
\caption{Confidence gate calibration per window.}
\label{tab:gate}
\begin{tabular}{lrrl}
\toprule
Window & $\tau^{\star}$ & Calibrated trade rate & Calibration metric \\
\midrule
2021 & 0.0000 & 1.0000 & Sortino \\
2022 & 0.2013 & 0.5041 & Sortino \\
2023 & 0.1613 & 0.3058 & Sortino \\
2024 & 0.0137 & 0.4000 & Sortino \\
OOT 2025 & 0.0000 & 1.0000 & Sortino \\
WF held-out union & 0.0000 & 1.0000 & Sortino \\
\bottomrule
\end{tabular}
\source{Per-window calibration of the confidence gate's abstention threshold $\tau^{\star}$. The threshold is selected on the last six months of each training window, held out from feature engineering and ranker training, by a one-dimensional grid search that maximizes the Sortino ratio on the held-out slice. At inference time, days whose confidence signal (top-1 minus top-2 candidate scores) falls below $\tau^{\star}$ produce zero positions for the day. The calibrated trade rate column reports the fraction of held-out days whose confidence signal exceeds $\tau^{\star}$. The final row reports the calibration on the union of the four walk-forward held-out slices, which produced the threshold applied to the 2025 out-of-time inference. The OOT 2025 row reports that threshold and the trade rate it implies on the 2025 slice.}
\end{table}

\FloatBarrier
\clearpage

\section{Feature survival across selection windows}
\label{sec:appendix_features}
\setcounter{table}{0}
\setcounter{figure}{0}

Section~\ref{subsec:features} discusses feature survival across the four
walk-forward training cycles and the out-of-time cycle, with the pipeline of
Algorithm~\ref{alg:feature_selection} run independently in each window.
Figure~\ref{fig:heatmap} shows the per-window survival pattern for the top 60
candidate features as a heatmap, and the 27 features selected in every window
appear as the contiguous block of always-on cells.

\begin{figure}[!ht]
    \centering
    \caption{Per-window feature survival heatmap.}
    \label{fig:heatmap}
    \includegraphics[width=\textwidth,height=0.62\textheight,keepaspectratio]{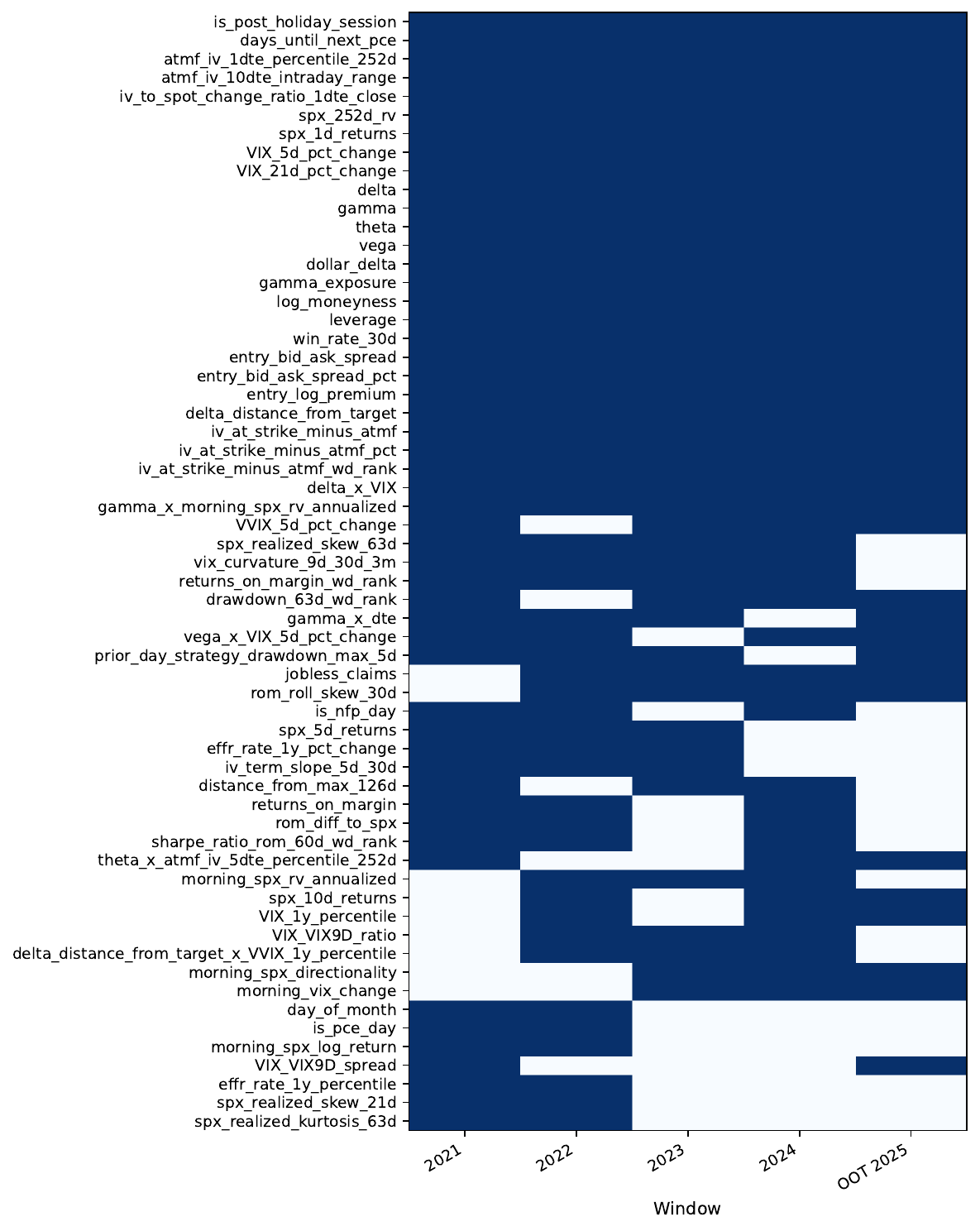}
    \source{Per-window survival pattern across the top 60 candidate features (rows) and the five selection windows (columns: the four walk-forward windows 2021--2024, and the 2025 out-of-time window). A filled cell indicates that the feature survives the selection pipeline in that window. The 27 always-surviving features appear as the contiguous block of always-on cells.}
\end{figure}

\endgroup

\end{document}